\documentclass[aps,nofootinbib,showpacs,preprintnumbers,amsmath,amssymb]{revtex4}
\def\be{\begin{equation}}
\def\ee{\end{equation}}
\def\ba{\begin{eqnarray}}
\def\ea{\end{eqnarray}}
\newcommand{\A}{{\mathcal{A}}}
\newcommand{\tA}{{\widetilde {\mathcal{A}}}}
\newcommand{\ta}{{\widetilde a}}
\newcommand{\tlA}{{\mathfrak A}}
\usepackage{graphics}
\usepackage{graphicx}% Include figure files
\usepackage{dcolumn}% Align table columns on decimal point
\usepackage{bm}% bold math
\usepackage{epsfig}
\usepackage{graphicx,color}
\usepackage{multirow}
\begin{document}

\preprint{USM-TH-267}

\title{Reconciling the analytic QCD with the ITEP operator product expansion philosophy}

\author{Gorazd Cveti\v{c}}
 \email{gorazd.cvetic@usm.cl}
\affiliation{Department of Physics and Valparaiso Center for Science and Technology, Universidad T{\'e}cnica Federico Santa Mar{\'\i}a, Valpara{\'\i}so, Chile}

\author{Reinhart K\"ogerler}
\email{koeg@physik.uni-bielefeld.de}
\affiliation{Department of Physics, Universit\"at Bielefeld, 33501 Bielefeld, Germany}

\author{Cristi\'an Valenzuela}
 \email{cvalenzuela@fis.puc.cl}
\affiliation{Department of Physics, Pontificia Universidad Cat\'olica de Chile, Santiago 22, Chile}

\date{\today}

\begin{abstract}
Analytic QCD models are those versions of QCD in which the
running coupling parameter $a(Q^2)$ has the same analytic properties
as the spacelike physical quantities, i.e., no singularities in the 
complex $Q^2$ plane except on the timelike semiaxis. In such models, 
$a(Q^2)$ usually differs from its perturbative analog by power terms 
$\sim (\Lambda^2/Q^2)^k$ for large momenta, introducing thus nonperturbative terms 
$\sim (\Lambda^2/Q^2)^k$ in spacelike physical quantities 
whose origin is the UV regime. Consequently, it contradicts the 
ITEP operator product expansion philosophy which states that such terms can come only from the
IR regimes. We investigate whether it is possible to construct analytic 
QCD models which respect the aforementioned ITEP philosophy and, at the same time, 
reproduce not just the high-energy QCD observables, but also the low-energy 
ones, among them the well-measured semihadronic $\tau$ decay ratio.
\end{abstract}
\pacs{12.38.Cy, 12.38.Aw,12.40.Vv}

\maketitle

\section{Introduction}
\label{sec:intr}

Today one of the main goals in strong interaction theory
is to technically enlarge the applicability of QCD to processes involving
lower momentum transfer $q^2$. Thereby several
obstacles have to be overcome. One of them is
that the running QCD coupling $a (Q^2) ={\alpha_s (Q^2)}/{\pi}$,
when calculated within the perturbative (``pt'') renormalization group 
formalism (we call it $a_{\rm pt}$), 
in the usual (``perturbative'') renormalization schemes,
yields singularities of $a_{\rm pt}(Q^2)$ at $Q^2>0$, usually called
Landau singularities. Consequently, spacelike
observables expressed in terms of powers of $a_{\rm pt}(Q^2)$ obtain singularities 
on the spacelike semiaxis $0 \le Q^2 \le \Lambda^2$
($Q^2 = -q^2$, with $q$ denoting the typical momentum transfer within
a given physical process or quantity). This is not acceptable due to
general principles of local quantum field theory \cite{BS}.
Furthermore, studies of ghost-gluon vertex and gluon self-energy 
using Schwinger-Dyson equations \cite{SDEs} and 
large-volume lattice calculations \cite{latt}, result in 
QCD coupling $a(Q^2)$ without Landau singularities
at $Q^2>0$ and even with a finite value at $Q=0$.
Consequently, the behavior of the coupling $a(Q^2)$ at low values of 
$Q^2$ should be corrected relative to that given by perturbative reasoning.

Several attempts at achieving such corrections have been recorded during
the last 14 years starting from 
(what we call) the minimal analytic (MA) QCD of
Shirkov and Solovtsov \cite{ShS}. Here, the trick lay in simply
omitting the wrong (spacelike) part of the branch cut within the
dispersion relation formula for $a (Q^2)$. Consequently, the resulting
analytized coupling $\A_1^{\rm (MA)}(Q^2) \equiv a^{\rm (MA)} (Q^2)$ is analytic
in the whole Euclidean part of the $Q^2$ plane except the nonpositive
semiaxis: $Q^2 \in \mathbb{C} \backslash (-\infty, 0]$.
Furthermore, for evaluation of physical observables which are represented,
in ordinary perturbation theory, as a (truncated) series of powers
of $a_{\rm pt}(Q^2)$, one also has to extend the analytization procedure
to $a_{\rm pt}^n$ ($n \geq 2$). In MA this was performed in Ref.~\cite{MSS}
(see also Ref.~\cite{Sh}) and resulted in the replacement of $a_{\rm pt}^n$
by nonpower expressions $\A_n^{\rm (MA)}(Q^2)$. This specific procedure was
dubbed by the authors of \cite{MSS,Sh} analytic perturbation theory (APT);
whereas we will refer to it generally as minimal analytic (MA) QCD.

Other analytic models for $a (Q^2)$ satisfy certain 
different or additional constraints 
at low and/or at high $Q^2$ 
\cite{Webber:1998um,Srivastava:2001ts,Simonov,Nesterenko,Nesterenko2,Alekseev:2005he,CV1,CV2,CCEM}.
Analytic QCD models have been used also in
the physics of mesons \cite{mes1,mes2} within the
Bethe-Salpeter approach, and in calculation of analytic
analogs of noninteger powers $a_{\rm pt}^{\nu}$ \cite{Bakulev} within the
MA model (for reviews of various analytic QCD models, and further references, see Refs.~\cite{Prosperi:2006hx,Shirkov:2006gv,Cvetic:2008bn}). 
We note that the MA couplings $\A_n^{\rm (MA)}$ ($n \geq 1$) defined 
here are the MA couplings of Refs.~\cite{ShS,Sh,Shirkov:2006gv} 
divided by $\pi$. 

All of these versions of analytic QCD have one common feature:
their (analytized) coupling $a(Q^2)$ differs from
the perturbative coupling even at higher energies by a power term:
\be
\vert \delta a(Q^2) \vert \equiv \vert a(Q^2) - a_{\rm pt} (Q^2) \vert 
\sim ( {\Lambda^2}/{Q^2} )^k \qquad (Q^2 \gg \Lambda^2) \ ,
\label{powerk}
\ee
where $k$ is a positive integer (usually $k=1$;
for the models of Refs.~\cite{Alekseev:2005he,CCEM}: $k=3$).
How can these power corrections be interpreted? 
In a given (usual) renormalization scheme, where $a_{\rm pt}(Q^2)$ has (Landau) 
singularities on the positive axis $Q^2 \sim \Lambda^2 (\sim 0.1 \ {\rm GeV}^2) >0$, 
analytization of $a_{\rm pt}(Q^2)$ can be understood to be achieved by a 
modification of the discontinuity (``spectral'') function
$\rho_1^{\rm pt}(\sigma) \equiv {\rm Im} a_{\rm pt}(Q^2=-\sigma - i \epsilon)$
at energies $|\sigma| \alt \Lambda^2$, 
%and by requiring $\rho_1 = 0$ for $\sigma < 0$ (i.e., for $Q^2 > 0$), 
thereby subtracting
the Landau singularities from $a_{\rm pt}(Q^2)$. It is this subtraction,
in the given renormalization scheme,
which leads to the power deviations Eq.~(\ref{powerk}) and, as a
consequence, to terms $\sim (\Lambda^2/Q^2)^k$ in all spacelike
physical quantities. But such contributions are definitely 
of nonperturbative origin,
since they are proportional to $\exp(-K/a_{\rm pt}(Q^2))$ which is
nonanalytic at $a_{\rm pt}=0$ [cf.~Eq.~(\ref{powt1}) in Sec.~\ref{sec:beta1}].

Whether such terms, produced in spacelike observables ${\cal D}(Q^2)$,
can be interpreted as being of ultraviolet (UV) origin or not, 
is not entirely clear. Interpretations of such terms in the literature 
differ from each other. For example, Ref.~\cite{Shirkov:1999hm} suggests
that the Landau pole is not of (entirely) UV origin because the Landau
pole persists in the renormalization group resummed expression for
$a_{\rm pt}(Q^2)$ even if one uses, instead of UV logs, the mass-dependent
polarization expression (with a sufficiently small gluon mass). 
On the other hand, the authors of Ref.~\cite{DMW} argue that
the aforementioned terms $\sim (\Lambda^2/Q^2)^k$ are of UV origin due
to the following consideration: 
If one considers the leading-$\beta_0$ summation of an inclusive
spacelike observable ${\cal D}(Q^2)$ (cf.~Appendix \ref{app:LB})
\be
{\cal D}^{\rm (LB)}(Q^2) \equiv 
\int_0^\infty \frac{dt}{t}\:  F_{\cal D}(t) \: 
a(t Q^2 e^{\overline {\cal C}}) \ ,
\label{LBint}
\ee
where $F_{\cal D}(t)$ is a characteristic function of the
observable and ${\overline {\cal C}}=-5/3$, 
then the quantity $t Q^2 e^{\overline {\cal C}}$
indicates the magnitude of the (squares of) internal loop momenta
appearing in the resummation. In the UV regime of these momenta,
e.g., for $t > 1$ (see also Ref.~\cite{Cvetic:2007ad}), 
the deviation (\ref{powerk}) then leads to
power terms of apparently UV origin in the observable
\be
\delta  {\cal D}^{\rm (LB)}(Q^2) \sim 
({\Lambda^2}/{Q^2} )^k \int_1^\infty \frac{dt}{t^{k+1}}\:  F_{\cal D}(t)
\sim ({\Lambda^2}/{Q^2} )^k \ .
\label{powerk2}
\ee

Considering all these arguments, we come to the conclusion that the
aforementioned $(\Lambda^2/Q^2)^k$ contributions in physical quantities
are at least partially due to UV effects. The existence of nonperturbative
contributions stemming from the UV regime is not in accordance
with the operator product expansion (OPE) philosophy as advocated by
the ITEP group \cite{Shifman:1978bx,DMW}. 
This philosophy rests on the assumption that the
OPE, which has originally been derived in perturbation theory (PT), is
valid in general (i.e., even when including the nonperturbative
contributions) and consequently allows for a separation of
short-range from long-range contributions to (inclusive)
QCD observables. While the short-range contributions can be
calculated perturbatively and lead to expressions for the
OPE coefficient functions, the long-range contributions
show up as matrix elements of local operators
and can be parametrized in terms of condensates (not accessible by PT).
And it is this long-range part which leads to power corrections
reflecting the contributions of nonperturbative origin to the
observable. Therefore, according to the ITEP interpretation, 
the power term corrections stem from the IR region. 
This ITEP-OPE approach rests on intuitive physical arguments, 
and has led to the success of QCD sum rules. 

In this work we will adopt the aforementioned ITEP philosophy when
analytizing perturbative QCD and,
consequently, we will request that the analytic coupling parameter 
$\A_1(Q^2) \equiv a(Q^2)$ differ
from the usual perturbative one at high $Q^2$ by less than any
power of $\Lambda^2/Q^2$.

We wish to stress, however, that there is nothing in quantum field
theory (QFT) that would impose on us the ITEP interpretation of the OPE. In
this context, we mention that the essential singularity at $a=0$
[such as $\exp(-K/a)$] has quite a general and mysterious genesis  -
first mentioned in QFT by Dyson \cite{Dyson1952} on specific
physical grounds, and later by many authors on more formal grounds
(for an overview, see \cite{KSh1980} and references therein).

An additional feature of most versions of analytized QCD is that they
fail to reproduce the correct value for the most important 
(since most reliably measured) QCD
observable at low energies, namely $r_\tau$, the strangeless semihadronic $\tau$
decay ratio, whose present-day experimental
value is (cf.~Appendix \ref{app:rtauexp}): $r_\tau({\rm exp.}) = 0.203 \pm 0.004$.
Most of the analytic QCD models are either unable to predict 
unambiguously $r_{\tau}$ value, or they predict significantly smaller values
(e.g., in MA, Ref.~\cite{MSS,MSSY}), unless unusual additional assumptions 
are made, e.g., in MA that the light quark masses are much higher than 
the values of their current masses \cite{Milton:2001mq}.\hfill\break
This finding (loss in the size of $r_\tau$) in MA appears to be connected 
with the elimination of the unphysical (Euclidean) part 
of the branch cut contribution
of perturbative QCD. Since $r_\tau$ is the most precisely measured
inclusive low momentum QCD observable, its reproduction in analytic QCD
models is of high importance. The apparent failure of the MA model with
light quark current masses to reproduce the correct value of $r_{\tau}$ had 
even led to the suggestion 
that the analytic QCD should be abandoned \cite{Geshkenbein:2001mn}.

Here, we are investigating whether a modified version of QCD can be defined
which simultaneously fulfills the following requirements:
\begin{itemize}
\item[(i)] It is compatible with all analyticity requirements of Quantum
Field Theory. In particular, it must not lead to Landau singularities 
of $a(Q^2)$, and furthermore 
we expect (see Sec.~\ref{sec:beta1}) that $a (Q^2)$ is analytic at
$Q^2 = 0$, and thus IR finite, with $a (Q^2 = 0) \equiv a_0 < \infty$.
\item[(ii)] It is in accordance with the ITEP-OPE philosophy which means
that the UV behavior of $a (Q^2)$ is such that 
$\vert a (Q^2) - a_{\rm pt} (Q^2) \vert < (\Lambda^2/Q^2)^k$ for any integer
$k$ at large $Q^2$.
\item[(iii)] The theory reproduces the experimental values for $r_\tau$
(and other low energetic observables, e.g. the Bjorken polarized sum rule
at low $Q^2$).
\end{itemize}
We will show that such a theory is attainable, but only at a certain
(acceptable, we think) price. Some of the main results of the present
work have been presented, in a summarized form, in Ref.~\cite{CKV1}.

We are approaching our aim in an indirect way, namely by properly 
modifying the $\beta$ function $\beta (x)$ $[x=a(Q^2)]$ of QCD. This
approach, which has been used first by R\c{a}czka \cite{Raczka}
in a somewhat different context,
means that the starting point in the construction is
the beta function $\beta(a)$, rather than the coupling parameter 
$a(Q^2)$ itself or its discontinuity function $\rho_1(\sigma) =
{\rm Im} a(Q^2=-\sigma - i \epsilon)$. 
The ITEP-OPE condition can be implemented in such an approach in a
particularly simple way (see below).
Consequently, we are trying to augment
$\beta (a)$ which, in general, is only specified by its perturbation series
around the point $a=0$
\be
\beta (a) = - \beta_0 a^2( 1 + c_1 a + c_2 a^2 + c_3 a^3 +{\cal O}(a^4) ) \ ,
\label{btexp1}
\ee
where $\beta_0$ and $c_1=\beta_1/\beta_0$ are two universal constants.
This should be done in such a way that the augmented beta function leads 
(via the renormalization group equation RGE) to an effective analytic coupling
$a(Q^2)$ which also enables the correct evaluation of 
low-energy QCD observables in a perturbative way.

The abovementioned requirements for $a (Q^2)$ imply the following 
constraints on the modified beta-function $\beta (a)$:
\begin{itemize}
\item[(1)] The $\beta$ function must be such that the RGE
gives a running coupling $a(Q^2)$ analytic in the entire complex 
plane of $Q^2$, with the possible exception of the nonpositive semiaxis:
$Q^2  \in \mathbb{C} \backslash (-\infty, 0]$.
\item[(2)] For small $|a|$, ${\beta} (a)$ has Taylor expansion (\ref{btexp1}) in
powers of $a$, i.e., the perturbative QCD (pQCD) behavior of $\beta(a)$, 
with universal $\beta_0$ and $c_1$, at high $Q^2$ is attained.
\item[(3)] ${\beta} (a)$ is an analytic (holomorphic) function 
of $a$ at $a = 0$ in order to ensure 
$\vert a(Q^2) - a_{\rm pt} (Q^2) \vert < (\Lambda^2/Q^2)^k$
for any $k > 0$ at large $Q^2$ (see Sec.~\ref{sec:beta1}), 
thus respecting the ITEP-OPE postulate that
powerlike corrections can only be IR induced.
At high $Q^2$, those pQCD
values $a_{\rm pt}(Q^2)$ which reproduce the known high-energy QCD
phenomenology are attained by $a(Q^2)$.  
\item[(4)] It turns out to be difficult or impossible to
achieve analyticity (holomorphy) of $a(Q^2)$ in the Euclidean complex
plane $Q^2  \in \mathbb{C} \backslash (-\infty, 0]$ unless the point
$Q^2=0$ is also included as a point of analyticity of $a(Q^2)$.
This then implies that $a(Q^2) \to a_0$ when $Q^2 \to 0$, where
$a_0$ is finite positive, and that $\beta(a)$ has Taylor expansion
around $a=a_0$ with Taylor coefficient at the first term being unity:
${\beta} (a) = (a - a_0) + {\cal O}((a-a_0)^2)$. Then, $\beta(a)$
is a nonsingular unambiguous function of $a$ in the positive
interval $a \in [0,a_0]$. Note that analyticity of $a(Q^2)$ at $Q^2=0$
is in full accordance with the general requirement that hadronic
transition amplitudes have only the singularities which are
enforced by unitarity.
\end{itemize}

We proceed in this work in the following way.
In Sec.~\ref{sec:beta1} we construct various classes of beta functions
which give analytic $a(Q^2)$ at all $Q^2 \in \mathbb{C} \backslash
(- \infty,0)$ and fulfill the ITEP-OPE condition. We relegate to 
Appendix \ref{app:impl} details of the analytic expressions
for the implicit solution of RGE and their implications for the
(non)analyticity of $a(Q^2)$. In Sec.~\ref{sec:beta2} we point out
the persistent problem of such models giving too low values of $r_{\tau}$.
In Sec.~\ref{sec:beta3} we present further modification of the 
aforementioned beta functions, such that, in addition, the correct value 
of $r_{\tau}$ is reproduced. In Appendix \ref{app:rtauexp} we present
the extraction of the massless and strangeless $r_{\tau}$ value from
experimental data.
We relegate to Appendixes \ref{app:hoanQCD}, \ref{app:LB} and
\ref{app:bLB} the presentation of formalisms for the evaluation, 
in any analytic QCD (anQCD) model, of massless observables, 
such as the Bjorken polarized sum rule (BjPSR), the Adler function and 
the related $r_{\tau}$. Appendix \ref{app:hoanQCD} presents
construction of the higher order anQCD couplings;
Appendix \ref{app:LB} presents a formalism of resummation
of the leading-$\beta_0$ (LB) contributions in anQCD; Appendix \ref{app:bLB}
presents a calculation of the beyond-the-leading-$\beta_0$ (bLB)
contributions in anQCD.
Section \ref{sec:concl} contains conclusions and outlines prospects
for further use of the obtained anQCD models.

\section{Beta functions for analytic QCD}
\label{sec:beta1}

Our starting point will be the construction of certain classes of
beta functions $\beta(a)$ for the coupling $a(Q^2)$ such that
ITEP-OPE conditions
\be
\vert a(Q^2) - a_{\rm pt}(Q^2) \vert < \left( \frac{\Lambda^2}{Q^2}
\right)^k  \ , \qquad (k=1,2,\ldots) \ ,
\label{ITEP}
\ee
are fulfilled and that, at the same time, they lead to an analytic
QCD (anQCD), i.e., the resulting $a(Q^2)$ is an analytic function
for all $Q^2 \in \mathbb{C} \backslash (-\infty,0]$.
This procedure is in contrast to other anQCD models which are
usually constructed either via a direct 
construction of $a(Q^2)$, or via specification of the discontinuity 
function $\rho_1(\sigma) \equiv {\rm Im} a(Q^2=-\sigma - i \epsilon)$ and
the subsequent application of the dispersion relation
to construct $a(Q^2)$
\be
a(Q^2) = \frac{1}{\pi} \int_0^{+\infty} \ d \sigma 
\frac{ \rho_1(\sigma) }{(\sigma + Q^2)} \ .
\label{dispa}
\ee
In such approaches, it appears to be difficult to fulfill the ITEP-OPE
conditions (\ref{ITEP})\footnote{
Instanton effects can modify the conditions (\ref{ITEP}) in the sense that
these conditions remain valid only for $k=1,2,\ldots,k_{\rm max}$ 
where $2 k_{\rm max}$ is
the largest dimension of condensates not affected by the 
small-size instantons.
Scenarios of instanton-antiinstanton gas give $k_{\rm max} < 4 \beta_0$
($=9$ for $n_f=3$), cf.~Ref.~\cite{DMW}. In this work we do not consider
such possible instanton effects.}
, and difficult or impossible
to extract the beta function $\beta(a)$ as a function of $a$.

On the other hand, starting with the construction of a 
beta function $\beta(a)$, which appears in the RGE
\be
Q^2 \frac{d a(Q^2)}{d Q^2} = \beta \left( a(Q^2) \right) \ ,
\label{RGE1}
\ee
it turns out to be simple to fulfill conditions  (\ref{ITEP})
(cf.~Ref.~\cite{Raczka}). Namely, if one requires that $\beta(a)$ 
be an analytic function of $a$ at $a=0$, then the corresponding 
$a(Q^2)$ respects the ITEP-OPE conditions (\ref{ITEP}).

This statement can be demonstrated in the following indirect way:
assuming that the conditions (\ref{ITEP}) do not hold, we will show
that $\beta(a)$ must then be nonanalytic at $a=0$. 
In fact, if the conditions (\ref{ITEP}) 
do not hold, then a positive $n_0$ exists such that
\be
a(Q^2) \approx
a_{\rm pt}(Q^2)  + \kappa (\Lambda^2/Q^2)^{n_0} \, 
\label{nonITEP}
\ee
for $Q^2 \gg \Lambda^2$. Asymptotic freedom of QCD implies that
at such large $Q^2$ the perturbative $a_{\rm pt}(Q^2)$ has
the expansion
(if the conventional, ${\overline {\rm MS}}$, scale 
$\Lambda={\overline \Lambda}$ \cite{Buras:1977qg,Bardeen:1978yd} is used)
\be
a_{\rm pt}(Q^2) = \frac{1}{\beta_0 \ln (Q^2/\Lambda^2)} -
\frac{c_1}{\beta_0^2}
\frac{\ln \ln (Q^2/\Lambda^2)}{\ln^2 (Q^2/\Lambda^2)} + 
{\cal O} \left( 
\frac{\ln^2 (\ln (Q^2/\Lambda^2))}{ \ln^3 (Q^2/\Lambda^2)} \right) \ ,
\label{apt2l}
\ee
and consequently the power term can be written as
\be
(\Lambda^2/Q^2)^{n_0} = \exp \left( -K/a_{\rm pt}(Q^2) \right) 
(\beta_0 a_{\rm pt})^{- K^{\prime}}
\left( 1 + {\cal O}(a \ln^2 a) \right) \ ,
\label{powt1}
\ee
where $K = n_0/\beta_0$ and $K^{\prime}=n_0 c_1/\beta_0$. 
Applying $d/d \ln Q^2$ to the relation (\ref{nonITEP}) and using 
expression (\ref{powt1}), we obtain
\be
\beta(a(Q^2)) \approx
\beta_{\rm pt}(a_{\rm pt}(Q^2)) - n_0 \kappa \exp \left( -K/a_{\rm pt}(Q^2) \right) 
(\beta_0 a_{\rm pt})^{- K^{\prime}}
\left( 1 + {\cal O}(a \ln^2 a) \right) \ .
\label{difbet1}
\ee
Replacing $a(Q^2)$ in the first beta function
in Eq.~(\ref{difbet1}) by the right-hand side (rhs) of Eq.~(\ref{nonITEP}),
using Eq.~(\ref{powt1}), and Taylor expanding the $\beta(a(Q^2))$ function 
around $a_{\rm pt}(Q^2)$ ($\not= 0$), gives 
\ba
\lefteqn{
\!\!\!\!\!\!\!\!\!\!\!\!\!\!\!\!
\!\!\!\!\!\!\!\!\!\!\!\!\!\!\!\!
\beta (a_{\rm pt}) + \kappa \exp ( -K/a_{\rm pt}) 
(\beta_0 a_{\rm pt})^{- K^{\prime}} \left( 1 + {\cal O}(a \ln^2 a) \right)
\times \frac{d \beta(a)}{d a} {\big |}_{a=a_{\rm pt}}
+ {\cal O} \left( \exp ( -2 K/a_{\rm pt})  a_{\rm pt}^{- 2 K^{\prime}}
\right) }
\nonumber\\
&& \approx \beta_{\rm pt}(a_{\rm pt}) -  n_0 \kappa \exp ( -K/a_{\rm pt}) 
(\beta_0 a_{\rm pt})^{- K^{\prime}}\left( 1 + {\cal O}(a \ln^2 a) \right) \ .
\label{difbet2}
\ea
In this relation, valid for small values of $|a_{\rm pt}|$,
the term with derivative $d \beta (a)/d a \sim a_{\rm pt}$ 
on the left-hand side (lhs) can be neglected 
in comparison with the corresponding term on
the rhs. Therefore, Eq.~(\ref{difbet2}) obtains the form
(with notation $a_{\rm pt} \mapsto a$)
\be
\beta(a) \approx \beta_{\rm pt}(a) - n_0 \kappa \exp ( -K/a) 
(\beta_0 a)^{- K^{\prime}} \left( 1 + {\cal O}(a \ln^2 a) \right)
\ .
\label{difbet3}
\ee
We note that $\beta_{\rm pt}(a)$, being a polynomial, is analytic at $a=0$.
The term proportional to $\exp(-K/a)$ is nonanalytic at $a=0$,
because $\exp(-K/a)$ has an essential singularity there. This shows that
nonfulfillment of the ITEP-OPE conditions (\ref{ITEP})
implies nonanalyticity of $\beta(a)$ at $a=0$, and the
demonstration is concluded.

This proof shows that nonfulfillment of ITEP-OPE conditions
implies nonfulfillment of $a=0$ analyticity of $\beta(a)$.
Or equivalently, fulfillment of $a=0$ analyticity of $\beta(a)$
implies fulfillment of the ITEP-OPE conditions
(\ref{ITEP}). This does not mean the equivalence of 
$a=0$ analyticity of $\beta(a)$ with the ITEP-OPE conditions.
But that will suffice for our purpose, since in the following we will
simply restrict the {\it Ans\"atze\/} for the $\beta$ function which
are analytic at $a=0$, thus having the ITEP-OPE conditions secured.

Integration of RGE (\ref{RGE1}) must be performed for all complex $Q^2$. 
To achieve this, we first need an initial condition
[equivalent to the fixing of $\Lambda^2$ scale 
($\sim 0.1 \ {\rm GeV}^2$)].
This is a subtle point within our approach, due to two reasons.
First, when we choose a specific form of the beta function $\beta(a)$,
we automatically choose a specific renormalization scheme (RSch)
as well, as represented by the coefficients $c_j \equiv \beta_j/\beta_0$
($j \geq 2$) of the power expansion of $\beta(a)$, Eq.~(\ref{btexp1}).
The running of the corresponding $a(Q^2)$ 
can be in general significantly different
from the running $a(Q^2; {\overline {\rm MS}})$ in 
${\overline {\rm MS}}$ RSch. Secondly, this running is also
influenced by the number of active quark flavors and
by flavor threshold effects. In our analyses of RGE
with our specific $\beta$ functions, we will consider the number
of active quark flavors to be $n_f=3$, i.e., the flavors of the
three (almost) massless quarks $u$, $d$ and $s$. 
We do not know how to include in a consistent way the massive quark degrees
($n_f \geq 4$) in anQCD. On the other hand, the ITEP-OPE conditions (\ref{ITEP})
tell us that the considered anQCD theories become practically
indistinguishable from pQCD at reasonably high energies $Q^2 \gg \Lambda^2$.
Therefore, we wish to keep $n_f=3$ in the RGE running to 
as high values of $|Q^2|$ as possible,
and to replace the theory at higher $|Q^2|$ by pQCD, in the
RSch dictated by the specific beta function.
Furthermore, in pQCD the threshold for $n_f=3 \mapsto n_f=4$ 
can be chosen at $Q^2 \sim (k m_c)^2$ with
$k \approx 1-3$ \cite{BW,RS,LRV,CKS}, where $m_c$ denotes the mass of the
charmed quark. We will use $k=3$,
i.e., at $|Q^2| \geq (3 m_c)^2$ ($\approx 14.5 \ {\rm GeV}^2$) the
anQCD theory will be replaced by pQCD theory.

In order to find the value of $a((3 m_c)^2) \equiv a_{\rm in}$
which will define our initial condition, we start from the 
experimentally best known value of the coupling parameter,
namely $a(M_Z^2,{\overline {\rm MS}})$. It is deduced, within
pQCD, from all relevant experiments at high 
$|Q^2| \agt 10^1 \ {\rm GeV}^2$ and found to be
$a(M_Z^2,{\overline {\rm MS}}) \approx 0.119/\pi$, Ref.~\cite{PDG2008}. 
We RGE run this value, in ${\overline {\rm MS}}$ RSch, down to the scale
$(3 m_c)^2$, and incorporate the quark threshold matching
conditions at the three-loop level according to
Ref.~\cite{CKS} at $Q^2= 3 m_q^2$ ($q=b,c$). We obtain\footnote{
For ${\overline \beta}(a) \equiv \beta(a,{\overline {\rm MS}})$ 
we used Pad\'e $[2/3](a)$ based on the known ${\overline {\rm MS}}$
$c_j$-coefficients: ${\bar c}_2$ and ${\bar c}_3$. 
Using truncated (polynomial) series up to $- \beta_0 {\bar c}_3 a^5$ instead,
changes the results almost insignificantly, by less than 1 per mil.
For the quark mass values we use: $m_c=1.27$ GeV and $m_b=4.20$ GeV
(cf.~Ref.~\cite{PDG2008}).}
${\overline a} \equiv a( (3 m_c)^2,{\overline {\rm MS}}, n_f=3) = 0.07245$. 
The value $a_{\rm in}= a( (3 m_c)^2 )$, at the same renormalization scale 
(RScl) but in the RSch 
as defined by our $\beta(a)$ function, is then obtained from the
aformentioned ${\overline {\rm MS}}$ value 
${\overline a} \equiv a( (3 m_c)^2,{\overline {\rm MS}}, n_f=3)$ 
by solving numerically the integrated RGE 
in its subtracted form (Ref.~\cite{CK1}, Appendix A there)
\ba
\frac{1}{a} + c_1 \ln \left( \frac{c_1 a}{1\!+\!c_1 a} \right)
+ \int_0^a dx \left[ \frac{ \beta(x) + \beta_0 x^2 (1\!+\!c_1 x)}
{x^2 (1\!+\!c_1 x) \beta(x) } \right]
& = &
\frac{1}{{\overline a}} + c_1 \ln \left( \frac{c_1 {\overline a}}{1\!+\!c_1 {\overline a}} \right)
+ \int_0^{\overline a} dx \left[ \frac{ {\overline \beta}(x) + \beta_0 x^2 (1\!+\!c_1 x)}
{x^2 (1\!+\!c_1 x) {\overline \beta}(x) } \right],
\label{match}
\ea
where $a \equiv a( (3 m_c)^2) = a_{\rm in}$ and
${\overline a} \equiv a( (3 m_c)^2,{\overline {\rm MS}}) = 0.07245$,
both with $n_f=3$;
further, ${\overline \beta}$ is the beta function of 
the ${\overline {\rm MS}}$ scheme. We note that in Eq.~(\ref{match})
our beta functions have expansions around $a=0$
[cf.~Eq.~(\ref{btexp1})],
with the RSch coefficients $(c_2, c_3, \ldots)$ which may be considerably 
different from the  ${\overline {\rm MS}}$ coefficients $({\overline c}_2,
{\overline c}_3, \ldots)$. Therefore, in Eq.~(\ref{match}) expansions
of $\beta$ in powers of $x$ are in general not justified.

\begin{figure}[htb] %\unitlength=1mm
%\centering\includegraphics[width=140mm]{Qzplane.eps}
%\centering\includegraphics[width=140mm]{figxl1.eps}
%\centering\includegraphics[width=20mm]{figxl1.pdf}
\centering\includegraphics[width=140mm]{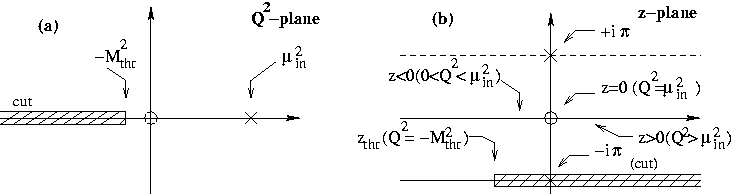}
\vspace{-0.4cm}
 \caption{\footnotesize  (a) Complex $Q^2$ plane; (b) complex
$z$ plane where $z=\ln(Q^2/\mu_{\rm in}^2)$; the physical
stripe is $-\pi \leq {\rm Im}z < + \pi$.}
\label{Qzplane}
 \end{figure}

Having the initial value $a_{\rm in}= 
a(Q_{\rm in}^2 \equiv \mu_{\rm in}^2=(3 m_c)^2)$ fixed,
RGE (\ref{RGE1}) can be solved numerically in the $Q^2$-complex plane.
It turns out that the numerical integration can be performed
more efficiently and elegantly
if, instead of $Q^2$, a new complex variable is introduced: 
$z = \ln(Q^2/\mu_{\rm in}^2)$.
Then the entire $Q^2$-complex plane (the first sheet) corresponds to
the semiopen stripe $- \pi \leq {\rm Im}z < + \pi$ in the complex $z$ plane.
The Euclidean part $Q^2 \in \mathbb{C} \backslash (-\infty,0]$
where $a(Q^2)$ has to be analytic corresponds to the open
stripe $- \pi < {\rm Im}(z) < + \pi$; the Minkowskian
semiaxis $Q^2 \leq 0$ is the $z$ line ${\rm Im}z = - \pi$;
the point $Q^2=0$ corresponds to $z=-\infty$; $Q^2=\mu_{\rm in}^2$
($= (3 m_c)^2 \approx 14.5 \ {\rm GeV}^2$) corresponds to $z=0$;
see Fig.~\ref{Qzplane}.
If we denote $a(Q^2) \equiv F(z)$, RGE (\ref{RGE1}) can be rewritten
\be
\frac{d F(z)}{d z} = \beta(F(z)) \ ,
\label{RGEz}
\ee
in the semiopen stripe $- \pi \leq {\rm Im}z < + \pi$. The analyticity
requirement for $a(Q^2)$ now means analyticity of $F(z)$ ($ \Rightarrow
\partial F/\partial {\bar z} = 0$) in the open stripe 
$- \pi < {\rm Im}(z) < + \pi$, and we expect (physical) singularities
solely on the line ${\rm Im} (z) = - \pi$.
Writing $z=x+i y$ and $F=u + i v$,
and assuming analyticity ($\partial F/\partial {\bar z} = 0$),
we can rewrite RGE (\ref{RGEz}) as a coupled system of
partial differential equations for $u(x,y)$ and $v(x,y)$
\ba
\frac{\partial u(x,y)}{\partial x} &=& {\rm Re} \beta (u + i v) \ ,
\quad 
\frac{\partial v(x,y)}{\partial x} = {\rm Im} \beta (u + i v) \ ,
\label{RGEx}
\\
\frac{\partial u(x,y)}{\partial y} &=& - {\rm Im} \beta (u + i v) \ ,
\quad 
\frac{\partial v(x,y)}{\partial y} = {\rm Re} \beta (u + i v) \ .
\label{RGEy}
\ea
Thus, beta functions $\beta(F)$ are analytic at $F=0$
[ITEP-OPE condition (\ref{ITEP})], and the expansion of
$\beta(F)$ around $F=0$ [cf.~Eq.~(\ref{btexp1})]
must reproduce the two universal
parameters $\beta_0$ and $c_1 = \beta_1/\beta_0$
(``pQCD condition,'' where $\beta_0=9/4$ and $c_1=16/9$ for $n_f=3$),
% \be
%\beta(F) = - \beta_0 F^2 \left[ 1 + c_1 F + \cdots \right] \ ,
%\label{c1con}
%\ee
and solution $F(z) = u(x,y)+i v(x,y)$ of RGEs (\ref{RGEx})-(\ref{RGEy})
satisfies the initial condition $F(0)=a_{\rm in}$ where $a=a_{\rm in}$ 
is determined by Eq.~(\ref{match}). 

We implement high precision numerical
integration of RGEs (\ref{RGEx})-(\ref{RGEy}) with MATHEMATICA
\cite{math}, for various {\it Ans\"atze\/} of $\beta(F(z))$ satisfying the
aforementioned ITEP-OPE and pQCD conditions. Numerical analyses
indicate that it is in general very difficult to obtain analyticity of 
$F(z)$ in the entire open stripe $-\pi < {\rm Im}(z) < + \pi$,
equivalent to the analyticity of $a(Q^2)$ for all complex $Q^2$
except $Q^2 \in (-\infty,0]$.
On the other hand, if we, in addition, require also
analyticity of $a(Q^2)$ at $Q^2=0$ ($\Leftrightarrow z=-\infty$),
certain classes of $\beta(a)$ functions do give us
$F(z)$ with the correct analytic behavior.
This $Q^2=0$ analyticity condition in general implies
\be
a(Q^2) = a_0 + a_1 (Q^2/\Lambda^2) + {\cal O}[ (Q^2/\Lambda^2)^2] \ ,
\label{aQ0}
\ee
where $0 < a_0 \equiv a(Q^2=0) = F(z=-\infty) < \infty$ and
$a_1 \not=0$.  Application
of $d/d \ln Q^2 = d/dz$ to Eq.~(\ref{aQ0}) then implies that in the Taylor
expansion of $\beta(F)$ around $F=a_0$ the first coefficient is unity
\be
\beta(F) = 1 \times (F - a_0) + {\cal O}[ (F - a_0)^2 ] \ ,
\label{anQ0a}
\ee
or equivalently\footnote{
If we assumed analyticity of $a(Q^2)$ in a special way,
with $a_1=0$ in Eq.~(\ref{aQ0}), then we would have
$a(Q^2) = a_0 +  {\cal O}[ (Q^2/\Lambda^2)^n]$ with 
$n \geq 2$ and $\beta^{\prime}(F) \vert_{F=a_0} = n$.
This would imply $a_0 \beta_0 f(1) = n$ ($\geq 2$).
From considerations in Appendix \ref{app:impl}
[cf.~Eqs.~(\ref{impl4})-(\ref{zpgen})] it follows then
that in such a case the RGE solution $F(z)$ has poles at
${\rm Im}z = \pm \pi/n$, i.e., Landau poles.}
\be
\beta^{\prime}(F) \vert_{F=a_0} = +1 \ .
\label{anQ0b}
\ee
We write our $\beta(F)$ {\it Ans\"atze\/} in the form
\be
\beta(F) = - \beta_0 F^2 (1 - Y) f(Y) \vert_{Y \equiv F/a_0} \ ,
\label{ans1}
\ee
with function $f(Y)$ fulfilling the three 
aforementioned conditions
\ba
f(Y) \ && {\rm analytic \ at} \ Y=0  \qquad {\rm (ITEP-OPE)} \ ,
\label{ITEPf}
\\
f(Y) & = & 1 + (1 + c_1 a_0) Y + {\cal O}(Y^2)  \qquad {\rm (pQCD)} \ ,
\label{pQCDf}
\\
a_0 \beta_0 f(1) &=& 1 \qquad (Q^2=0 \ {\rm analyticity}) \ .
\label{Q20f}
\ea
We always consider $a_0$ [$\equiv a(Q^2=0)$] to be
positive [note: $a = (g_s/2/\pi)^2  > 0$].

We will argue in more detail why and how this additional
constraint [analyticity of $a(Q^2)$ at $Q^2=0$]
improves the analytic behavior of $a(Q^2) \equiv F(z)$ 
in the entire $Q^2$ plane ($z$ stripe), 
in the sense of avoiding Landau singularities. 
For this, it is helpful to consider some simple classes of 
beta functions which, on the one hand, allow for an 
implicit analytic solution $z = G(F)$ of RGE (\ref{RGEz}) and, 
on the other hand, are representative because larger classes 
of beta functions can be successively approximated by them.
Specifically, we consider $f(Y)$ in Eq.~(\ref{ans1}) to be
either a polynomial or a rational function\footnote{
In the following we characterize such functions by the
corresponding Pad\'e-notations.}
\ba
f(Y) &=& 1 + \sum_{k=1}^R r_k Y^k = {\rm P}[R/0]_f(Y) \ ,
\label{P}
\\
f(Y) &=& (1 + \sum_{k=1}^M m_k Y^k )/(1 + \sum_{\ell=1}^N n_{\ell} Y^{\ell}) = 
{\rm P}[M/N]_f(Y) \ .
\label{R}
\ea
Here, the degrees ($R; M,N$) are in principle arbitrary,
and the coefficients ($r_k; m_k,n_{\ell}$) as well.
Such {\it Ans\"atze\/} apparently can fulfill all constraints
(\ref{ITEPf})-(\ref{Q20f}). It is also intuitively clear
that they can approximate large classes of other $\beta$ functions
that fulfill the same constraints.

Now we undertake the following procedure. Formal 
integration of RGE (\ref{RGEz}) leads to the solution
\be
z = G(F) \ , \quad G(F(z)) = \int_{a_{\rm in}}^{F(z)} 
\frac{d {\widetilde F}}{\beta({\widetilde F})} \ ,
\label{impl1}
\ee
where $a_{\rm in}$ is the aforementioned initial value
$a_{\rm in} = a(Q^2=\mu^2_{\rm in}) = F(0)$.
Equation (\ref{impl1}) represents an implicit (inverted) equation
for $F=F(z)=G^{-1}(z)$. In both cases,
Eqs.~(\ref{P}) and (\ref{R}), the integration
in Eq.~(\ref{impl1}) can be performed explicitly.
This is performed in Appendix \ref{app:impl}.

Here we quote, for orientation, the results
for two simple examples of $f(Y)$, a quadratic\footnote{
A linear polynomial has at first only one free parameter $r_1
=(1+ c_1 a_0)$ by the condition (\ref{pQCDf}); however, this
$a_0$ gets fixed by the $Q^2=0$ analyticity condition (\ref{Q20f}):
$a_0 \approx 0.1904$.}
polynomial ${\rm P}[2/0]_f$ and a rational function 
${\rm P}[1/1]_f$.

In the case of quadratic polynomial we have
\be
f(Y) = 1 + r_1 Y + r_2 Y^2 \ ,
\label{P2o1}
\ee
where $r_1=(1 + c_1 a_0)$ due to the pQCD condition (\ref{pQCDf}).
The (positive) quantity $a_0 \equiv a(Q^2=0)$ is then
obtained as a function of the only free parameter
$r_2$ by the $Q^2=0$ analyticity condition (\ref{Q20f})
\be
a_0(r_2) = \frac{1}{2 c_1} \left[ - (2 + r_2) +
\sqrt{ (2 + r_2)^2 + 4 c_1/\beta_0 } \right] \ .
\label{a0P2}
\ee
For the integration (\ref{impl1}), we need to rewrite the
polynomial (\ref{P2o1}) in a factorized form
\ba
f(Y=1/t) &=& \frac{1}{t^2} (t - t_1) (t - t_2) \ ,
\label{ft}
\\
\left(
\begin{array}{c}
t_1(r_2) \\
t_2(r_2) 
\end{array}
\right)
& = & \frac{1}{2} \left[ - r_1 \pm
\sqrt{ r_1^2 - 4 r_2 } \right] \ , \qquad
\left( r_1 = 1 + c_1 a_0(r_2) \right) \ .
\label{t1t2P2}
\ea
Integration (\ref{impl1}) then gives the following
implicit equation for $F(z) \equiv a(Q^2)$: 
\be
z =  \left\{
\frac{(-1)}{\beta_0} \left( \frac{1}{a_{\rm in}} -
\frac{1}{F(z)} \right) +
\ln \left( \frac{a_0/F(z) -1}{a_0/a_{\rm in} -1} \right)
+ \frac{1}{\beta_0 a_0} \sum_{j=1}^2 B_j
\ln \left( \frac{a_0/F(z) -t_j}{a_0/a_{\rm in} - t_j} \right)
\right\} \ ,
\label{implP2}
\ee
where
\ba
B_1 & = & \frac{t_1^3}{(t_1-1)(t_1-t_2)} \ ,
\qquad
B_2 = \frac{t_2^3}{(t_2-1)(t_2-t_1)} \ .
\label{B1B2}
\ea
In this solution we took into account that the coefficient
$B_0/(\beta_0 a_0))=1/((1-t_1)(1-t_2)(\beta_0 a_0))$ 
in front of the first logarithm 
in Eq.~(\ref{implP2}) is simply unity by the $Q^2=0$ analyticity
condition (\ref{Q20f}).
The poles $z_{\rm p}$, at which $F(z_{\rm p}) = \infty$,
are obtained from Eq.~(\ref{implP2}) by simply replacing
$1/F(z)$ by zero
\ba
z_{\rm p} & = & \left\{
\ln \left( \frac{(-1)}{a_0/a_{\rm in} -1} \right)
- \frac{1}{\beta_0 a_{\rm in}} 
+ \frac{1}{\beta_0 a_0} \sum_{j=1}^2 B_j
\ln \left( \frac{-t_j}{a_0/a_{\rm in} - t_j} \right)
\right\} \ .
\label{implP2p}
\ea
It turns out that $a_0 > a_{\rm in}$ (typically,
$a_0 \approx 0.1$-$0.2$ and $a_{\rm in} < 0.1$).
If, in addition, $0<r_2<r_1^2/4$, then Eqs.~(\ref{t1t2P2})
imply $t_1, t_2 <0$. Therefore, when $0<r_2<r_1^2/4$, all
the arguments in logarithms in Eq.~(\ref{implP2p})
are positive, except in the first logarithm where
$\ln(-1) = \pm i \pi$ and thus the only poles of $F(z)$
in the physical stripe ($-\pi \leq {\rm Im}z < \pi$) have
\be
{\rm Im} z_{\rm p} = - \pi \ .
\label{zppi}
\ee
This implies that for $0<r_2<r_1^2/4$ the considered singularity 
must lie on the timelike axis ($Q^2 < 0$) and hence does not represent
a Landau pole. We stress that for such a conclusion, the
$Q^2=0$ analyticity condition (\ref{Q20f}) is of central importance,
since it fixes the coefficient in front of $\ln(-1)$ in Eq.~(\ref{implP2p})
to be unity.\footnote{This also explains why it is nearly impossible
to obtain an analytic $a(Q^2)$ if we abandon the $Q^2=0$ analyticity
condition (\ref{Q20f}).}
We can derive from Eq.~(\ref{implP2p})
the location of the pole in the $Q^2$ plane at
\ba
Q^2_{\rm p.} &= & \mu_{\rm in}^2 \exp(z_{\rm p}) = 
- \mu_{\rm in}^2 \exp({\rm Re} z_{\rm p})
\nonumber\\
& = & - \mu_{\rm in}^2 \exp \left( - \frac{1}{\beta_0 a_{\rm in}} \right)
\left( \frac{a_0}{a_{\rm in}} - 1 \right)^{-1} \prod_{j=1}^2
\left( \frac{a_0/a_{\rm in}-t_j}{-t_j} \right)^{-B_j/(\beta_0 a_0)} \ .
\label{Q2p}
\ea
On the other hand, if the aforementioned conditions are not
fulfilled, we obtain $-\pi < {\rm Im} z_{\rm p} < \pi$,
representing a pole inside the physical $z$ stripe and thus a Landau singularity.
Specifically, when $r_2 < 0$, we have $t_1>0$ and $t_2<0$ by Eqs.~(\ref{t1t2P2});
numerically, we can check that in this case always $a_0/a_{\rm in} - t_1 > 0$ and,
consequently the $j=1$ logarithm in Eq.~(\ref{implP2p}) becomes nonreal and
$-\pi < {\rm Im}z_p < \pi$, i.e., Landau pole. 

\begin{figure}[htb] %\unitlength=1mm
\begin{minipage}[b]{.49\linewidth}
% \centering\includegraphics[width=85mm]{pl3dabsbtP20r2_0.jpg}
%\centering\includegraphics[width=85mm]{figxl2a.eps}
%\centering\includegraphics[width=85mm]{figxl2a.pdf}
\centering\includegraphics[width=85mm]{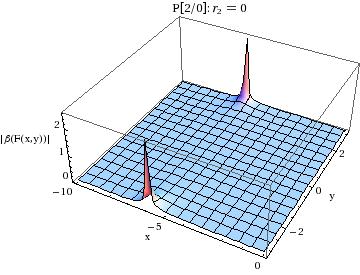}
\end{minipage}
\begin{minipage}[b]{.49\linewidth}
% \centering\includegraphics[width=85mm]{pl3dabsbtP20r2__2.jpg}
%\centering\includegraphics[width=85mm]{figxl2b.eps}
%\centering\includegraphics[width=85mm]{figxl2b.pdf}
\centering\includegraphics[width=85mm]{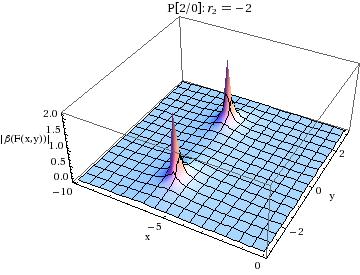}
\end{minipage}
\vspace{-0.4cm}
 \caption{\footnotesize  $|\beta(F(z))|$ as a function of
$z=x+i y$ for the beta-function (\ref{ans1}) with $f(Y)$
having the form (\ref{P2o1}) with (a) $r_2=0$;
(b) $r_2=-2$.}
\label{figP20bt}
 \end{figure}

\begin{figure}[htb] %\unitlength=1mm
\begin{minipage}[b]{.49\linewidth}
% \centering\includegraphics[width=85mm]{pl3rho1P20r2_0.jpg}
%\centering\includegraphics[width=85mm]{figxl3a.eps}
\centering\includegraphics[width=85mm]{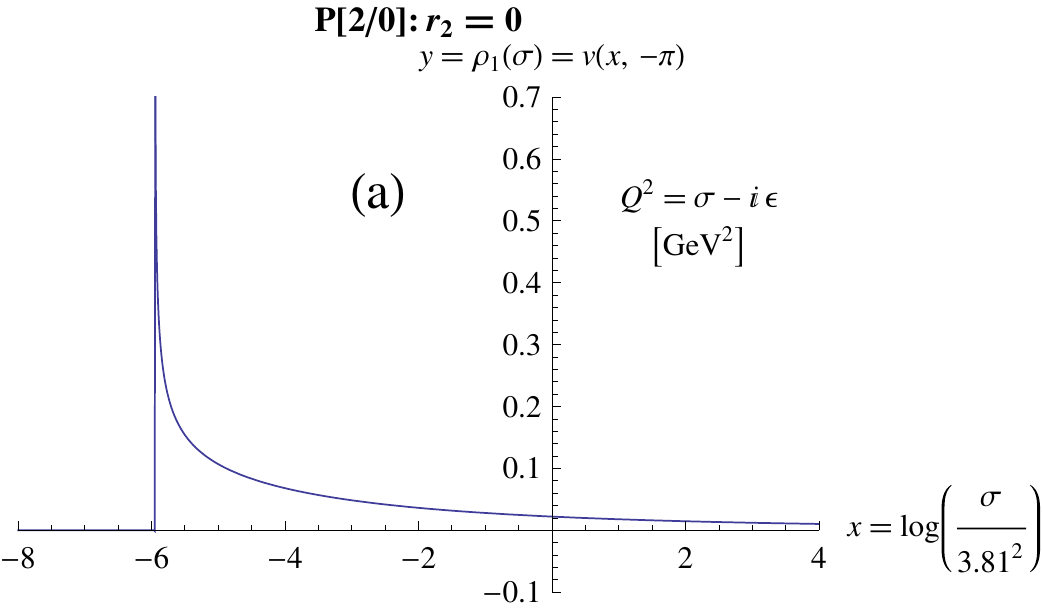}
%\centering\includegraphics[width=85mm]{figxl3a.jpg}
\end{minipage}
\begin{minipage}[b]{.49\linewidth}
% \centering\includegraphics[width=85mm]{pl3urho1P20r2_0.jpg}
%\centering\includegraphics[width=85mm]{figxl3b.eps}
\centering\includegraphics[width=85mm]{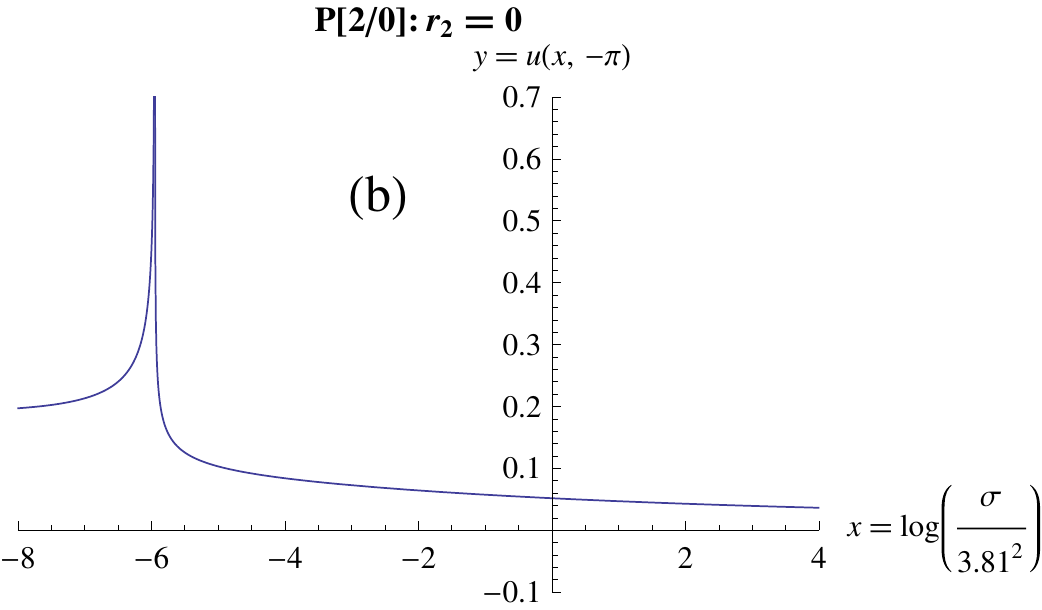}
%\centering\includegraphics[width=85mm]{figxl3b.jpg}
\end{minipage}
\vspace{-0.4cm}
 \caption{\footnotesize  (a) The discontinuity function
$\rho_1(\sigma) = {\rm Im} a(Q^2=-\sigma - i \epsilon) =
{\rm Im} F(z=x - i \pi) = v(x,y=-\pi)$ as a function of
$x = {\rm Re}(z) = \ln(\sigma/\mu^2_{\rm in})$, for the
case when $f(Y)$ has the form (\ref{P2o1}) with $r_2=0$,
i.e., linear polynomial;
(b) same as in (a), but for 
${\rm Re} a(Q^2=-\sigma - i \epsilon)= {\rm Re} F(z=x - i \pi) = u(x,y=-\pi)$.}
\label{figP20rhor2_0}
 \end{figure}

\begin{figure}[htb] %\unitlength=1mm
\begin{minipage}[b]{.49\linewidth}
% \centering\includegraphics[width=85mm]{pl3rho1P20r2__2.jpg}
%\centering\includegraphics[width=85mm]{figxl4a.eps}
\centering\includegraphics[width=85mm]{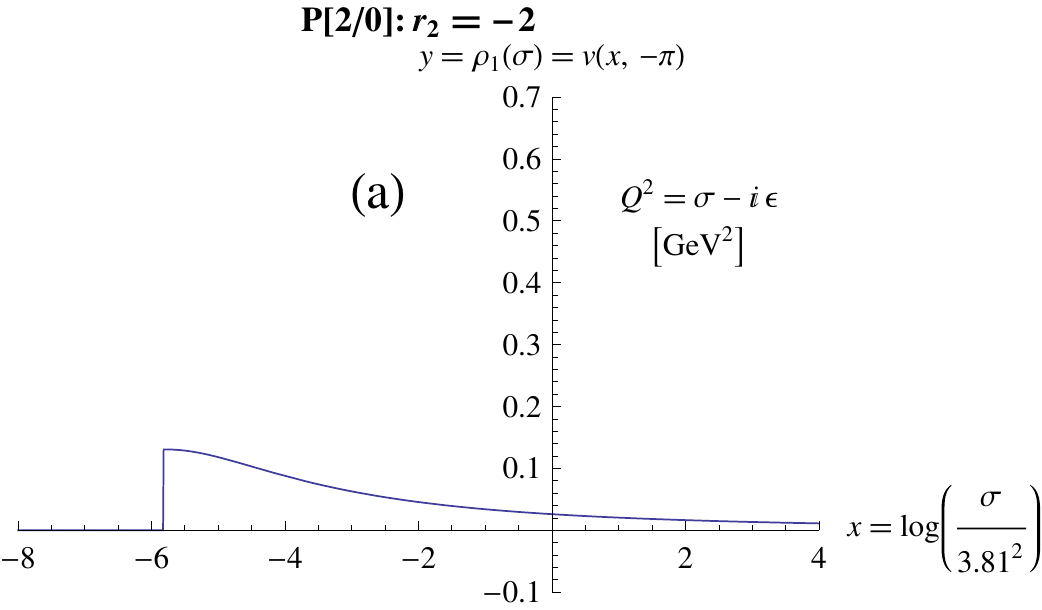}
%\centering\includegraphics[width=85mm]{figxl4a.jpg}
\end{minipage}
\begin{minipage}[b]{.49\linewidth}
% \centering\includegraphics[width=85mm]{pl3urho1P20r2__2.jpg}
%\centering\includegraphics[width=85mm]{figxl4b.eps}
\centering\includegraphics[width=85mm]{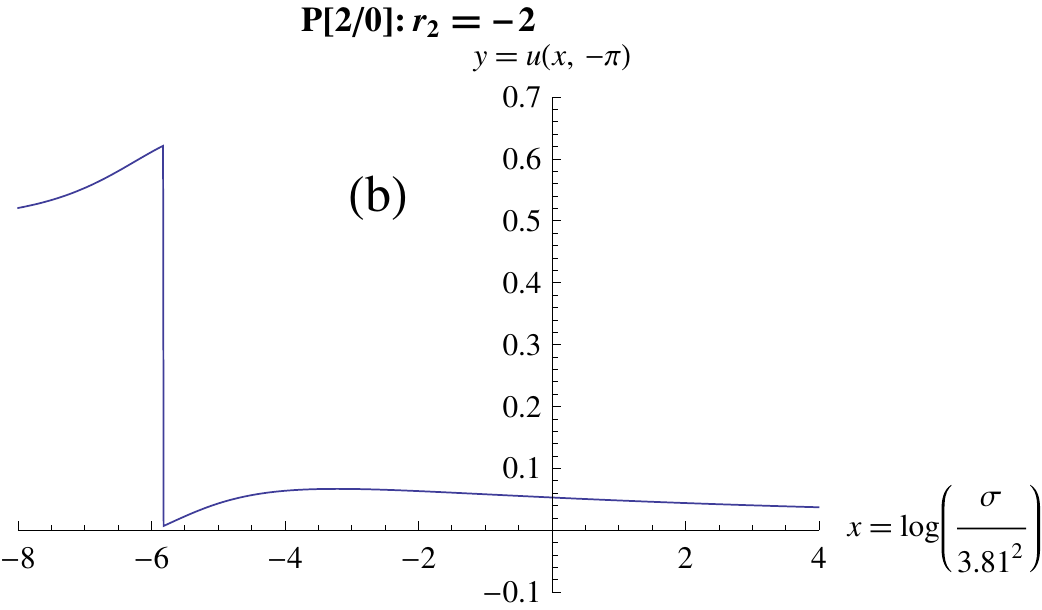}
%\centering\includegraphics[width=85mm]{figxl4b.jpg}
\end{minipage}
\vspace{-0.4cm}
\caption{\footnotesize  Same as in Figs.~\ref{figP20rhor2_0},
but this time $r_2=-2$.}
\label{figP20rhor2__2}
 \end{figure}

To observe in more detail the occurrence and the shape of these singularities,
we pursued the numerical solution of RGE (\ref{RGEz}), i.e.,
RGEs (\ref{RGEx})-(\ref{RGEy}), accounting for the initial condition
at $\mu_{\rm in}^2 = (3 m_c)^2$ in the aforementioned way. 
In order to see the appearance of singularities of $F(z) \equiv F(x+i y)$ in the
physical $z$ stripe, it is convenient to inspect the behavior of
$|\beta(F(z))|$ which should show similar singularities. The numerical results
for $|\beta(F(z))|$, in the case of $r_2=0$ and $r_2=-2$ are
given in Figs.~\ref{figP20bt}(a), (b), respectively.
In these figures, we see clearly that the singularities are
on the timelike edge ${\rm Im} z = \pm \pi$ in the case
of $r_2=0$ where we have $a_0=1.901$, $t_1 \approx -1.338$
[$t_2$ is not present as $f(Y)$ is a linear polynomial]. 
The pole moves inside the $z$ stripe (i.e., become
Landau singularities) in the case of $r_2=-2$, where we have
$a_0=0.5$, $t_1 \approx 0.756$ and $t_2 \approx -2.645$.
In Fig.~\ref{figP20rhor2_0}(a) we present the
numerical results for the discontinuity function 
$\rho_1(\sigma) = {\rm Im} a(Q^2=-\sigma - i \epsilon) =
{\rm Im} F(z=x - i \pi) = v(x,y=-\pi)$ as a function of
$x = {\rm Re}(z) = \ln(\sigma/\mu^2_{\rm in})$, for the case $r_2=0$.
In  Fig.~\ref{figP20rhor2_0}(b) the analogous curve for 
${\rm Re} a(Q^2=-\sigma - i \epsilon) = {\rm Re} F(z=x - i \pi) = u(x,y=-\pi)$ 
is presented, for the same $r_2=0$ case.
In Figs.~\ref{figP20rhor2__2} (a), (b), the corresponding
curves for the $r_2=-2$ case are depicted.
% see file AfP20m2.out, at the end - the cases r2=0, r2=-2.

\begin{figure}[htb] %\unitlength=1mm
\begin{minipage}[b]{.49\linewidth}
% \centering\includegraphics[width=85mm]{pl3dabsbtEEy1_01.jpg}
%\centering\includegraphics[width=85mm]{figxl5a.eps}
\centering\includegraphics[width=85mm]{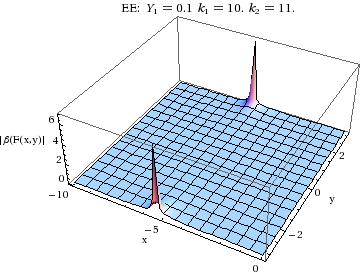}
\end{minipage}
\begin{minipage}[b]{.49\linewidth}
% \centering\includegraphics[width=85mm]{pl3dabsbtEEk1_6_y1_11.jpg}
%\centering\includegraphics[width=85mm]{figxl5b.eps}
\centering\includegraphics[width=85mm]{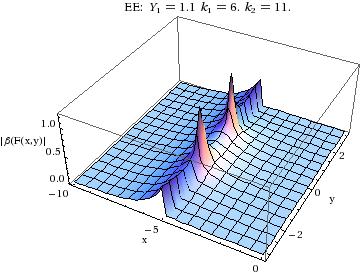}
\end{minipage}
\vspace{-0.4cm}
 \caption{\footnotesize  (a) $|\beta(F(z))|$ as a function of
$z=x+i y$, where $\beta$ has the form (\ref{ans1}) with $f(Y)$
having the EE form (\ref{EE}) 
with the values of free parameters $y_1, k_1, and k_2$
as indicated;
(b) same as in (a), but with different values of parameters $y_1$ and $k_1$.}
\label{figEEbt}
 \end{figure}

\begin{figure}[htb] %\unitlength=1mm
\begin{minipage}[b]{.49\linewidth}
% \centering\includegraphics[width=85mm]{pl3rho1EEy1_01.jpg}
%\centering\includegraphics[width=85mm]{figxl6a.eps}
\centering\includegraphics[width=85mm]{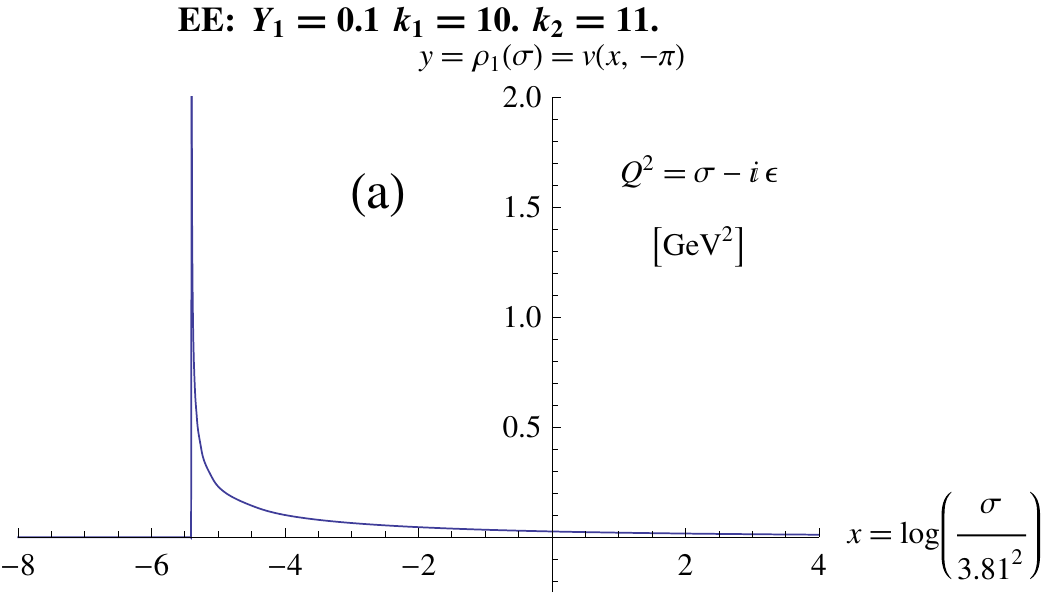}
%\centering\includegraphics[width=85mm]{figxl6a.jpg}
\end{minipage}
\begin{minipage}[b]{.49\linewidth}
% \centering\includegraphics[width=85mm]{pl3urho1EEy1_01.jpg}
%\centering\includegraphics[width=85mm]{figxl6b.eps}
\centering\includegraphics[width=85mm]{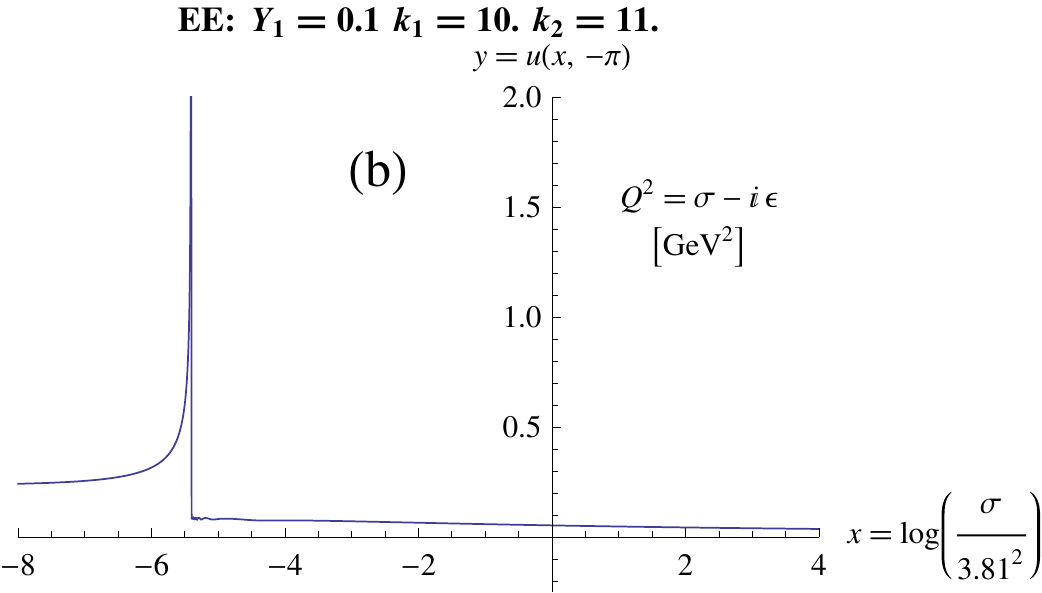}
%\centering\includegraphics[width=85mm]{figxl6b.jpg}
\end{minipage}
\vspace{-0.4cm}
 \caption{\footnotesize  (a) The discontinuity function
$\rho_1(\sigma) = {\rm Im} a(Q^2=-\sigma - i \epsilon) =
{\rm Im} F(z=x - i \pi) = v(x,y=-\pi)$ as a function of
$x = {\rm Re}(z) = \ln(\sigma/\mu^2_{\rm in})$, for the
case when $f(Y)$ is the exponential-related EE function
(\ref{EE}) with $y_1=0.1; k_1=10; k_2=11$;
(b) same as in (a), but for 
${\rm Re} a(Q^2=-\sigma - i \epsilon)= {\rm Re} F(z=x - i \pi) = u(x,y=-\pi)$.}
\label{figEErhoy1_01}
 \end{figure}

\begin{figure}[htb] %\unitlength=1mm
\begin{minipage}[b]{.49\linewidth}
% \centering\includegraphics[width=85mm]{pl3rho1EEk1_6_y1_11.jpg}
%\centering\includegraphics[width=85mm]{figxl7a.eps}
\centering\includegraphics[width=85mm]{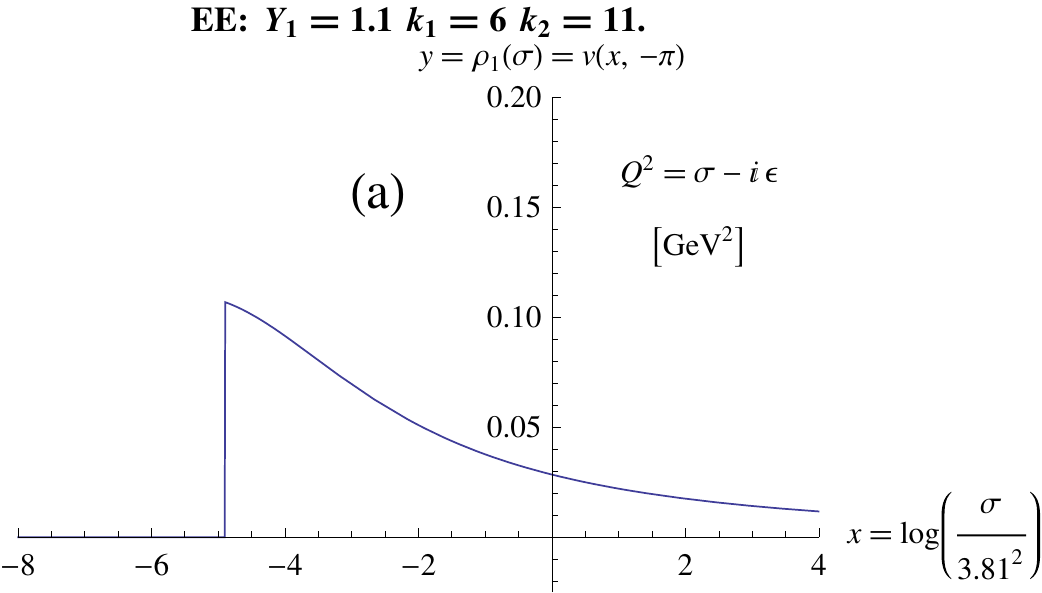}
%\centering\includegraphics[width=85mm]{figxl7a.jpg}
\end{minipage}
\begin{minipage}[b]{.49\linewidth}
% \centering\includegraphics[width=85mm]{pl3urho1EEk1_6_y1_11.jpg}
%\centering\includegraphics[width=85mm]{figxl7b.eps}
\centering\includegraphics[width=85mm]{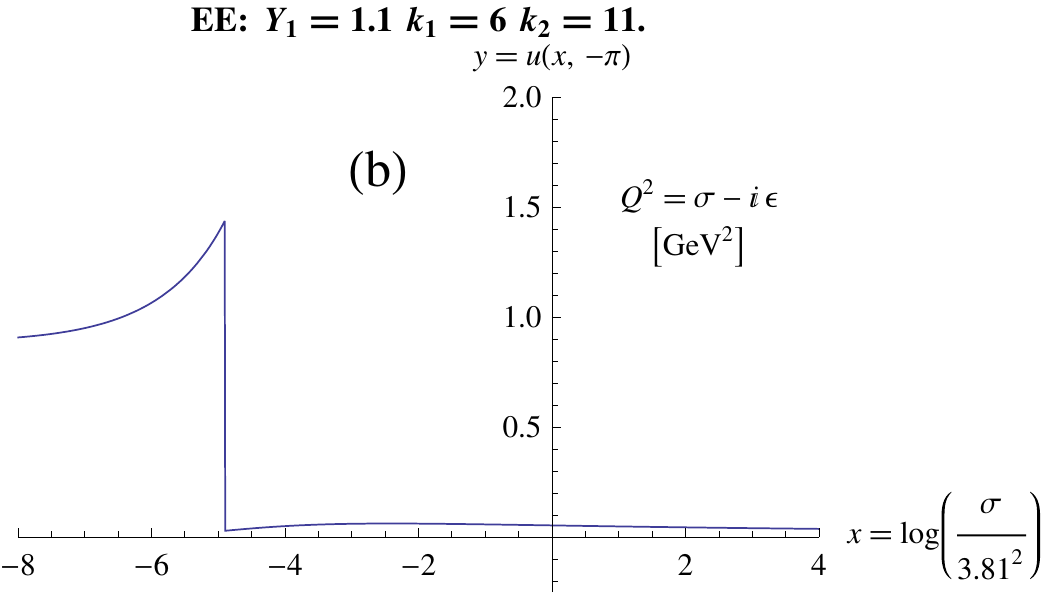}
%\centering\includegraphics[width=85mm]{figxl7b.jpg}
\end{minipage}
\vspace{-0.4cm}
 \caption{\footnotesize  Same as in Figs.~\ref{figEErhoy1_01},
but this time $y_1=1.1$ and $k_1=6$.}
\label{figEErhoy1_11}
 \end{figure}

We can try many other $f(Y)$ functions, for example, the following
set of functions involving (rescaled and translated) functions
$(e^Y-1)/Y$ and $Y/(e^Y-1)$:
\be
{\rm EE:} \qquad
f(Y) =  \frac{ \left( \exp[- k_1 (Y - Y_1)] -1 \right) }
{ [ k_1 (Y - Y_1) ] }
\frac{ [ k_2 (Y - Y_2) ] }{ \left( \exp[- k_2 (Y - Y_2)] -1 \right) }
\times {\cal K}(k_1,Y_1,k_2,Y_2) \ ,
\label{EE}
\ee 
where the constant ${\cal K}$ ensures the required normalization
$f(Y=0)=1$.
In this ``EE'' case we have, at first, five real parameters:
$a_0 \equiv a(Q^2=0)$ and
four parameters for translation and rescaling
($Y_1$, $k_1$, $Y_2$, and $k_2$). 
Two of the parameters, e.g., $Y_2$ and $a_0$, are eliminated
by conditions (\ref{pQCDf}) and (\ref{Q20f}). We need
$0 < k_1 < k_2$ to get physically acceptable behavior
and fulfill the aforementioned two conditions. 
It turns out that, in general, increasing the value of $Y_1$
tends to create Landau poles.
We consider two typical cases: (1) $y_1=0.1; k_1=10; k_2=11$;
(2) $y_1=1.1; k_1=6; k_2=11$.
The numerical results for $\beta(F(z))$ for two cases
are presented in Figs.~\ref{figEEbt}(a), (b), respectively.
We see that the first case shows no sign of
Landau poles, while the second case strongly indicates Landau poles. 
In Figs.~\ref{figEErhoy1_01} and \ref{figEErhoy1_11} we present
the behavior of the imaginary ($v$) and real ($u$) parts of the coupling
$F(z=x - i \pi) = a(Q^2=-\sigma - i \epsilon)$ along the
timelike axis of the $Q^2$ plane for the aforementioned two EE cases.
% see file AfEEm2.out, at the end - the cases EE (1), and EE (2).

There is one interesting feature which can be seen most clearly in
Figs.~\ref{figP20rhor2_0}(a) and \ref{figEErhoy1_01}(a):
the discontinuity function $\rho_1(\sigma) \equiv {\rm Im} 
a(Q^2=-\sigma - i \epsilon)$ is zero at negative
$Q^2$-values above a ``threshold'' value: 
$(-M^2_{\rm thr} \equiv)$ $- \sigma_{\rm thr} < Q^2 < 0$.
For the two cases cited there (``${\rm P}[1/0]$'' which is
``${\rm P}[2/0]$'' with $r_2=0$, and EE with $Y_1=0.1$),
we obtain $x_{\rm thr} = -5.948$ and $-5.403$, respectively,
leading to the threshold masses $M_{\rm thr} = 195$ MeV
and $256$ MeV, respectively.
These threshold masses are nonzero and comparable to the
low QCD scale $\Lambda_{\rm QCD}$ or pion mass, a behavior
that appears physically reasonable.\footnote{
Furthermore, analytic couplings with nonzero $M_{\rm thr.}$ have
the mathematical property of being
Stieltjes functions, and therefore 
their (para)diagonal Pad\'e approximants are guaranteed, 
by convergence theorems, to converge to them
as the Pad\'e index increases \cite{Cvetic:2009mq}.}
This nonzero threshold behavior (see also Fig.~\ref{Qzplane})
for the discontinuity function
$\rho_1(\sigma)$ appears because of the $Q^2=0$ analyticity
requirement for $a(Q^2)$, Eq.~(\ref{Q20f}). 
On the other hand, earlier, we saw that the condition Eq.~(\ref{Q20f}) 
is practically a necessary condition to avoid the
appearance of Landau poles of $a(Q^2)$.

While Figs.~\ref{figP20bt} and \ref{figEEbt} provide 
only a visual indication of whether the coupling $a(Q^2)$ is 
analytic, there is a more quantitative, numerical test for
the analyticity. Namely, application of the Cauchy theorem
implies for an analytic $a(Q^2)$, with cut along the
negative axis $Q^2 \leq - M^2_{\rm thr}$, the well-known
dispersion relation (\ref{dispa}) 
where the integration starts effectively at 
$\sigma=\sigma_{\rm thr}= M^2_{\rm thr}$
\be
a(Q^2) = \frac{1}{\pi} \int_{M_{\rm thr}^2}^{+\infty} \ d \sigma 
\frac{ \rho_1(\sigma) }{(\sigma + Q^2)} \ ,
\label{dispa2}
\ee
where $\rho_1(\sigma) = {\rm Im} a(Q^2=-\sigma - i \epsilon)$.
The high precision numerical solution of RGE (\ref{RGEz})
gives us $a(Q^2) = F(z)$ in the entire complex $Q^2$ plane,
including the negative semiaxis. This allows us to compare numerical
values of the lhs and rhs of dispersion relation (\ref{dispa2}),
for various values of $Q^2$.

It turns out that, for low positive $Q^2 \leq 1 \ {\rm GeV}^2$,
the numerical uncertainties of the obtained results for the rhs
of Eq.~(\ref{dispa2}) are of the order of per cent (using
64-bit MATHEMATICA \cite{math} for Linux), and they
slowly increase with increasing $Q^2$. If the deviation
of the rhs from the lhs is more than a few percent, then this represents
a strong indication that the resulting $a(Q^2)$ is not analytic.
In Table \ref{tabdev} we present the relative deviations
for the aforementioned two ${\rm P}[2/0]$ and the two EE cases.
Inspecting these deviations, we can clearly see that $a(Q^2)$
in the P[2/0] case with $r_2=-2$ and the EE case with $Y_1=1.1$
is nonanalytic; in the other two cases, the table gives
strong indication that $a(Q^2)$ is analytic.
\begin{table}
\caption{The relative deviation $R[Q^2]=({\rm rhs/lhs} -1)$ for the lhs and
the rhs of dispersion relation (\ref{dispa2}) as obtained numerically, 
for various low positive $Q^2$ ($Q^2=0., 0.1, 1.0 \ {\rm GeV}^2$), 
for the aforementioned cases of the beta function.}
\label{tabdev}  
\begin{ruledtabular}
\begin{tabular}{lclll}
$f(Y)$ & parameters & $R[Q^2=0.0]$ & $R[Q^2=0.1]$ & $R[Q^2=1.0]$ 
\\ 
\hline
P[2/0] & $r_2=0.0$ & $3.3 \times 10^{-3}$ & $4.6 \times 10^{-3}$ &  
$7.0 \times 10^{-3}$ 
\\
P[2/0] & $r_2=-2.0$ & $-0.62$ & $-0.38$ & $-0.09$  \\
EE & $Y_1=0.1, k_1=10.0, k_2=11.0$ & $4.7 \times 10^{-3}$ & $4.8 \times 10^{-3}$ & 
$6.5 \times 10^{-3}$ 
\\
EE & $Y_1=1.1, k_1=6.0, k_2=11.0$ & $-0.82$ & $-0.68$ & $-0.19$ 
\end{tabular}
\end{ruledtabular}
\end{table} 

\section{Evaluation of low-energy observables}
\label{sec:beta2}

The semihadronic $\tau$ decay ratio $R_{\tau}$ is
the most precisely measured low-energy QCD quantity
to date. The measured value of the ``QCD-canonical'' part
$r_{\tau}= a + {\cal O}(a^2)$, 
with the strangeness and quark mass effects subtracted, is
$r_{\tau}^{\rm (exp)} = 0.203 \pm 0.004$ 
(cf.~Appendix \ref{app:rtauexp}). Experimental values of other
low-energy observables, such as (spacelike) sum rules,
among them the Bjorken polarized sum rule (BjPSR) $d_{\rm Bj}(Q^2)$,
are known with far less precision. 
The minimal analytic (MA) 
model \cite{ShS,MSS,Sh,Shirkov:2006gv}, with the value of
${\overline {\Lambda}}$ such that high-energy QCD observables
are reproduced, turns out to
give for this quantity too low values $r_{\tau} \approx 0.14$
\cite{MSS,MSSY} unless the (current) masses of the light quarks
are taken to be unrealistically large 
($m_q \approx 0.25$-$0.45$ GeV) or strong
threshold effects are introduced \cite{Milton:2001mq}. 
Further, MA does not fulfill the ITEP-OPE condition (\ref{ITEP})
since $|a^{\rm (MA)}(Q^2) - a_{\rm pt}(Q^2)| \sim (\Lambda^2/Q^2)$.

The approach described in the previous Sec.~\ref{sec:beta1} automatically
fulfills the ITEP-OPE condition (\ref{ITEP}); however, the analyticity
of $a(Q^2)$, i.e., the absence of Landau poles, is achieved only for
limited regions of the otherwise free parameters of the
$\beta$ function. For general anQCD models, the evaluation 
of massless spacelike observables ${\cal D}(Q^2)$ such as BjPSR
and Adler function, and for the timelike observable $r_{\tau}$,
is presented in the sequence of Appendixes \ref{app:hoanQCD},
\ref{app:LB}, \ref{app:bLB}, particularly Eqs.~(\ref{man2tr})-(\ref{an3tr})
for spacelike and (\ref{rtman2tr})-(\ref{rtan3tr}) for $r_{\tau}$.
In the cases considered in this work, the beta function $\beta(a)$
is analytic at $a=0$ (due to the ITEP-OPE condition), and therefore
the higher order analogs $\A_{n+1}$ in those Appendixes are
simply $\A_{n+1} = a^{n+1}$, cf.~Eq.~(\ref{ourAn}). Furthermore, here we
use all the time the notation $\A_1 \equiv a$ for the analytic coupling,
and $\tA_{n+1} \equiv {\widetilde a}_{n+1}$ for the
logarithmic derivatives of $a$ [cf.~Eq.~(\ref{tAn})].

In Table \ref{tabinput1} we present the resulting values of RSch parameters
$c_2$, $c_3$ and $c_4$ [cf.~Eq.~(\ref{btexp1})], for some typical choices 
of input parameters in four forms of $f(Y)$: P[1/0], P[3/0], P[1/1], and EE. 
Here, ${\rm P}[M/N]$ is the general
notation for Pad\'e form Eq.~(\ref{fPade}) in Appendix \ref{app:impl};
${\rm P}[M/0]$ is thus a polynomial of degree $M$;
EE is the {\it Ansatz\/} (\ref{EE}) involving exponential functions.
The otherwise free parameters (``input'') of the models are chosen such 
that the analyticity is maintained, i.e., no Landau poles.
The case P[1/0] is in fact the aforementioned case of P[2/0]
with $r_2=0$, cf.~Eq.~(\ref{P2o1}), and it has no free parameters.
The cases P[3/0] and P[1/1] have each one free input parameter;
for P[3/0] the first root $t_1$ is the specified input, and for P[1/1]
the first pole $u_1$, where the notation (\ref{fPade}) of
Appendix \ref{app:impl} is used. The case EE is given in
Eq.~(\ref{EE}), and has three free parameters. We recall that
an apparently additional parameter in the {\it Ans\"atze\/} for $f(Y)$
is fixed by the pQCD condition (\ref{pQCDf}).
In addition, we present the values of $a(Q^2)$ at the initial
condition scale $\mu_{\rm in}^2 = (3 m_c)^2$ ($m_c=1.27$ GeV)
and at $Q^2=0$; and the threshold value $x_{\rm thr}$ of
the discontinuity function $\rho_1(\sigma) = {\rm Im} a(-\sigma - i \epsilon)$,
where: $z_{\rm thr} = x_{\rm thr} - i \pi$, $\sigma_{\rm thr} = 
(3 m_c)^2 \exp(x_{\rm thr})$. Further, the corresponding threshold mass 
$M_{\rm thr}$ is given [$M_{\rm thr} = 3 m_c \exp(x_{\rm thr}/2)$].
\begin{table}
\caption{Four cases of $\beta$ function $(f(Y))$, with chosen
input parameters. Given are the resulting RSch parameters
$c_n$ ($n=2,3,$ and $4$), and the values of $a(Q^2)$ at $Q^2=(3 m_c)^2$ and $Q^2=0$.
Further, the resulting threshold parameter $x_{\rm thr}$ and
the threshold mass $M_{\rm thr}$ (in GeV) are given. Recall that 
$a((3 m_c)^2,{\overline {\rm MS}}) = 0.07245$.}
\label{tabinput1}  
\begin{ruledtabular}
\begin{tabular}{ll|rrrrrrl}
$f$ & Input & $c_2$ & $c_3$ & $c_4$ & $a((3 m_c)^2)$ & $a_0 \equiv a(0)$ &
$x_{\rm thr}$ & $M_{\rm thr}$ [GeV]
\\
P[1/0] & -- & -37.02 & 0  & 0 & 0.06047 & 0.1901 & -5.948 & 0.195
\\
P[3/0] & $t_1=1+i 0.45$ & -39.55 & 115.88 & -105.80 & 0.06066 & 0.4562 & -11.092 & 0.015
\\
P[1/1] & $u_1=-0.1$ & -37.54 &  18.84 & -9.46 & 0.06048 & 0.1992 & -6.060 & 0.184
\\
EE  & $Y_1=0.1$, $k_1=10.0$, $k_2=11.0$ & -10.80 & -157.62 & -644.32 & 0.06544 & 0.2360 &
-5.403 & 0.256
\end{tabular}
\end{ruledtabular}
\end{table}

\begin{figure}[htb] %\unitlength=1mm
\begin{minipage}[b]{.49\linewidth}
% \centering\includegraphics[width=85mm]{plfP10.jpg}
%\centering\includegraphics[width=85mm]{figxl8a.eps}
\centering\includegraphics[width=85mm]{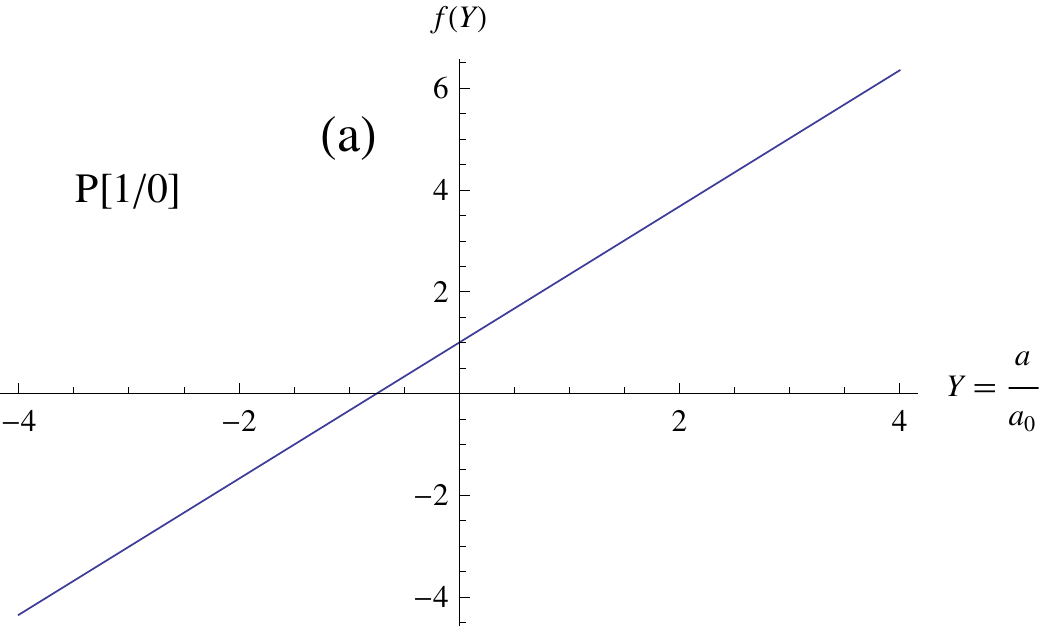}
%\centering\includegraphics[width=85mm]{figxl8a.jpg}
\end{minipage}
\begin{minipage}[b]{.49\linewidth}
% \centering\includegraphics[width=85mm]{plbfP10.jpg}
%\centering\includegraphics[width=85mm]{figxl8b.eps}
\centering\includegraphics[width=85mm]{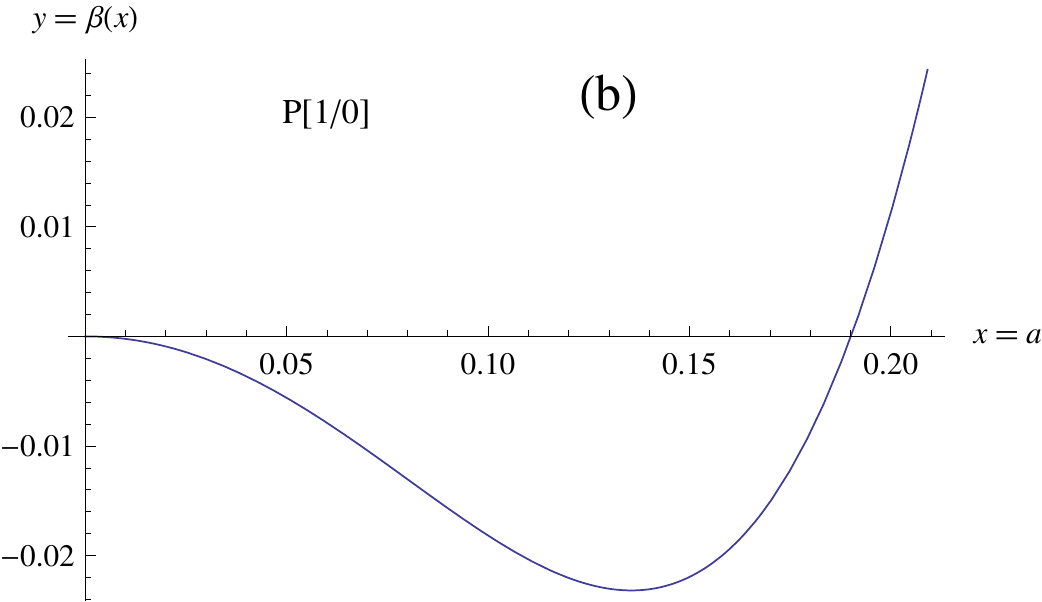}
%\centering\includegraphics[width=85mm]{figxl8b.jpg}
\end{minipage}
\vspace{-0.4cm}
 \caption{\footnotesize  (a) $f(Y)$ function as defined by 
Eq.~(\ref{ans1}), for real values of $Y \equiv a/a(0)$,
for the case of $f$ being P[1/0] linear function
($\Leftrightarrow$ P[2/0] with $r_2=0$); (b) $\beta(x)$ function
for the same case, for positive $x \equiv a$.}
\label{betaP10}
 \end{figure}

\begin{figure}[htb] %\unitlength=1mm
\begin{minipage}[b]{.49\linewidth}
% \centering\includegraphics[width=85mm]{plfEEy1_01.jpg}
%\centering\includegraphics[width=85mm]{figxl9a.eps}
\centering\includegraphics[width=85mm]{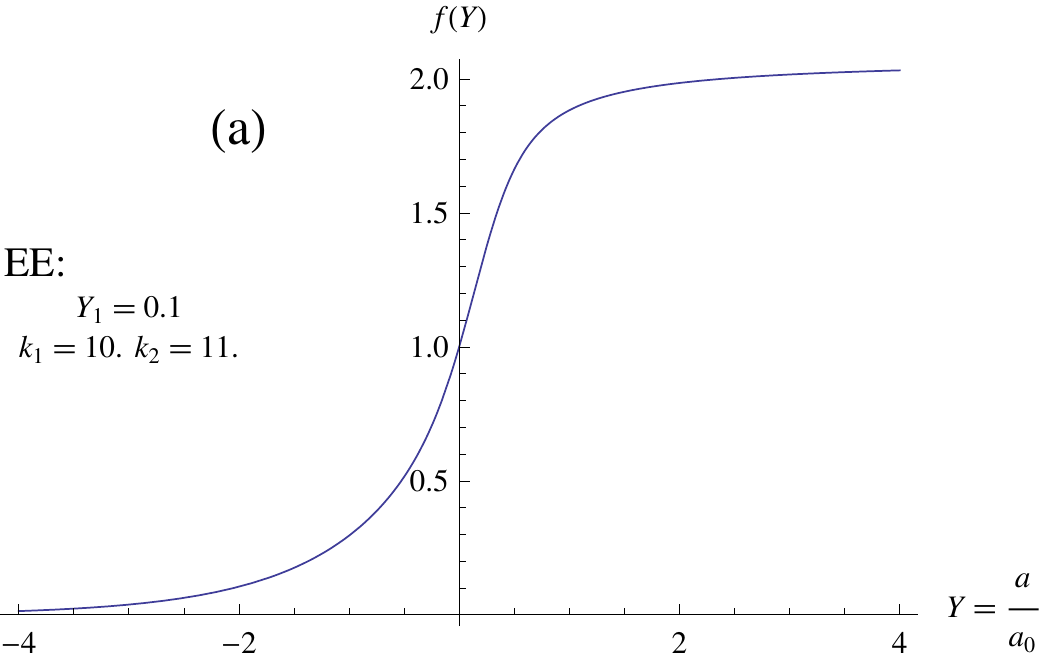}
%\centering\includegraphics[width=85mm]{figxl9a.jpg}
\end{minipage}
\begin{minipage}[b]{.49\linewidth}
% \centering\includegraphics[width=85mm]{plbfEEy1_01.jpg}
%\centering\includegraphics[width=85mm]{figxl9b.eps}
\centering\includegraphics[width=85mm]{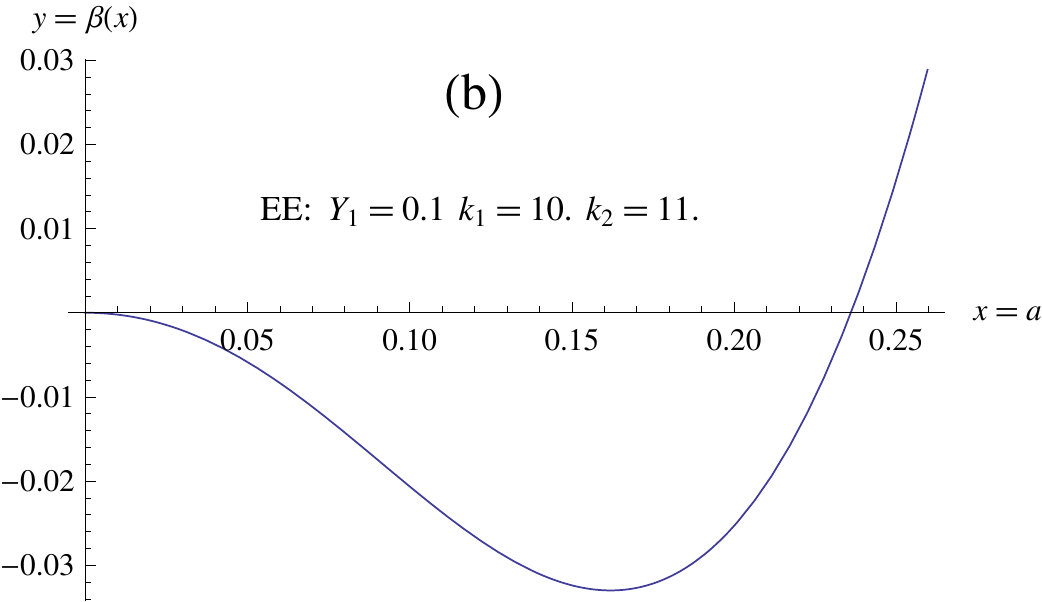}
%\centering\includegraphics[width=85mm]{figxl9b.jpg}
\end{minipage}
\vspace{-0.4cm}
 \caption{\footnotesize Same as in Fig.~\ref{betaP10}, but this
time $f(Y)$ being the exponential-related function EE, Eq.~(\ref{EE}).}
\label{betaEE}
 \end{figure}

\begin{figure}[htb] %\unitlength=1mm
\begin{minipage}[b]{.49\linewidth}
% \centering\includegraphics[width=85mm]{plaP10.jpg}
%\centering\includegraphics[width=85mm]{figxl10a.eps}
\centering\includegraphics[width=85mm]{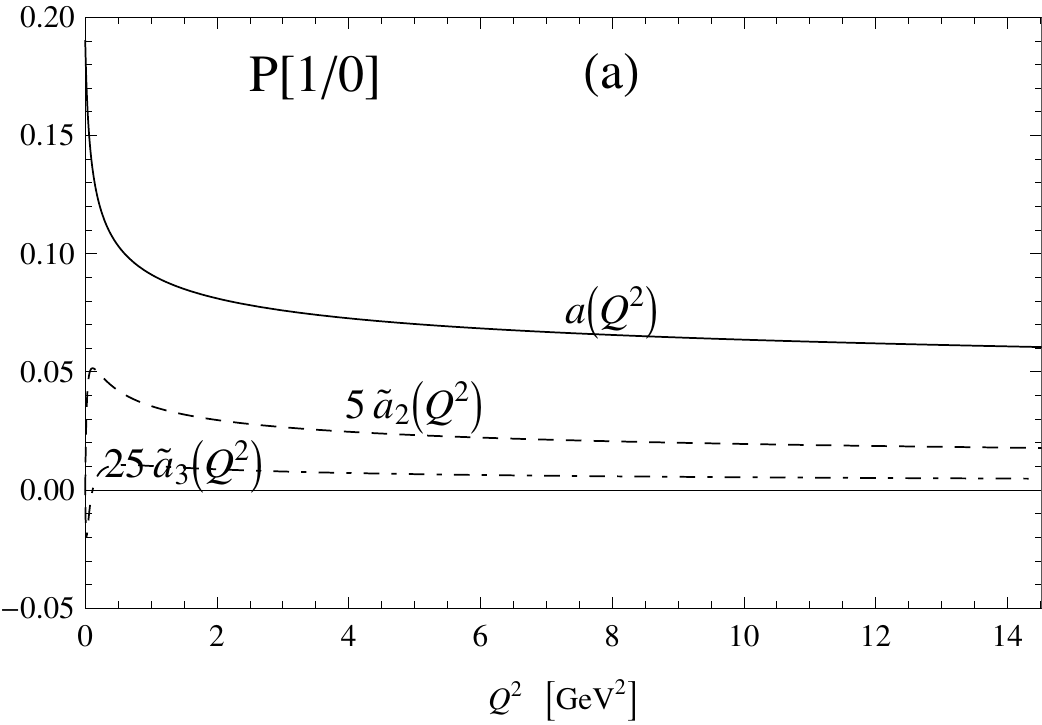}
%\centering\includegraphics[width=85mm]{figxl10a.jpg}
\end{minipage}
\begin{minipage}[b]{.49\linewidth}
% \centering\includegraphics[width=85mm]{plaP10low.jpg}
%\centering\includegraphics[width=85mm]{figxl10b.eps}
\centering\includegraphics[width=85mm]{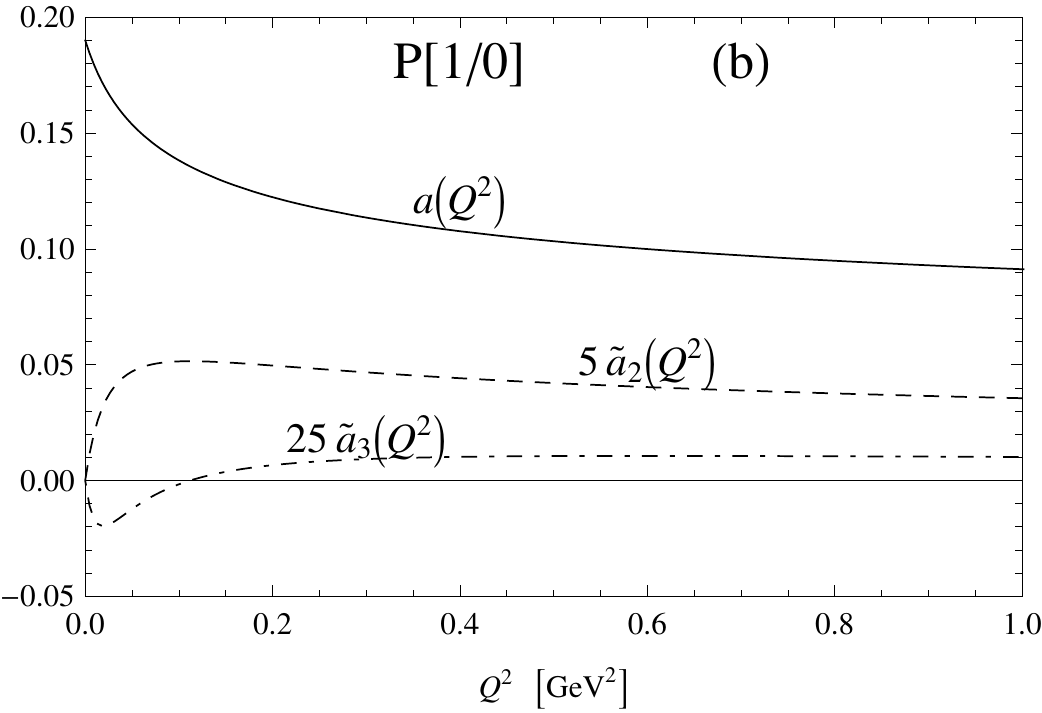}
%\centering\includegraphics[width=85mm]{figxl10b.jpg}
\end{minipage}
\vspace{-0.4cm}
 \caption{\footnotesize  (a) Analytic coupling $a(Q^2)$ and its
higher order analogs ${\ta}_{n+1}$ ($n=1,2$) as defined in Eq.~(\ref{tAn}),
for positive $Q^2$, for the model P[1/0]. 
For better visibility, the higher order analogs
are scaled by factors of $5$ and $5^2$, respectively.
(b) Same as in (a), but at lower $Q^2$. We recall that,
formally: $\ta_{n+1} = a^{n+1} + {\cal O}(a^{n+2})$.}
\label{plaP10}
 \end{figure}

\begin{figure}[htb] %\unitlength=1mm
\begin{minipage}[b]{.49\linewidth}
% \centering\includegraphics[width=85mm]{plaEEy1_01.jpg}
%\centering\includegraphics[width=85mm]{figxl11a.eps}
\centering\includegraphics[width=85mm]{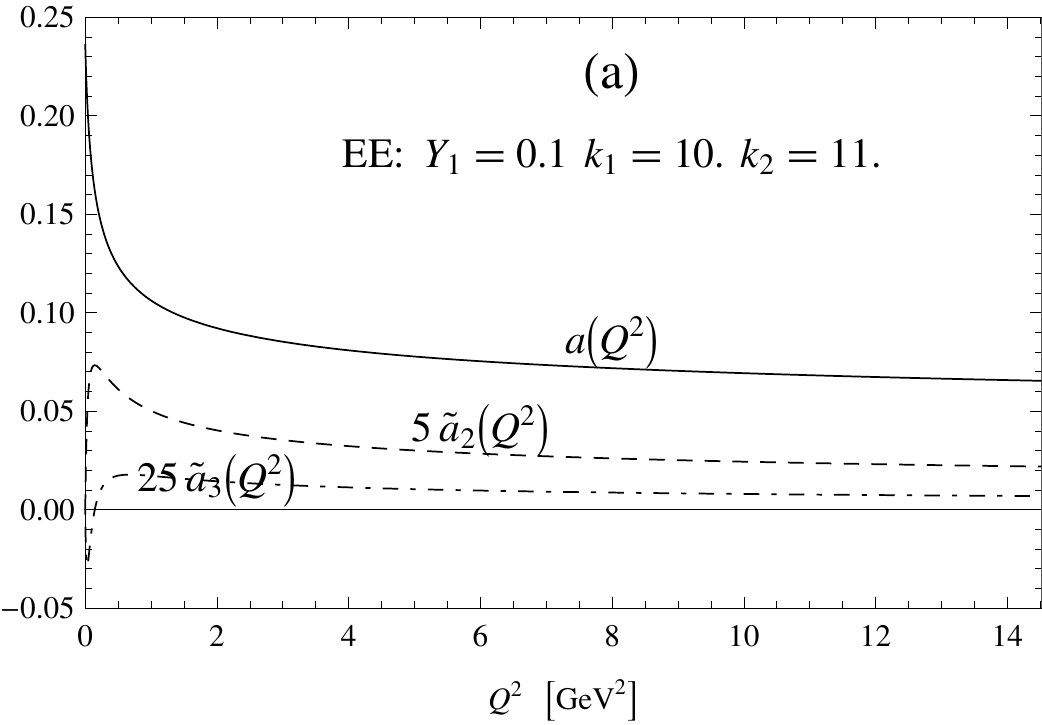}
%\centering\includegraphics[width=85mm]{figxl11a.jpg}
\end{minipage}
\begin{minipage}[b]{.49\linewidth}
% \centering\includegraphics[width=85mm]{plaEEy1_01low.jpg}
%\centering\includegraphics[width=85mm]{figxl11b.eps}
\centering\includegraphics[width=85mm]{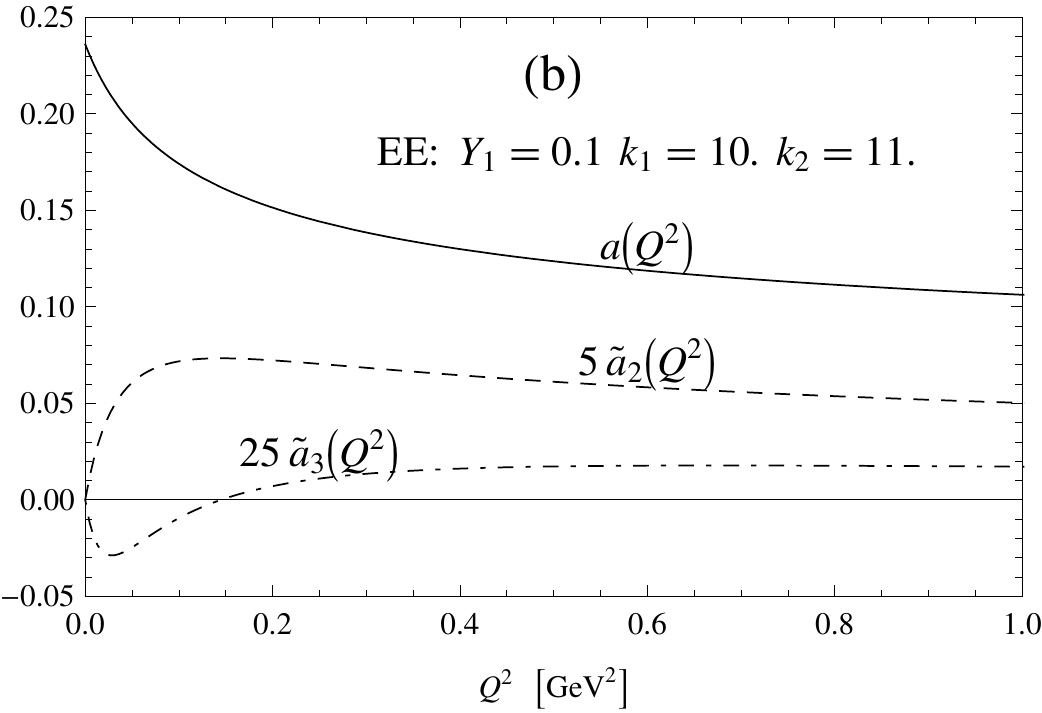}
%\centering\includegraphics[width=85mm]{figxl11b.jpg}
\end{minipage}
\vspace{-0.4cm}
 \caption{\footnotesize  Same as in Fig.~\ref{plaP10}, but for the
model EE, Eq.~(\ref{EE}).}
\label{plaEE}
 \end{figure}

For two of these models (P[1/0], and EE), we depict in
Figs.~\ref{betaP10} and \ref{betaEE}
the form of $f(Y)$ and $\beta(x)$ functions 
for real values of $Y=a/a_0$ and positive values of $x \equiv a > 0$,
respectively. In Figs.~\ref{plaP10}-\ref{plaEE} we
present the running coupling $a(Q^2)$ as a function of $Q^2$ for
positive $Q^2$ in the two models; there we include, in addition,
the higher order analytic couplings $\ta_{n+1}$ ($n=1,2$).
%\end{document}

The model with $f =$ P[1/0] is, at first sight, very similar
to the model of Ref.~\cite{MS} which was obtained on the
basis of the principle of minimal (renormalization scheme) sensitivity
(PMS) \cite{Stevenson} applied to the QCD part of $R_{e^+ e^-}(s)$
ratio. There, the beta function is also a polynomial of the fourth
degree, i.e., $f(Y)$ is linear, and it has a finite positive
value of $a(Q^2=0) \equiv a_0$. It turns out that for
the beta function of Ref.~\cite{MS} the conditions (\ref{ITEPf}) and
(\ref{pQCDf}) are fulfilled, but not the condition
of $Q^2=0$ analyticity Eq.~(\ref{Q20f}). As argued in the
present paper, such beta function will give unphysical (Landau) poles,
although in this case not on the positive $Q^2$ axis. Specifically,
for $n_f=2$ and $n_f=3$ the $Q^2=0$ analyticity condition
(\ref{Q20f}) yields in the P[1/0] case the values
$a_0=0.1761$ and $0.1901$, respectively, while the values
of $a_0$ in Ref.~\cite{MS} are $a_0=0.263$ and $0.244$, respectively.
We checked numerically that this PMS solution leads to
(Landau) poles of $a(Q^2)$ at $Q^2 \approx (-0.027 \pm i 0.065) \ {\rm GeV}^2$ for
$n_f=2$, and at $Q^2 \approx (-0.031 \pm i 0.032) \ {\rm GeV}^2$ for
$n_f=3$ (massless quarks).

Let us now apply these results to calculating low-energy
QCD observables.

We start with $r_{\tau}$.

In Table \ref{tabrtau1} we present the predicted values of $r_{\tau}$
for the choices of $\beta$ functions and input parameters
given in Table \ref{tabinput1}.
\begin{table}
\caption{The four terms in truncated analytic expansions
(\ref{rtman2tr}) and (\ref{rtan2tr}) for $r_{\tau}$,
i.e., with LB contributions resummed and the three bLB
terms organized in contour integrals of 
$\tA_{n+1} \equiv {\widetilde a}_{n+1}$ (first line)
and of $\A_{n+1} = a^{n+1}$ (second line of each model). 
In parentheses are the 
corresponding results when no LB resummation is performed,
i.e., the truncated analytic expansions Eqs.~(\ref{rtman3tr}) and
(\ref{rtan3tr}), respectively. The RScl parameter is ${\cal C}=0$.
The last column contains variations of these truncated sums when
the RScl parameter ${\cal C}$ increases from $0$ to $\ln 2$.}
\label{tabrtau1}  
\begin{ruledtabular}
\begin{tabular}{l|lllllc}
$f$ & $r_{\tau}:$ LB (LO) & NLB (NLO) & 
${\rm N}^2{\rm LB}$ (${\rm N}^2{\rm LO}$) &
${\rm N}^3{\rm LB}$ (${\rm N}^3{\rm LO}$) &
Sum (sum) & $\delta$ (${\cal C}$ dependence)
\\
\hline
\multirow{2}{10mm}{P[1/0]} & 0.1135(0.0940) & 0.0006(0.0123) & 0.0139(0.0214)& 0.0007(0.0012) & 0.1287(0.1289) & $-0.2\% (-0.4\%)$
\\
 & 0.1135(0.0940) & 0.0007(0.0137) & 0.0209(0.0340)& 0.0091(0.0113) & 0.1442(0.1529) & $-2.8\% (-2.7\%)$
\\
\hline
\multirow{2}{10mm}{P[3/0]} & 0.1200(0.0954) &  0.0007(0.0131) &  0.0184(0.0275) &  -0.0009(0.0000) & 0.1381(0.1360) & $-0.3\% (-0.8\%)$
\\
 & 0.1200(0.0954) & 0.0007(0.0141) & 0.0233(0.0369) & 0.0067(0.0087) & 0.1507(0.1550) &  $-2.4\% (-2.9\%)$
\\
\hline
\multirow{2}{10mm}{P[1/1]} & 0.1142(0.0941) & 0.0006(0.0124) & 0.0146(0.0224) & 0.0005(0.0011) & 0.1300(0.1300) & $-0.2\% (-0.5\%)$
\\
  &  0.1142(0.0941) & 0.0007(0.0138) & 0.0213(0.0344) & 0.0088(0.0109) & 0.1450(0.1532) & $-2.8\%(-2.7\%)$
\\
\hline
\multirow{2}{10mm}{EE}  & 0.1348(0.1088) & 0.0009(0.0173) & 0.0025(0.0156) & 0.0048(0.0061) & 0.1466(0.1478) & $-0.8\%(-1.2\%)$
\\
 &  0.1348(0.1088) & 0.0009(0.0180) & 0.0033(0.0224) & 0.0102(0.0173) & 0.1528(0.1666) & $-2.8\%(-3.7\%)$
\end{tabular}
\end{ruledtabular}
\end{table}
Therein we separately give (in each line) the four terms of the truncated
analytic series for $r_{\tau}$ and then quote their sum.
Furthermore, for each model of $f(Y)$ we present the results for
basically two different ways of treating the higher orders. In the first
row of each model, the results of the series (\ref{rtman2tr})
are presented, which performs leading-$\beta_0$ (LB) resummation
and adds the (three) beyond-the-leading-$\beta_0$ (bLB) terms
organized in contour integrals of logarithmic derivatives
$\ta_{n+1}$ ($n=1,2,3$). In the second line, the analogous results are
presented, where now the (three) bLB terms are contour integrals
of powers $\A_{n+1} \equiv a^{n+1}$, Eq.~(\ref{rtan2tr}).
At each of the entries, the corresponding terms are given when
no LB resummation is performed, cf.~Eqs.~(\ref{rtman3tr}), (\ref{rtan3tr}).
The RScl parameter used is ${\cal C}=0$, i.e., the radius of the contour
in the $Q^2$ plane is $m_{\tau}^2$. In the last column,
the relative variation of the sum is given when the RScl parameter is
increased from ${\cal C}=0$ to $\ln 2$, i.e., the radius of the
contour integration is increased to $2 m_{\tau}^2$.
The results using
the powers $a^{n+1}$ for the bLB (or: higher order) contributions
show significantly less stability under the RScl variation; the
reason for this lies in two numerical facts:
\begin{itemize} 
\item
The expansion coefficient $(t_{\rm Adl})_3$ of the latter series
is usually larger than the corresponding coefficient $(T_{\rm Adl})_3$ 
of the series containing $\ta_{n+1}$: $|(t_{\rm Adl})_3| > |(T_{\rm Adl})_3|$;
this seems to be true in all the RSch's dictated by the presented 
$\beta$ functions.
\item
Apparently in all cases we have $|\ta_{n+1}| < |a^{n+1}|$,
although formally $\ta_{n+1} = a^{n+1} + {\cal O}(a^{n+2})$.
\end{itemize}
Furthermore, the variations of the result under variations of RScl
are generally smaller when LB resummation is performed.
Therefore, we will consider as our preferred choice the
evaluated values of the first lines (not in parentheses) of each model in 
Table \ref{tabrtau1}, i.e., the evaluations using
$\ta_{n+1}$ for the higher order contributions,
i.e., Eq.~(\ref{rtman2tr}).

We note that the obtained values of $r_{\tau}$ 
(see the ``sum'' in Table \ref{tabrtau1})
are all much too low when compared with the experimental value
$r_{\tau}^{\rm (exp)} = 0.203 \pm 0.004$ 
(cf.~Appendix \ref{app:rtauexp}). 
In fact, the free parameters in the {\it Ans\"atze\/} for 
$f(Y)$ of the beta function were chosen in Tables 
\ref{tabinput1}-\ref{tabrtau1} in such a way as to (approximately) maximize
the result for $r_{\tau}$ while still maintaining analyticity of $a(Q^2)$
(i.e., no Landau singularities).\footnote{
When $f(Y)$ is P[2/0], it turns out that the largest evaluated value
of $r_{\tau}$ is obtained when $r_2=0$ in Eq.~(\ref{P2o1}), i.e.,
when $f(Y)$ reduces to a linear function P[1/0].}
We can see that the preferred evaluation method,
i.e., the first line of each case, gives us always a
value $r_{\tau} < 0.15$. We tried many
choices for the function $f(Y)$ of Eq.~(\ref{ans1}), 
fulfilling all conditions (\ref{ITEPf})-(\ref{Q20f}),
and scanning over the remaining free parameters in $f(Y)$. It turned out
that $r_{\tau} < 0.16$ always as long as Landau poles were absent.\footnote{
In some cases, e.g., when increasing the value of $Y_1$ in
the case EE, the preferred evaluation method, Eq.~(\ref{rtman2tr}),
gives us values of $r_{\tau}$ between 0.15 and 0.16. However, in such
cases, it is not any more clear that the analyticity is maintained;
increasing $Y_1$ even further leads to clear appearance of Landau poles.}
Only when free parameters were chosen such that Landau poles appeared,
was it possible to increase $r_{\tau}$ beyond $0.16$.

As the second example we consider the Bjorken polarized
sum rule (BjPSR) $d_{\rm Bj}(Q^2)$.

In Table \ref{tabBj1} we present results for 
$d_{\rm Bj}(Q^2)$ in the aforementioned cases,
at three of those low values of $Q^2$ where experimental results are
available: $Q^2 = 1.01$, $2.05$, and $2.92 \ {\rm GeV}^2$.
\begin{table}
\caption{Bjorken polarized sum rule (BjPSR) results $d_{\rm Bj}(Q^2)$
for the four considered $\beta$ {\it Ans\"atze\/}, 
evaluated with the truncated analytic expansions
(\ref{man2tr}) and (\ref{an2tr}),
i.e., with LB contributions resummed and the three bLB
terms $\propto \tA_{n+1} \equiv \ta_{n+1}$ (first line)
and $\propto \A_{n+1} = a^{n+1}$ (second line). In parentheses are the 
corresponding results when no LB resummation is performed,
i.e., truncated analytic expansions Eqs.~(\ref{man3tr}) and
(\ref{an3tr}), respectively. The RScl parameter is ${\cal C}=0$.
In brackets, the corresponding
variations of the results under the RScl variation are given
(see the text for details). For explanation of the experimental values in the
last (four) lines, see the text for details.} 
\label{tabBj1} 
\begin{ruledtabular}
\begin{tabular}{llll}
$f$ & $d_{\rm Bj}(Q^2):\  Q^2=1.01 \ {\rm GeV}^2$ & $Q^2=2.05 \ {\rm GeV}^2$ &
$Q^2=2.92 \ {\rm GeV}^2$
\\
\hline 
\multirow{2}{10mm}{P[1/0]} 
& $0.1343[+0.3\%]$ $(0.1420[-1.9\%])$ & $ 0.1208[-0.1\%]$ $(0.1255[-0.5\%])$ & $0.1140[-0.2\%]$ $(0.1173[-0.7\%])$
\\
& $0.1535[-4.1\%]$ $(0.1974[-5.1\%])$ & $ 0.1313[+2.8\%]$ $(0.1552[-4.1\%])$ & $0.1218[-2.4\%]$ $(0.1393[-3.6\%])$ 
\\
\hline
\multirow{2}{10mm}{P[3/0]} 
& $0.1609[-0.4\%]$ $(0.1630[-1.9\%])$ & $0.1366[-0.4\%]$ $(0.1361[-2.0\%])$ & $0.1261[-0.4\%]$ $(0.1249[-1.9\%])$ 
\\
& $0.1773[-3.7\%]$ $(0.2053[-6.3\%])$ & $0.1456[-2.6\%]$ $(0.1587[-4.6\%])$ & $0.1329[-2.2\%]$ $(0.1417[-4.0\%])$ 
\\
\hline
\multirow{2}{10mm}{P11} 
& $0.1373[+0.2\%]$ $(0.1450[-1.5\%])$ & $0.1226[-0.1\%]$ $(0.1270[-0.6\%])$ & $0.1154[-0.2\%]$ $(0.1184[-0.9\%])$ 
\\
& $0.1561[-4.0\%]$ $(0.1985[-5.3\%])$ & $0.1329[-2.8\%]$ $(0.1557[-4.2\%])$ & $0.1231[-2.4\%]$ $(0.1396[-3.7\%])$ 
\\
\hline
\multirow{2}{10mm}{EE}  
& $0.1507[+0.3\%]$ $(0.1659[-3.7\%])$ & $0.1338[+0.1\%]$ $(0.1434[-1.0\%])$ & $0.1256[+0.1\%]$ $(0.1324[-1.0\%])$ 
\\
 & $0.1436[+0.7\%]$ $(0.2300[-6.8\%])$ & $0.1304[+0.5\%]$ $(0.1725[-5.1\%])$ & $0.1232[+0.4\%]$ $(0.1521[-4.4\%])$
\\
\hline
Exp. (a): & $0.23 \pm 0.18$ & $0.11 \pm 0.11$ & $0.09 \pm 0.07$ 
\\
$\mu_4^{\rm p-n}=  -0.040 \pm 0.028$ & $0.23 \pm 0.12 \pm 0.13$ & $0.11 \pm 0.09 \pm 0.06$ &
$0.09 \pm 0.05 \pm 0.05$
\\
\hline
Exp. (b): & $0.30 \pm 0.18$ & $0.15 \pm 0.11$ & $0.11 \pm 0.07$
\\ 
$\mu_4^{\rm p-n}=  -0.024 \pm 0.028$ & $0.30 \pm 0.12 \pm 0.13$ & $0.15 \pm 0.09 \pm 0.06$ &
$0.11 \pm 0.05 \pm 0.05$
\end{tabular}
\end{ruledtabular}
\end{table}
As in the previous Table \ref{tabrtau1}, the first line 
of each model contains the
results with our preferred method, i.e., LB resummation and usage of
$\ta_{n+1}$ for the bLB contributions, Eq.~(\ref{man2tr}); the second line
represents the results of LB resummation and the usage of $a^{n+1}$ powers
for the bLB contributions, Eq.~(\ref{an2tr}). In the parentheses, the
corresponding results are given when no LB resummation is performed,
Eqs.~(\ref{man3tr}) and (\ref{an3tr}), respectively. In the corresponding
brackets, the variations of the results are given when the
RScl parameter varies either from ${\cal C}=0$ ($\mu^2=Q^2$) to
${\cal C}=\ln 2$ ($\mu^2 = 2 Q^2$), or from  ${\cal C}=0$ to
${\cal C}=\ln (1/2)$ ($\mu^2 = Q^2/2$) -- the larger of the variations
is given. As in the case of $r_{\tau}$, we see that the most stable evaluation 
under variations of RScl is the LB resummation and the usage 
of $\ta_{n+1}$ for the bLB contributions, Eq.~(\ref{man2tr}).

For comparison, we include in Table \ref{tabBj1} (last lines)
three sets of experimental
data based on the JLab CLAS EG1b (2006) measurements \cite{Deur2008}
of the $\Gamma_1^{\rm p-n}(Q^2)$ sum rule for spin-dependent proton and neutron
structure functions $g_1^{\rm p,n}$ \cite{Kataev:1994gd}.
$\Gamma_1^{\rm p-n}$ is connected to $d_{\rm Bj}$ in the following way:
\ba
\Gamma_1^{\rm p-n}(Q^2) &\equiv& \int_0^1 \; d x_{\rm Bj} \;
\left( g_1^{\rm p}(x_{\rm Bj},Q^2) -  g_1^{\rm n}(x_{\rm Bj},Q^2) \right) 
\label{Bjdef}
\\
&=&
\frac{g_A}{6} \left( 1 - d_{\rm Bj}(Q^2) \right) +
\sum_{j=2}^{\infty} \frac{ \mu_{2 j}^{\rm p-n} (Q^2) }{ (Q^2)^{j-1} } \ ,
\label{G1}
\ea
where $g_A = 1.267 \pm 0.004$ \cite{PDG2008} is the triplet axial charge,
$1 - d_{\rm Bj}(Q^2) = 1 - a(Q^2) + {\cal O}(a^2)$ is the nonsinglet leading-twist
Wilson coefficient, and $\mu_{2 j}^{\rm p-n}/Q^{2 j -2}$ ($j \geq 2$)
are the higher-twist contributions.
If we take into account the data with the elastic contribution excluded,
we can restrict ourselves to the first higher-twist term 
$\mu_4^{\rm p-n}/Q^2$. The elastic contribution affects
largely only the other higher-twist terms $\sim 1/(Q^2)^{j-1}$
with $j \geq 3$, as has been noted in Refs.~\cite{Pase1,Pase2}.
Moreover, the exclusion of the elastic contribution leads to strongly
suppressed higher-twist terms $\sim 1/(Q^2)^{j-1}$ with $j \geq 3$
\cite{Pase1} in pQCD and MA (APT) approaches. 
The first experimental set (a) for $d_{\rm Bj}(Q^2)$ in
Table \ref{tabBj1} is obtained from the measured 
values of $\Gamma_1^{\rm p-n}(Q^2)$ (with the elastic part excluded)
by subtracting the $\mu_4^{\rm p-n}/Q^2$ contribution as obtained by
a 3-parameter pQCD fit \cite{Deur2008}: 
$\mu_4^{\rm p-n} \approx \mu_4^{\rm p-n}(Q=1{\rm GeV}) = -0.040 \pm 0.028$;\footnote{
Almost the same value was obtained by the authors of 
Refs.~\cite{Pase1,Pase2}:
$\mu_4^{\rm p-n}/M_{\rm p}^2 \approx -0.048$ corresponding to
$\mu_4^{\rm p-n} \approx  -0.042$ (Ref.~\cite{Pase1}), and
$\mu_4^{\rm p-n}/M_{\rm p}^2 \approx -0.042$ corresponding to
$\mu_4^{\rm p-n} \approx  -0.037$ (Ref.~\cite{Pase2}, accounting for the
$Q^2$-dependence of $\mu_4^{\rm p-n}$ due to RG evolution.). The interesting
aspect is that they applied MA (i.e., APT) model of Refs.~\cite{ShS,MSS}
in the fit of the aforementioned JLab data, then obtaining the $1/Q^2$-term
as the sum of the contribution from the MA (APT) series and the
contribution of the explicit $1/Q^2$-term (obtained through fit).
Such a sum of $1/Q^2$-terms, in their model, is not interpreted
as originating entirely from the IR regime since MA does not satisfy the
conditions of Eq.~(\ref{ITEP}).}
the second set (b) is obtained in the same way, but now by subtracting the 
$\mu_4^{\rm p-n}/Q^2$ contribution obtained by 
a 4-parameter pQCD fit \cite{Deur2008}:
$\mu_4^{\rm p-n} \approx \mu_4^{\rm p-n}(Q=1{\rm GeV}) =  -0.024 \pm 0.028$.
In the second line of each experimental set, the uncertainties were split
into the contribution coming from the uncertainty of the measured value of 
$\Gamma_1^{\rm p -n}(Q^2)$ and the one from the uncertainty of the
fitted value $\mu_4^{\rm p-n}$ \cite{Deur2008}.

We see from Table \ref{tabBj1} that the evaluated values for BjPSR
lie in general relatively close to the central experimental
values $d_{\rm Bj}(Q^2)_{\rm exp.}$:
$d_{\rm Bj}(Q^2)_{\rm exp.}= 0.23$ (or $0.30$) for $Q^2=1.01 \ {\rm GeV}^2$;
$0.11$ (or $0.15$) for $Q^2=2.05 \ {\rm GeV}^2$;
$0.09$ (or $0.11$) for $Q^2=2.92 \ {\rm GeV}^2$.
However, in contrast to $r_{\tau}$, the experimental uncertainties are
now much larger and the theoretical predictions lie well
within the large intervals of experimental uncertainties. 

\section{Tackling the problem of too low $r_{\tau}$}
\label{sec:beta3}

The problem of too low $r_{\tau}$, encountered in the previous Section, 
appears to be common to all or most of the anQCD models.
For example, in the MA of Shirkov, Solovtsov and Milton
\cite{ShS,MSS,Sh,Shirkov:2006gv,MSSY}, when
adjusting ${\overline \Lambda}$ to such a value as to
reproduce higher energy QCD observables ($Q^2 \agt 10^1 \ {\rm GeV}^2$),
i.e., ${\overline \Lambda} \approx 0.4$ GeV, the resulting\footnote{
The value ${\overline \Lambda}=0.4$ GeV corresponds to the 
$\Lambda$ value in the Lambert function \cite{Gardi:1998qr}
for the (MA) coupling $\A_1(Q^2)$ in the 't Hooft RSch $\Lambda_{\rm Lambert}
=0.551$ GeV. In general, it can be checked that the following relation
holds: $\Lambda_{\rm Lambert} \approx {\overline \Lambda} \exp(0.3205)$,
and this holds irrespective of whether we consider pQCD or MA couplings.}
value of (massless and strangeless) $r_{\tau}$ is about $0.140$-$0.141$
\cite{MSS,MSSY,CV2},
much too low. The results of the previous section indicate
that this problem persists even in anQCD models which, unlike MA,
fulfill the ITEP-OPE condition (\ref{ITEP}). The aspect of anQCD
models which appears to cause the tendency toward too low values of
$r_{\tau}$ is the absence of (unphysical) Landau cut along
the positive $Q^2$ axis ($0 \leq Q^2 < Q_{\rm LP}^2$).\footnote{
A somewhat similar reasoning can be found in Ref.~\cite{Geshkenbein:2001mn}.}
Therefore, we are apparently facing a strange situation: 
\begin{itemize}
\item
In pQCD the Landau cut of the coupling gives a numerically positive contribution
to $r_{\tau}$, and pQCD is able to reproduce the experimental value of
$r_{\tau}$ (cf.~Refs.~\cite{Geshkenbein:2001mn,Braaten:1988hc,Pivovarov:1991rh,
Le Diberder:1992te,Ball:1995ni,ALEPH2,ALEPH3,DDHMZ,Ioffe,Beneke:2008ad,Caprini:2009vf,DescotesGenon:2010cr,MY},
because of this (unphysical) feature of the theory.
\item
In anQCD the physically unacceptable low-energy (Landau) singularities
of the coupling are eliminated, but then the values of $r_{\tau}$ tend to
decrease too much.
\end{itemize}

Here we indicate one possible solution to this problem (cf.~also
our shorter version \cite{CKV1}). Table \ref{tabrtau1} indicates
that the LB-resummed contribution to $r_{\tau}$ cannot surpass
the values $0.14$-$0.15$. We performed many trials with various forms
of $f(Y)$ functions and were not able to obtain larger values of
$r_{\tau}^{\rm (LB)}$.
But the ${\rm N}^2 {\rm LB}$ term, which is the only 
nonnegligible bLB term in Table \ref{tabrtau1}, can 
be increased by increasing the coefficient $(T_{\rm Adl})_2$
of expansion (\ref{rtman2tr}) while maintaining,
at least approximately, the values of $a(Q^2)$ and
$\ta_{n+1}(Q^2)$ for most of the complex $Q^2$. It can be deduced
from the presentation in Appendix \ref{app:bLB} that the
RSch dependence of coefficient $(T_{\rm Adl})_2$ is
in the contribution $(-c_2 + {\overline c}_2)$.
Therefore, if we multiply the $f(Y)$ function by
a factor $f_{\rm fact}(Y)$, which is close to unity for most of
the values of $Y$ ($\equiv a/a_0$) but which
significantly decreases the RSch parameter $c_2$,
the value of $(T_{\rm Adl})_2$ will increase while
the values of of $a(Q^2)$ and $\ta_{n+1}(Q^2)$ will not change strongly 
for most of the complex $Q^2$ values.\footnote{
The next-to-leading-$\beta_0$ (NLB) term cannot be increased in this way, because the 
coefficient $(T_{\rm Adl})_1 = 1/12$ turns out to be RSch independent
(and small).}
This can be achieved by the following replacement:
 \ba
f_{\rm old}(Y) \mapsto f_{\rm new}(Y) &=&
f_{\rm old}(Y) f_{\rm fact}(Y) \ , 
\label{modf}
\\
{\rm with:} \; f_{\rm fact}(Y) &=& \frac{ (1 + B Y^2)}{(1 + (B+K) Y^2)} \ ,
\quad (1 \ll K \ll B) \ .
\label{ffact}
\ea
The function $f_{\rm fact}(Y)$ is really close to unity for
most $Y$'s because $K \ll B$; and it decreases
the $c_2$ RSch parameter to low negative values [cf.~Eq.~(\ref{btexp1})] 
because $1 \ll K$ ($c_2 \sim -K$). 
More specifically, expansion in powers of $Y \equiv a/a_0$ then gives
the RSch coefficients $c_n$ with large absolute values
$c_2 \approx - K/a_0^2 (\sim -K)$; $c_3 \approx - c_1 K/a_0^2 (\sim -K)$;
$c_4 \approx B K/a_0^4 (\sim B K)$; etc. 
This implies that the coefficients $(T_{\rm Adl})_n$,
$(t_{\rm Adl})_n$, $({\widetilde d_{\rm Adl}})_n$ and $(d_{\rm Adl})_n$
appearing in analytic expansions Eqs.~(\ref{rtman2})-(\ref{rtan3tr})
behave as $\approx -c_2 \sim K$ for $n=2$; $\sim \pm c_2, -c_3 \sim \pm K$
for $n=3$; $\sim - c_4 \sim - B K$ for $n=4$; etc.
Therefore, these coefficients are large for $n=2, 3$, and
even much larger for $n \geq 4$.
In fact, it turns out that the larger $B$ is, the less
the LB contribution $r_{\tau}^{\rm (LB)}$ decreases. However, then the 
absolute values of coefficients of analytic expansions 
Eqs.~(\ref{rtman2})-(\ref{rtan3tr}) increase explosively for $n \geq 4$.
On the other hand, when $B$ ($\gg 1$)
decreases, the aforementioned divergence of the series
(\ref{rtman2}) at $n \geq 4$ becomes less dramatic, but then
$r_{\tau}^{\rm (LB)}$ decreases and it becomes difficult to
reproduce the experimental value $r_{\tau} \approx 0.203$.
We chose the values of $B$ in each model such that, roughly,
$r_{\tau}^{\rm (LB)} \approx 0.10$ or above (if possible).

Further, it turns out that these modifications 
(i.e., inclusion of $f_{\rm fact}$) do not
destroy the analyticity of $a(Q^2)$. The (two- and three-dimensional)
diagrams presented in the figures of the previous section
change only little when the modification factor
(\ref{ffact}) is introduced in the
corresponding beta functions. 
 
The numerical results in the models of Tables \ref{tabinput1}, 
\ref{tabrtau1}, \ref{tabBj1} of the previous section, 
modified by replacements (\ref{modf})-(\ref{ffact}) in
the aforementioned way so that the preferred evaluation
method Eq.~(\ref{rtman2tr}) gives $r_{\tau} = 0.203$, 
are given in the corresponding Tables \ref{tabinput2}, 
\ref{tabrtau2}, \ref{tabBj2}.
\begin{table}
\caption{Four models of $\beta$ function $(f(Y))$ of the previous section, 
with modification Eqs.~(\ref{modf})-(\ref{ffact}), with inputs
as given in Table \ref{tabinput1}, and the values of the
additional input parameters $K$ and $B$ ($1 \ll K \ll B$) adjusted so
that the evaluation
method Eq.~(\ref{rtman2tr}) gives $r_{\tau} = 0.203$.
Given are the resulting RSch parameters
$c_n$ ($n=2,3,$ and $4$), and the values of $a(Q^2)$ at $Q^2=(3 m_c)^2$ and $Q^2=0$,
as well as the resulting threshold parameter $x_{\rm thr}$ and
the threshold mass $M_{\rm thr}$ (in GeV).}
\label{tabinput2}  
\begin{ruledtabular}
\begin{tabular}{ll|rrrrrrl}
$f_{\rm old}$ & Input $f_{\rm fact}$ & $c_2$ & $c_3$ & $c_4$ & $a((3 m_c)^2)$ & 
$a_0 \equiv a(0)$ &
$x_{\rm thr}$ & $M_{\rm thr}$ [GeV]
\\
P[1/0] & $B=4000,K=6.71$ & -222.06 & -329.13  & $2.047 \times 10^{7}$ & 0.05763 & 0.1904 & -6.331 & 0.161
\\
P[3/0] & $B=5000,K=44.5$ & -249.65 & -260.93  & $5.036 \times 10^{6}$ & 0.05430 & 0.4597 & -12.023 & 0.009
\\
P[1/1] & $B=4000,K=7.11$ & -216.04 &  -298.77 & $1.799 \times 10^{7}$ & 0.05761 & 0.1995 & -6.448 & 0.152
\\
EE  & $B=1000,K=5.4$ & -106.80 & -326.71 &  $1.721 \times 10^{6}$ & 0.06125 & 0.2370 &
-5.887 & 0.201
\end{tabular}
\end{ruledtabular}
\end{table}
\begin{table}
\caption{The evaluated quantity $r_{\tau}$ as in Table \ref{tabrtau1}, 
but now with modifications
Eqs.~(\ref{modf})-(\ref{ffact}), as given in Table \ref{tabinput2},
so that the evaluation method Eq.~(\ref{rtman2tr}) gives $r_{\tau} = 0.203$.}
\label{tabrtau2}  
\begin{ruledtabular}
\begin{tabular}{l|lllllc}
$f_{\rm old}=f/f_{\rm fact}$ & $r_{\tau}:$ LB (LO) & NLB (NLO) & 
${\rm N}^2{\rm LB}$ (${\rm N}^2{\rm LO}$) &
${\rm N}^3{\rm LB}$ (${\rm N}^3{\rm LO}$) &
Sum (sum) & $\delta$ (${\cal C}$ dependence)
\\
\hline
\multirow{2}{10mm}{P[1/0]} 
 & 0.1060(0.0880) & 0.0006(0.0110) & 0.0907(0.0974)& 0.0057(0.0063) & 0.2030(0.2026) & $-1.4\% (-1.5\%)$
\\
 & 0.1060(0.0880) & 0.0006(0.0121) & 0.1264(0.1373)& 0.0552(0.0438) & 0.2882(0.2812) & $-8.4\% (-10.1\%)$
\\
\hline
\multirow{2}{10mm}{P[3/0]} 
 & 0.0997(0.0815) &  0.0005(0.0099) &  0.0967(0.1029) &  0.0061(0.0068) & 0.2030(0.2011) & $-2.5\% (-2.7\%)$
\\
 & 0.0997(0.0815) & 0.0005(0.0104) & 0.1143(0.1230) & 0.0447(0.0347) & 0.2592(0.2496) &  $-7.6\% (-9.6\%)$
\\
\hline
\multirow{2}{10mm}{P[1/1]} 
  & 0.1064(0.0880) & 0.0006(0.0111) & 0.0902(0.0971) & 0.0058(0.0063) & 0.2030(0.2025) & $-1.6\% (-1.7\%)$
\\
  & 0.1064(0.0880) & 0.0006(0.0121) & 0.1229(0.1338) & 0.0532(0.0423) & 0.2832(0.2762) & $-8.3\% (-10.0\%)$
\\
\hline
\multirow{2}{10mm}{EE}  
 & 0.1247(0.0987) & 0.0007(0.0146) & 0.0678(0.0786) & 0.0097(0.0108) & 0.2030(0.2027) & $-2.4\%(-2.8\%)$
\\
 & 0.1247(0.0987) & 0.0008(0.0149) & 0.0787(0.0934) & 0.0432(0.0385) & 0.2474(0.2456) & $-8.8\%(-10.3\%)$
\end{tabular}
\end{ruledtabular}
\end{table}
\begin{table}
\caption{The evaluated quantity BjPSR $d_{\rm Bj}(Q^2)$ as in Table \ref{tabBj1}, 
but now with modifications Eqs.~(\ref{modf})-(\ref{ffact}), 
as given in Table \ref{tabinput2}.
The experimentally measured values are given in the last four lines of
Table \ref{tabBj1} (see the text there for details).}
\label{tabBj2} 
\begin{ruledtabular}
\begin{tabular}{llll}
$f_{\rm old}=f/f_{\rm fact}$ & $d_{\rm Bj}(Q^2):\  Q^2=1.01 \ {\rm GeV}^2$ & $Q^2=2.05 \ {\rm GeV}^2$ &
$Q^2=2.92 \ {\rm GeV}^2$
\\
\hline 
\multirow{2}{10mm}{P[1/0]} 
& $0.2138[ -2.9\%]$ $(0.2199[ -3.5\%])$ & $ 0.1895[-1.7\%]$ $(0.1927[-2.0\%])$ & $0.1761[-2.2\%]$ $(0.1782[-2.6\%])$
\\
& $0.3795[+15.0\%]$ $(0.3673[+15.2\%])$ & $ 0.2803[+12.4\%]$ $(0.2697[+12.9\%])$ & $0.2442[+11.1\%]$ $(0.2344[+11.8\%])$ 
\\
\hline
\multirow{2}{10mm}{P[3/0]} 
& $0.2485[-4.9\%]$ $(0.2476[-5.7\%])$ & $0.2008[-4.5\%]$ $(0.1991[-5.3\%])$ & $0.1813[-4.3\%]$ $(0.1795[-5.0\%])$ 
\\
& $0.3485[+14.8\%]$ $(0.3221[+16.0\%])$ & $0.2579[+11.5\%]$ $(0.2392[+12.8\%])$ & $0.2252[+10.2\%]$ $(0.2093[+11.5\%])$ 
\\
\hline
\multirow{2}{10mm}{P11} 
& $0.2185[-2.1\%]$ $(0.2244[-2.5\%])$ & $0.1909[-2.1\%]$ $(0.1938[-2.5\%])$ & $0.1767[-2.5\%]$ $(0.1785[-3.0\%])$ 
\\
& $0.3742[+15.2\%]$ $(0.3618[+15.4\%])$ & $0.2761[+12.3\%]$ $(0.2654[+13.0\%])$ & $0.2406[+11.0\%]$ $(0.2308[+11.8\%])$ 
\\
\hline
\multirow{2}{10mm}{EE}  
& $0.2166[-3.0\%]$ $(0.2281[-4.1\%])$ & $0.1879[-2.3\%]$ $(0.1938[-3.0\%])$ & $0.1728[-2.7\%]$ $(0.1765[-3.6\%])$ 
\\
& $0.3246[+18.4\%]$ $(0.3416[+18.3\%])$ & $0.2380[+13.6\%]$ $(0.2421[+14.6\%])$ & $0.2074[+11.6\%]$ $(0.2081[+12.8\%])$ 
\end{tabular}
\end{ruledtabular}
\end{table}

When comparing Table \ref{tabrtau2} with Table \ref{tabrtau1}, we see that
the modification (\ref{modf})-(\ref{ffact}) really results in
a significantly larger ${\rm N}^2{\rm LB}$ contribution (and a somewhat
larger ${\rm N}^3{\rm LB}$ contribution) to $r_{\tau}$, reaching in this
way the middle experimental value $r_{\tau}=0.203$. The variations
$\delta$ under the variations of RScl are now larger in Table \ref{tabrtau2}
than in \ref{tabrtau1}; nonetheless, 
the evaluation method of Eq.~(\ref{rtman2tr})
is still the most stable under the RScl variations. However, now the
series for $r_{\tau}$ is strongly divergent when terms ${\rm N}^4{\rm LB}$
and higher are included, for the reasons mentioned earlier in this section.
For example, the ${\rm N}^4{\rm LB}$ contribution to $r_{\tau}$, 
in the methods of Eqs.~(\ref{rtman2tr}) and (\ref{rtman3tr})
which use $\ta_{n+1}$ in higher order contributions,
is estimated to be $\sim - 10^0=-1$. Specifically, when the RScl parameter
is ${\cal C}=0$, these terms are estimated to be
-3.1 (P[1/0]); -2.0 (P[3/0]); -3.7 (P[1/1]); -1.0 (EE).\footnote{
When using evaluation methods of Eqs.~(\ref{rtan2tr}) and (\ref{rtan3tr})
which use powers $a^{n+1}$ instead, these estimated terms are: 
-22.9 (P[1/0]); -3.9 (P[3/0]);-20.1 (P[1/1]); -2.9 (EE). 
These terms have significantly higher absolute values than
those for the methods of Eqs.~(\ref{rtman2tr}) and (\ref{rtman3tr}),
although the estimated coefficients are the same in both cases.
The reason for this difference lies in the fact that $|a^{5}(Q^2)| >
|\ta_5(Q^2)|$ for most values of (complex) $Q^2$. It appears to be a general
numerical fact in all models presented in this work
that $|a^{n+1}(Q^2)| > |\ta_{n+1}(Q^2)|$ ($n \geq1$),
although formally $\ta_{n+1} = a^{n+1} + {\cal O}(a^{n+2})$.}

It remains unclear how to deal with such an analytic series, which has
relatively reasonable convergence behavior in its first four
contributions and behaves uncontrollably for $n \geq 4$.
One might consider this behavior as an indication of the
asymptotic series nature of the expansion (``precocious asymptoticity'').
Certainly, this divergence problem appears to be the price 
that is paid to achieve in anQCD
the correct value $r_{\tau} \approx 0.20$ via $\beta$ function modification
Eqs.~(\ref{modf})-(\ref{ffact}). The modified beta functions
$\beta(a)$ now acquire poles and zeros on the imaginary axis close
to the origin in the complex $a$ plane: $a_{\rm pole}=\pm i a(0)/\sqrt{B+K}$,
$a_{\rm zero}=\pm i a(0)/\sqrt{B}$. Consequently, the convergence
radius of the perturbation expansion of $\beta(a)$ in powers of $a$ becomes short:
$R=a(0)/\sqrt{B+K}$. Nonetheless, $\beta(a)$ remains an analytic function
of $a$ at $a=0$, fulfilling thus the ITEP-OPE condition (\ref{ITEP}).
We note that such a modification of the beta function
brings us into an RSch where the absolute values of the 
(perturbative) RSch parameters $c_n$ rise fast when $n$ increases.
There is no physical equivalence of such
RSch's with the usual RSch's such as ${\overline {\rm MS}}$
or 't Hooft RSch (where $c_n=0$ for $n \geq 2$). 
For example, in these two latter RSch's,
the coupling $a(Q^2)$ is not even analytic. Physical nonequivalence
can even be discerned between, on the one hand, the much ``tamer'' RSch's
of the previous Section which give analytic $a(Q^2)$ (see Table \ref{tabinput1})
and, on the other hand, the aforementioned 
nonanalytic RSch's ${\overline {\rm MS}}$ or 't Hooft.     

When comparing the evaluated BjPSR values for the beta functions 
modified by Eqs.~(\ref{modf})-(\ref{ffact}), as presented in Table \ref{tabBj2},
with those of unmodified beta functions as presented in Table \ref{tabBj1},
we note that the modification increases the values of BjPSR, 
generally to above the experimental middle values. 
Nonetheless, the results generally
remain inside the large intervals of experimental uncertainties.
The variations of the results under the variation of the RScl are now
larger. 

The evaluation methods of Eqs.~(\ref{man2tr}) and (\ref{man3tr}),
for spacelike observables such as BjPSR, and the analogous
methods of Eqs.~(\ref{rtman2tr}) and (\ref{rtman3tr}) for the
timelike $r_{\tau}$, which use logarithmic derivatives $\ta_{n+1}$, 
are significantly more stable 
under the variation of RScl than the methods of
Eqs.~(\ref{an2tr}), (\ref{an3tr}), (\ref{rtan2tr}) and (\ref{rtan3tr}),
which use powers $a^{n+1}$. This can be seen clearly by comparing
the variations (percentages) of the first and the second line
of each anQCD model in Tables \ref{tabrtau2} and \ref{tabBj2}.
In this sense, the method of Eqs.~(\ref{man2tr}) for spacelike,
and (\ref{rtman2tr}) for timelike observables, which performs
LB resummation and uses logarithmic derivatives $\ta_{n+1}$ for
the bLB contributions, remains the preferred method, as in the
previous section.

We wish to add a minor numerical observation. 
Unlike the results of the previous section where 
the LB resummation improved significantly
the stability under the RScl variation, this improvement
becomes less clear in the results of the present section,
as can be seen by comparing the variations (percentages) outside the
parentheses with the corresponding ones inside the parentheses.
This can be understood in the following way: the modification
of $\beta$ functions by Eqs.~(\ref{modf})-(\ref{ffact}) introduced,
via large values of $|c_n|$'s,
in the expansion coefficients ${\widetilde d}_{n+1}$ and $d_{n+1}$ of the 
(spacelike) observables (here the Adler function and BjPSR) numerically
large contributions $\approx -c_{n+1}/n$ which are not a large-$\beta_0$ (LB) part
of these coefficients. The latter is true because the LB part of 
${\widetilde d}_{n+1}$ and $d_{n+1}$ is $\sim \beta_0^{n+1}$
while $c_{n+1}=\beta_{n+1}/\beta_0 \sim \beta_0^n$ 
(cf.~Appendixes \ref{app:LB} and \ref{app:bLB}).
Therefore, the LB parts of the coefficients are now not dominant, and the
LB resummation cannot be expected to improve significantly the
RScl stability of the result.

\section{Conclusions}
\label{sec:concl}

In this work we tried to address two aspects which
are not addressed by most of the analytic QCD (anQCD) models
presented up to now in the literature:
\begin{itemize}
\item
Several anQCD models, in particular the most widely used anQCD model
(minimal analytic: MA) of Shirkov, Solovtsov, and Milton 
\cite{ShS,MSS,Sh,Shirkov:2006gv},
give significantly too low
values of the well-measured (QCD-canonical) semihadronic $\tau$-decay ratio
$r_{\tau}$
once the free parameter(s) (such as ${\overline {\Lambda}}$) are adjusted
so that the models reproduce the experimental values of high-energy
QCD observables ($|Q^2| \agt 10^1 \ {\rm GeV}^2$),
cf.~Refs.~\cite{MSS,MSSY}.
\item
In most of the anQCD models presented up to now, the ITEP-OPE condition 
(\ref{ITEP}) is not fulfilled.\footnote{\label{ftnt:ad}
In Ref.~\cite{Cvetic:2007ad} an anQCD coupling $\A_1$ was constructed
directly (not from a $\beta$ function {\it Ansatz\/}) which fulfills the
ITEP-OPE condition. The construction was performed in a specific
RSch and contains several adjustable parameters. Physical observables 
were not evaluated.} 
Hence such models give nonperturbative power contributions
$\sim ({\overline {\Lambda}}^2/Q^2)^k$ of ultraviolet origin in the
(leading-twist part of the) spacelike observables ${\cal D}(Q^2)$,
contravening the ITEP-OPE philosophy \cite{Shifman:1978bx,DMW} which
postulates that nonperturbative contributions have exclusively infrared origin.
If the latter philosophy is not respected by a model, 
application of the OPE evaluation method in such a model becomes questionable.
\end{itemize}
In this work, the second aspect (ITEP-OPE) was addressed via  
construction of the analytic coupling $a(Q^2) = \alpha_s^{\rm (an.)}(Q^2)/\pi$
by starting from beta functions $\beta(a)$ analytic at $a=0$ and
performing integration of the corresponding renormalization group
equation (RGE) in the complex $Q^2$ plane. 
It then turned out that, in order to avoid the occurrence of Landau singularities
of $a(Q^2)$, it was virtually necessary to impose on the coupling $a(Q^2)$
analyticity at $Q^2=0$. We tried the construction with many different 
$\beta$ functions which fulfill such conditions and which, at the same time,
give relatively tame perturbation renormalization scheme (RSch) coefficients
$c_n \equiv \beta_n/\beta_0$ ($n=2,3,\ldots$), i.e., where the sequence
$\{ |c_n|, n=2,3,\ldots \}$ is not increasing very fast. It turned out that
all such beta functions resulted either in analytic coupling $a(Q^2)$
which gave $r_{\tau} < 0.16$, significantly below the well-measured experimental
value $r_\tau({\rm exp.}) = 0.203 \pm 0.004$
of the (strangeless and massless) $r_{\tau}$,
or the coupling $a(Q^2)$ gave $r_{\tau} > 0.16$ at the price of
developing Landau singularities. 

This persistent problem was then addressed
by a specific modification of the aforementioned beta-functions,
Eqs.~(\ref{modf})-(\ref{ffact}),
introducing in $\beta(a)$ complex poles and zeros on the imaginary
axis of the complex $a$ plane close to the origin. In this way,
the correct value $r_{\tau}=0.203$ was reproduced, and the analyticity
of $a(Q^2)$ and the ITEP-OPE condition were maintained. However,
the sequence of perturbation RSch coefficients  $\{ |c_n|, n=2,3,\ldots \}$
in such cases increases very fast starting at $n=4$. As a consequence,
in such cases the analytic evaluation series of
QCD observables (including $r_{\tau}$) starts
showing strong divergent behavior when terms $\sim \ta_5 \sim a^5$ 
are included, because the coefficients at such terms become large. 
It remains unclear how to deal properly with this problem.

In this work we evaluated, in the aforementioned anQCD models, 
the (timelike) observable $r_{\tau}$ and the spacelike observable 
Bjorken polarized sum rule (BjPSR) $d_{\rm Bj}(Q^2)$ at low $Q^2$,
by evaluating only the leading-twist contribution, and accounting
for the chirality-violating higher-twist OPE terms by estimating
and subtracting those ``mass'' terms in the case of $r_{\tau}$
(see Appendix \ref{app:rtauexp}). 
This means that the chirality-conserving higher-twist contributions,
such as the gluon condensate contribution, were not taken into account.
While the values of the chirality-violating condensates are known with
relatively high degree of precision and are expected to be the same 
in perturbative QCD (pQCD+OPE) and in anQCD (anQCD+OPE), 
the values of the chirality-conserving condensates
have in pQCD+OPE very high levels of uncertainty.
For example, the dimension-four gluon condensate, which 
%appears to be the only 
is the numerically relevant chirality-conserving condensate 
with the lowest dimension
in the evaluation of $r_{\tau}$, acquires (in pQCD+OPE) value almost
compatible with zero: $\langle a G_{\mu \nu}^2 \rangle = 0.005 \pm 0.004
\ {\rm GeV}^4$ \cite{Ioffe}, obtained by fitting pQCD+OPE evaluations
of the current-current polarization operators with the 
corresponding integrals of the experimentally
measured spectral functions of the $\tau$-decay.
In anQCD models, before fitting, the value of $\langle a G_{\mu \nu}^2 \rangle$
is a free parameter. 
In principle, the inclusion of
this parameter, i.e., inclusion of the corresponding dimension-four 
term in the anQCD+OPE evaluation of $r_{\tau}$ can give us
the correct value of $r_{\tau}$ once the value of the
parameter is adjusted accordingly, without the need to
perform the modification (\ref{modf})-(\ref{ffact}) of the beta function.
It appears that the resulting value of this parameter
$\langle a G_{\mu \nu}^2 \rangle$ in such anQCD models will be large,
especially since it enters the dimension-four term for $r_{\tau}$
with an additional suppression factor $a$. 
Another, more systematic, approach \cite{wp}
would be to extract the
value of $\langle a G_{\mu \nu}^2 \rangle$, in anQCD models presented here,
by performing analyses similar to those of Refs.~\cite{Ioffe,MY},
involving $\tau$-decay spectral functions
and suppressing the OPE contributions with dimension larger than
four by employing specific (finite energy) sum rules.  
One of the attractive features of the anQCD models presented in this work is
that most of them give results very similar to each other
[for $a(0)$, $M_{\rm thr}$, $r_{\tau}$, BjPSR -- 
see Tables.~\ref{tabinput1}-\ref{tabBj1} for nonmodified,
and \ref{tabinput2}-\ref{tabBj2} for modified $\beta$ functions]
when the $f(Y)$ function appearing in the $\beta$ function has various 
different forms, of the type P[1/0], P[1/1], or EE.

\begin{acknowledgments}
\noindent
One of the authors (G.C.) thanks D.V.~Shirkov for valuable
comments.
This work was supported in part by FONDECYT Grant No.~1095196 
and Rings Project No.~ACT119 (G.C.),
DFG-CONICYT Bilateral Project No.~060-2008 (G.C. and R.K.), 
Conicyt (Chile) Bicentenario Project No.~PBCT PSD73
and FONDECYT Grant No.~1095217 (C.V.).

\end{acknowledgments}

\appendix

\section{Implicit solutions of RGE and singularity structure}
\label{app:impl}

It is evident that for an arbitrary choice of $\beta(F)$,
even when constrained by conditions (\ref{ans1})-(\ref{Q20f}),
RGE Eq.~(\ref{RGEz}) cannot be solved analytically and one
has to resort to numerical methods. On the other hand, if one
concentrates on the question of for which type of $\beta$ function
the resulting coupling may have no Landau singularities,
more general statements can be derived by analytic methods as shown below.

We suppose that the $\beta$ function has the form Eq.~(\ref{ans1})
of Sec.~\ref{sec:beta1}. 
We will show that, if $f(Y)$ of Eq.~(\ref{ans1}) is any rational function
(Pad\'e) of type ${\rm P}[M/N]$ (with real coefficients and $M \geq N-1$), 
with the $Q^2=0$ analyticity condition (\ref{Q20f}) fulfilled, 
then there exists in the physical $z$ stripe
of $F(z)$ of Fig.~\ref{Qzplane}
($-\pi \leq {\rm Im} z < \pi$) at least one pole $z_{\rm p}$ of $F(z)$
[$F(z_{\rm p}) = \infty$] such that ${\rm Im}(z_{\rm p}) = - \pi$.
The latter means that this is a physically acceptable pole 
of $a(Q^2)$ for $Q^2 < 0$, i.e., not a Landau pole.
The function $f(Y)$ being a Pad\'e of the type ${\rm P}[M/N](Y)$ means
\be
f(Y) = f(1/t) = \frac{ (1 -t_1/t) \cdots (1-t_M/t)}
{(1-u_1/t)\cdots(1-u_N/t)} \ ,
\label{fPade}
\ee
where the normalization condition $f(1)=1$, a consequence of the
pQCD condition Eq.~(\ref{pQCDf}), is evidently fulfilled.
The fact that this Pad\'e has real coefficients must be
reflected in the fact that the zeros $t_j$ are either real,
or (some of them) appear in complex conjugate pairs, the same being 
valid for the poles $u_j$.
When using the form (\ref{fPade}) in the $\beta$ function
(\ref{ans1}) and the latter in the integral (\ref{impl1})
of the implicit solution of RGE, we end up with the
following integral:
\ba
\frac{1}{\beta_0 a_0} \int_{a_0/a_{\rm in}}^{a_0/F(z)} dt \; 
t^{M-N+1} \frac{(t-u_1)\cdots(t-u_N)}{(t-t_0)(t-t_1)\cdots(t-t_M)}
 & = & z \ ,
\label{impl2}
\ea
where $t_0=1$ is the value coming from the first factor $(1-y)$
in the $\beta$ function Eq.~(\ref{ans1}).
When $M \geq N-1$, the integrand in Eq.~(\ref{impl2}) can be split into
a sum of simple partial fractions $1/(t-t_j)$
\ba
\frac{1}{\beta_0 a_0} \int_{a_0/a_{\rm in}}^{a_0/F(z)} dt 
\left\{
1 + \sum_{j=0}^M B_j \frac{1}{(t-t_j)} \right\}
& = & z \ ,
\label{impl3}
\ea
where
\be
B_j=\frac{N_j}{D_j} \ ,
\label{Bjs1}
\ee
with
\ba
N_j &=& t_j^{M-N+1} (t_j-u_1) \cdots (t_j-u_N) \qquad (j=0,1,\ldots M) \ ,
\label{Bjs2}
\\
D_j&=&(t_j-t_0) \cdots (t_j-t_{j-1})(t_j-t_{j+1})\cdots (t_j-t_M)
\qquad (j=1,\ldots M) \ ,
\label{Bjs3}
\\
D_0 &=& (t_0-t_1)(t_0-t_2) \cdots (t_0-t_M) \ .
\label{Bjs4}
\ea
These formulas can be obtained by direct algebraic manipulations,
or by using a symbolic software.
Integration in Eq.~(\ref{impl3}) then gives the following implicit
solution of the RGE for $F=F(z)$ in the form $z=G(F)$:
\ba
z&=& \left\{ 
\frac{1}{\beta_0} \left( \frac{1}{F(z)} - \frac{1}{a_{\rm in}} \right)
+ \frac{1}{\beta_0 a_0} \sum_{j=0}^M B_j  
\ln \left( \frac{a_0/F(z) -t_j}{a_0/a_{\rm in} - t_j} \right) 
\right\} \ .
\label{impl4}
\ea
Within the sum on the rhs of Eq.~(\ref{impl4}), the term with $j=0$ is
(using $t_0=1$)
\be
\frac{1}{\beta_0 a_0} B_0 \ln \left( \frac{a_0/F(z) -1}{a_0/a_{\rm in} - 1} \right) 
\quad {\rm with:} \quad B_0=\frac{(1-u_1)\cdots(1-u_N)}{(1-t_1)\cdots(1-t_M)} \ .
\label{B0a}
\ee
Comparing $B_0$ with $f(Y)$ in Eq.~(\ref{fPade}) we realize that
$B_0=1/f(1)$. Consequently, the $Q^2=0$ analyticity condition (\ref{Q20f})
yields $B_0= \beta_0 a_0$ [where $a_0 \equiv a(Q^2=0)$]. Therefore, the total
coefficient at the $j=0$ logarithm on the rhs of Eq.~(\ref{impl4}) is
equal exactly to 1
\be
\frac{1}{\beta_0 a_0}  B_0 = 1 \ .
\label{B0b}
\ee
On the other hand, this implies that the pole locations
$z_{\rm p}$ at which $F(z_{\rm p})=\infty$ are given by
\ba
z_{\rm p} & = & \left\{ -\frac{1}{\beta_0 a_{\rm in}} +
\ln (-1) - \ln \left( \frac{a_0}{a_{\rm in}}-1 \right) 
+ \frac{1}{\beta_0 a_0} \sum_{j=1}^M B_j  
\ln \left( \frac{-t_j}{a_0/a_{\rm in} - t_j} \right) 
\right\} \ .
\label{zpgen}
\ea
Let us now investigate where these poles can be localized in the $z$-plane.
In the cases considered here, we have $0 < a_{\rm in} < a_0$ [$\equiv a(Q^2=0)$],
because otherwise (i.e., if $0 < a_0 < a_{\rm in}$) the
resulting coupling would give significantly too low values
of low-energy QCD observables such as the semihadronic
$\tau$ decay ratio\footnote{
It can be deduced from Appendix \ref{app:LB}, Eq.~(\ref{LBrt2}) and
Fig.~\ref{tFtauplz} there, that ${\widetilde F}_r(t) < 1$ and thus
the leading-$\beta_0$ (LB) contribution to $r_{\tau}$ is
$r_{\tau}^{\rm (LB)} < a_0$. On the other hand, 
$a_{\rm in} \equiv a((3 m_c)^2) < 0.075$.
Hence, when $0 < a_0 < a_{\rm in}$, we have $r_{\tau}^{\rm (LB)} < 0.075$,
significantly too low to achieve $r_{\tau} \approx 0.20$.} 
($r_{\tau}$) or the Bjorken polarized sum rule
(BjPSR) at low positive $Q^2$'s. Therefore, $a_0/a_{\rm in} > 1$.
In the following, we discuss several scenarios for locations
of poles $z_{\rm p}$:

\begin{enumerate}

\item
If, on the one hand, the roots $t_j$ are all real negative,
then in the sum over $j$'s ($j \geq 1$) on
the rhs of Eq.~(\ref{zpgen}) all logarithms 
$\ln [-t_j/(a_0/a_{\rm in} - t_j)]$ are unique and real, 
as are the coefficients $B_j$. Hence, this sum is real.
The only nonreal term on the rhs of Eq.~(\ref{zpgen}) is
$\ln(-1) = - i \pi +i 2 \pi n$.
Therefore,\footnote{Note: $-\pi \leq
{\rm Im} z < \pi$ is the physical considered stripe in the
complex $z$-plane.}
${\rm Im} z_{\rm p}= - \pi$. This means that in such a case
there is only one pole and this pole lies on the
timelike $Q^2$-axis ($Q^2 < 0$); hence, no Landau poles.
One of such cases is the one illustrated in
Fig.\ref{figP20bt}(a) of Sec.~\ref{sec:beta1}, i.e., 
the case of $f(Y)$ being ${\rm P}[1/0]$ ($r_2=0$; $M=1$, $N=0$)
with $t_1 \approx -1.338$.

\item
If, on the other hand, some of the roots $t_j$ appear as 
complex conjugate pairs,
the sum over $j$'s ($j \geq 1$) on the rhs of Eq.~(\ref{zpgen}) 
can be real and the same conclusion would apply. 
However, that sum can turn out to be nonreal
and we end up with Landau poles. How can this occur?
If, for example, $t_{j+1} = t_j^*$, then Eqs.~(\ref{Bjs1})-(\ref{Bjs4}) imply
$B_{j+1}=B_j^*$. However, the corresponding logarithms
for $j$ and $j+1$ in the sum on the rhs of Eq.~(\ref{zpgen})
are not necessarily complex conjugate to each other,
but can have a modified relation due to nonuniqueness of
logarithms of complex arguments
\be
\ln \left( \frac{-t_j}{a_0/a_{\rm in} - t_j} \right) = 
\left[ \ln \left( \frac{-t_{j+1}}{a_0/a_{\rm in} - t_{j+1}} \right) \right]^*
+ i 2 \pi n_j \ .
\label{lns}
\ee
Here, integers $n_j$ can be nonzero, but their values must be
such that the requirement is fulfilled so that $z_{\rm p}$ is within the 
physical stripe: $-\pi \leq {\rm Im} z_{\rm p} < \pi$. Thus, in this case,
we can get several poles, some of them with 
$-\pi < {\rm Im} z_{\rm p} <  \pi$, i.e., Landau poles.
This case is illustrated
in the case of $f(Y)$ being cubic polynomial (${\rm P}[3/0]$)
in Figs.~\ref{figP30bt} (a), (b), for the case of two
different complex values of roots $t_1$: $t_1= 1 + i 0.5$ and
$t_1=1+i 0.4$. Here, the root $t_2$ is then complex conjugate of $t_1$;
and $t_3$ is determined by the pQCD condition (\ref{pQCDf}) and turns out
to be negative. We can see that in the case $t_1=1+i 0.5$ there are
no Landau poles, just a pole at $z_{\rm p}=-11.6312  - i \pi$.
The numerical test with the use of dispersion relation (\ref{dispa2})
of Sec.~\ref{sec:beta1} (cf.~also Table \ref{tabdev})
also confirms that $a(Q^2) \equiv F(z)$ is analytic in this case. 
However, in the case $t_1=1+i 0.4$ there are, beside the pole at
$z_{\rm p}=-10.5023 - i \pi$, Landau poles at $z=-6.32336 \pm i 2.6005 $.
% Look at: rtau_AfP30C.m, line with (* zth[... *) comments 
\begin{figure}[htb] %\unitlength=1mm
\begin{minipage}[b]{.49\linewidth}
% \centering\includegraphics[width=85mm]{pl3dabsbtP30t1_1i05.jpg}
%\centering\includegraphics[width=85mm]{figxl12a.eps}
%\centering\includegraphics[width=85mm]{figxl12a.pdf}
\centering\includegraphics[width=85mm]{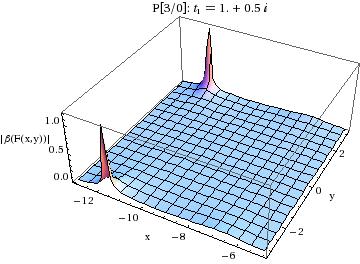}
\end{minipage}
\begin{minipage}[b]{.49\linewidth}
% \centering\includegraphics[width=85mm]{pl3dabsbtP30t1_1i04.jpg}
%\centering\includegraphics[width=85mm]{figxl12b.eps}
%\centering\includegraphics[width=85mm]{figxl12b.pdf}
\centering\includegraphics[width=85mm]{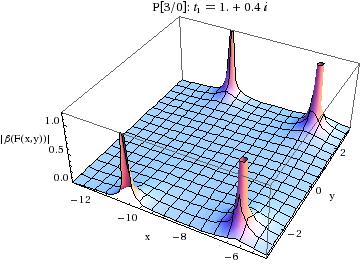}
\end{minipage}
\vspace{-0.4cm}
 \caption{\footnotesize  (a) $|\beta(F(z))|$ as a function of
$z=x+i y$ for the beta-function (\ref{ans1}) with $f(Y)$
being cubic polynomial with $t_1=1+i 0.5$ ($t_2=1-i 0.5, t_3=-3.67591$);
(b) the same as in (a), but with $t_1=1+i 0.4$ ($t_2=1-i 0.4, t_3=-3.98969$).}
\label{figP30bt}
 \end{figure}
This can be understood in the following way. The expression for
the location of poles $z_{\rm p}$ is given by Eq.~(\ref{zpgen}), with the
sum there over $j=1,2,3$. Usually softwares such as MATHEMATICA give for
logarithms $\ln U$ of complex arguments $U$ expressions with imaginary part
$-\pi < {\rm Im} (\ln U) \leq \pi$. In this case, if only the term $\ln(-1)$
in Eq.~(\ref{zpgen}) gets replaced by $[\ln(-1) - i 2 \pi] = - i \pi$,
the resulting $z_{\rm p}$ has ${\rm Im}z_{\rm p} = - i \pi$,
in both cases $t_1=1 + i 0.5$ and $t_1=1+i 0.4$.
Namely, $z_{\rm p} = -11.6312 - i \pi$ and 
$z_{\rm p} = -10.5023 - i \pi$, respectively. However, if we,
in addition, replace $\ln [-t_2/( a_0/a_{\rm in} - t_2 )]$
by $\ln [-t_2/( a_0/a_{\rm in} - t_2 )] + i 2 \pi$, we get 
in the case of $t_1=1 + i 0.4$ a pole location $z_{\rm p}$ inside the
physical stripe  $-\pi \leq {\rm Im} z < \pi$:
$z_{\rm p} = -6.32336 - i 2.6005$, which is the location of
one of the Landau poles seen in Fig.~\ref{figP30bt}(b);
the other Landau pole is at $z_{\rm p}-6.32336 + i 2.6005$.

In general, by adding to each of the logarithms of 
complex arguments in Eq.~(\ref{zpgen}) multiples of
$i 2 \pi$, we end up with a set of possible pole locations $z_{\rm p}$.
Only those values which lie within the physical stripe
$-\pi \leq {\rm Im}z < \pi$ are candidates for the
location of (Landau) poles. However, in practice,
only some of them represent poles $F(z_{\rm p}) = \infty$,
while others may have finite values of $F(z_{\rm p})$.
This is so because the RGE integration, for the physical stripe
of $z$'s, with a specific initial condition at $z=0$,
will not cover all the possibilities of these multiples.

\item
Yet another possibility is to have some roots $t_j$ real positive.
Since we have $a_0 \equiv a(Q^2=0)$ by our notation, 
the value $a=a_0$ is a root of the
beta function $\beta(a)$, and there are no other roots of $\beta(a)$
in the positive interval $0 < a < a_0$ [note that $\beta(0)=0$ by asymptotic
freedom]. Therefore, we are not allowed to have $t_j > 1$ since this
would imply that $a_j=a_0/t_j < a_0$ is a root of $\beta(a)$; hence if $t_j$
is positive it must lie in the interval $0 < t_j < 1$. Such $t_j$'s
then fulfill the relations $(0 < t_j < 1 < a_0/a_{\rm in})$ and hence
give a nonreal value of the logarithm $\ln(-t_j/(a_0/a_{\rm in}-t_j))$
in Eq.~(\ref{zpgen}); the value of $B_j$ is real. Therefore,
in such a case we generally obtain ${\rm Im} z_{\rm p} \not= - \pi$,
i.e., we generally obtain a Landau pole. 

\item
We may obtain Landau poles, or Landau singularities, in several
other cases, e.g., when some of the poles $u_k$ of the beta function
are larger than unity. However, a systematic (semi-)analytic
analysis of these problems appears to be too difficult here.
We just mention, as an aside, that the appearance of
Landau singularities [e.g., finite discontinuities of $F(z)$] usually
implies the appearance of Landau poles [infinities of $F(z)$].

\end{enumerate}

When $M \leq N-2$, the implicit solution of the type (\ref{impl4})
obtains additional terms on the rhs: $\ln(F(z))$,
$F(z)$, ..., $F(z)^{N-M-2}$ (if $M \leq N-3$)
[if $M=N-2$: only $\ln F(z)$]. In this case the poles 
$|F(z_{\rm p})|=\infty$ are reached at $z_{\rm p} = - \infty$,
i.e., $Q^2=0$. This implies that in such cases 
the condition $a(Q^2=0) \equiv a_0 < \infty$ cannot be fulfilled.

\section{Massless part of the strangeless tau decay ratio}
 \label{app:rtauexp}

At present, the most precisely measured low-energy observable
referring to an inclusive process is the ratio $R_{\tau}(\Delta S=0)$, 
which is proportional to the
branching ratio of $\tau$-decays into nonstrange hadrons.
Consequently, it plays a central role for testing the validity
of our anQCD approach. However, for a a careful comparison
of the available experimental result with our theoretical prediction
it is essential to extract from the quantity $R_{\tau}(\Delta S=0)$ the 
pure massless QCD-canonic part $r_{\tau} \equiv r_{\tau}(\Delta S=0, m_q=0)$. This
analysis has already been presented in Appendix E of Ref.~\cite{CV2}.
Here we redo it, but with updated experimental values of $R_{\tau}(\Delta S=0)$,
of the Cabibbo-Kobayashi-Maskawa (CKM) matrix element $|V_{ud}|$
and of higher-twist contributions.
The strangeless (V+A)-decay ratio
extracted from measurements by the ALEPH Collaboration \cite{ALEPH2,ALEPH3}
and updated in Ref.~\cite{DDHMZ} is
\begin{eqnarray}
R_{\tau}(\triangle S\!=\!0)  &\equiv& 
\frac{ \Gamma (\tau^- \to \nu_{\tau} {\rm hadrons} (\gamma) )}
{ \Gamma (\tau^- \to \nu_{\tau} e^- {\overline {\nu}_e} (\gamma))}
- R_{\tau}(\triangle S\!\not=\!0)
\label{Rtaudef}
\\
& = & 
%\frac{(1 - B_e - B_{\mu} )}{B_e} - R_{\tau}(\triangle S\!\not=\!0) = 
3.479  \pm 0.011 \ .  
\label{Rtauexp}
\end{eqnarray}
The canonic massless quantity $r_{\tau}(\triangle S=0, m_q=0)$ 
is obtained from the above quantity by removing the
non-QCD [CKM and electroweak (EW)] factors and contributions, as well as
chirality-violating (quark mass) contributions
\ba
r_{\tau}(\triangle S=0, m_q=0) 
&=& \frac{ R_{\tau}(\triangle S=0) }
{ 3 |V_{ud}|^2 (1 + {\delta}_{{\rm EW}}) } -
(1 + \delta_{\rm EW}^{\prime} ) - 
\delta r_{\tau}(\triangle S=0, m_{u,d}\not=0) \ .
\label{rtgen2}
\ea
This quantity is massless QCD-canonic,
i.e., its pQCD expansion is 
$r_{\tau}(\triangle S=0, m_q=0)_{\rm pt} = a + {\cal O}(a^2)$.
The updated value of the CKM matrix element $|V_{ud}|$
is \cite{PDG2008}
\be
|V_{ud}| = 0.97418 \pm 0.00027 \ .
\label{Vud}
\ee
The EW correction parameters are
$1+\delta_{\rm EW} = 1.0198 \pm 0.0006$
\cite{ALEPH2,ALEPH3} and $\delta_{\rm EW}^{\prime} = 0.0010$
\cite{Braaten:1990ef}.
The (V+A)-channel corrections
$\delta r_{\tau}(\triangle S=0, m_{u,d}\not=0)$
due to the nonzero quark masses are
\cite{Braaten:1988hc,ALEPH3}
the sum of corrections $(\delta_{ud,V}^{(D)} + \delta_{ud,A}^{(D)})/2$
with dimensions $D=2,4,6,$ and $8$.
It appears that, among the chirality-nonviolating 
$D \geq 2$ contributions, the only possibly nonnegligible \cite{Ioffe}
is the $D=4$ contribution  
$\delta_{\langle GG \rangle} = 
(11/4) \alpha_s^2(m_{\tau}^2) \langle a G G \rangle/m_{\tau}^4$
from gluon condensate. 
The authors of Ref.~\cite{DDHMZ} obtained from their fit the 
gluon condensate value $\langle a G G \rangle = (-1.5 \pm 0.3) \times 10^{-2}
\ {\rm GeV}^4$,
giving thus $\delta_{\langle GG \rangle} \approx -5 \times 10^{-4}$;
their entire value of higher dimension contributions ($2 \leq D \leq 8$) to
$r_{\tau}(\triangle S=0, m_{u,d}\not=0)$ is
$(-6.3 \pm 1.4) \times 10^{-3}$. On the other hand, the
value of the gluon condensate may be compatible with zero;
e.g., the $\tau$-decay analysis of Ref.~\cite{Ioffe} based on sum rules
gives $ \langle a G G \rangle = (0.005 \pm 0.004) \ {\rm GeV}^4$
which is almost compatible with zero. In our analysis
we assume that this is the case, i.e., zero value of the gluon condensate.
With this assumption, the higher dimension contributions
to $r_{\tau}(\triangle S=0, m_{u,d}\not=0)$ are only the 
chirality-violating (i.e., due to nonzero quark mass) terms,
their value being thus 
\be
\delta r_{\tau}(\triangle S=0, m_{u,d}\not=0) = (-5.8 \pm 1.4) \times 10^{-3} \ .
\label{drtaumq}
\ee
Using the aforementioned results in Eq.~(\ref{rtgen2}) leads to
\be
r_{\tau}(\triangle S=0, m_q=0)_{\rm exp.} =
0.203 \pm 0.004 \ ,
\label{rtauexp}
\ee
where the experimental uncertainties were added in quadrature.
The uncertainty here is dominated by the experimental
uncertainty $\delta R_{\tau} = \pm 0.011$, Eq.~(\ref{Rtauexp}).
The central value (\ref{rtauexp}) would increase to $0.204$ if the
gluon condensate value $\langle a G G \rangle = (-1.5 \pm 0.3) \times 10^{-2}
\ {\rm GeV}^4$ of Ref.~\cite{DDHMZ} was taken.
The central value $0.203$ of Eq.~(\ref{rtauexp}) is also obtained
by using the analysis and results of Ref.~\cite{Ioffe}, but
with the updated values $R_{\tau}(\triangle S\!=\!0)$ of Eq.~(\ref{Rtauexp})
and $|V_{ud}|$ of Eq.~(\ref{Vud}).

\section{Higher order terms in analytic QCD}
 \label{app:hoanQCD}

Here we summarize the general approach to calculate higher order
corrections in analytic QCD (anQCD) models, as described first in
our earlier works \cite{CV1,CV2}. In order not to confuse
the general analytic coupling $a(Q^2)$ with pQCD coupling $a_{\rm pt}(Q^2)$,
we will use in this Appendix the notation $\A_1(Q^2)$ for the
analytic coupling. 

First we note that the analytic coupling
$\A_1(Q^2)$ does not fulfill the ITEP-OPE conditions (\ref{ITEP})
in any of the anQCD models that have appeared
in the literature up to now.\footnote{
Except for Ref.~\cite{CKV1} where some of the main results of the present
work have already been summarized, and Ref.~\cite{Cvetic:2007ad}
where a direct construction of an analytic coupling $\A_1$ 
with several parameters was performed 
(cf.~footnote \ref{ftnt:ad} in this work).
The anQCD model of Ref.~\cite{Alekseev:2005he} fulfills this condition 
approximately.} 
Nonfulfillment of ITEP-OPE conditions implies that the
respective beta function $\beta(\A_1) \equiv \partial \A_1(Q^2)/\partial \ln Q^2$
is not analytic in $\A_1$ (cf.~arguments in Sec.~\ref{sec:beta1}).
Consequently, in these models the beta function, which is usually
not known explicitly, cannot be Taylor expanded around $\A_1=0$,
and therefore the powers $\A_1^n$ cannot be expected to be the analytized
analogs of $a_{\rm pt}^n$. In fact, they usually are not.
The construction of $\A_n(Q^2)$, the analytic analogs of $a_{\rm pt}(Q^2)^n$
($n \geq 2$), is yet another important ingredient in anQCD.

A spacelike massless observable ${\cal D}(Q^2)$, in
its canonical form, has the
following perturbation series:
\be
{\cal D}(Q^2)_{\rm pt} = a_{\rm pt} +
d_1 a_{\rm pt}^2 + d_2 a_{\rm pt}^3 +\cdots \ ,
\label{PS}
\ee
and the corresponding truncated perturbation series (TPS) is
\be
{\cal D}(Q^2)_{\rm pt}^{[N]} = a_{\rm pt} +
d_1 a_{\rm pt}^2 + \cdots d_{N-1} a_{\rm pt}^N \ .
\label{TPSN}
\ee
Here, $a_{\rm pt}$ and $d_j$'s have given renormalization scale (RScl)
and scheme (RSch) dependences. Analytization means, in the first
instance, to replace in the first term $a_{\rm pt}$ by $\A_1(Q^2)$.
For treating the higher order terms, there are, in principle,
several options at hand. For instance, one could replace all
powers of $a_{\rm pt}$ by the corresponding powers of $\A_1$
($a_{\rm pt}^n \mapsto \A_1^n$). Or, as is done in MA,
one could subject each $a_{\rm pt}^n$ to an analogous analytization
procedure as $\A_1$ (if such an analogous procedure unambiguously exists),
yielding additional analytic couplings $a_{\rm pt}^n \mapsto \A_n$, where,
in general, $\A_n \not= \A_1^n$. 
In MA such a prescription unambiguously exists.
The advantage of such a prescription in MA
lies in the fact that the RGEs governing the running of $\A_n^{\rm (MA)}$'s,
as well as the RSch dependence of $\A_n^{\rm (MA)}$'s,
are identical to the corresponding pQCD RGEs and RSch dependence once the
replacements $a_{\rm pt}^n \mapsto \A_n^{\rm (MA)}$ are performed there
\cite{Magradze}. We consider this property as physically important,
especially because there is a clear hierarchy $\A_1^{\rm (MA)}
> |\A_2^{\rm (MA)}| > |\A_3^{\rm (MA)}| \cdots$ at all positive
$Q^2$ values. Among other things, this hierarchy implies that
the MA-analytized version of the TPS Eq.~(\ref{TPSN})
\be
{\cal D}(Q^2)_{\rm (MA)}^{[N]} = \A_1^{\rm (MA)} +
d_1 \A_2^{\rm (MA)} + \cdots d_{N-1} \A_N^{\rm (MA)} \ ,
\label{MATPSN}
\ee
becomes systematically more RScl and RSch independent when
the truncation index $N$ increases
\be
\frac{\partial {\cal D}(Q^2;{\rm RS})_{\rm (MA)}^{[N]}}{\partial ({\rm RS})}
= k_N \A_{N+1}^{\rm (MA)} + {\cal O}(\A_{N+2}^{\rm (MA)}) \ .
\label{RSMATPSN}
\ee 
Here, ``RS'' stands for logarithm $\ln \mu^2$ of RScl $\mu$, or 
for any RSch parameter $c_j = \beta_j/\beta_0$ ($j\geq 2$).

However, when constructing anQCD models beyond MA, by changing
the discontinuity function $\rho_1(\sigma) = 
{\rm Im} a_{\rm pt}(-\sigma - i \epsilon)$ appearing in the
dispersion relation (\ref{dispa}) for $\A_1^{\rm (MA)}(Q^2)$
\cite{Nesterenko2,CV1,CV2}, or by different constructions of $\A_1(Q^2)$ 
(cf.~\cite{Webber:1998um,Srivastava:2001ts,Simonov,Nesterenko,Alekseev:2005he}
and references therein),
the meaning of ``analogous analytization'' of higher powers $a_{\rm pt}^n$
becomes unclear or, at best, ambiguous. 
On the other hand, it is almost imperative to maintain
relations (\ref{RSMATPSN}) in any anQCD model with
hierarchy $\A_1 > |\A_2| > |\A_3| \cdots$, because
then the physical condition of RScl and RSch independence
of the evaluated observables is guaranteed to
be increasingly well fulfilled at any $Q^2$ when the number of terms
increases.

Furthermore, it is preferable to have the higher power analogs
$a_{\rm pt}^n \mapsto \A_n$ not simply constructed as $\A_n \equiv (\A_1)^n$,
but rather by application of linear (in $\A_1$) operations on $\A_1$,
such as, e.g., derivatives and linear combinations thereof. The
underlying reason is the compatibility with linear integral transformations
(such as Fourier and Laplace) \cite{Shirkov:1999np}. In linear transformations,
the image of a power of a function  is not the power of the image of the
function.\footnote{Such a construction of $\A_n(Q^2)$, 
as a linear operation applied on $\A_1(Q^2)$, was presented in anQCD 
in Refs.~\cite{CV1,CV2,Shirkov:2006gv}.}

The construction of higher order analogs
$\A_n$ (applicable to any anQCD model) which obey all these conditions
was first presented in Refs.~\cite{CV1,CV2}.
The procedure proposed there for obtaining $\A_n$ from a given
anQCD coupling $\A_1$, in a given RSch, is the following:
First we define the logarithmic derivatives of $\A_1(\mu^2)$
(where $\mu^2= \kappa Q^2$ is any chosen RScl), i.e., we
define
\be
\tA_{n+1}(\mu^2) \left( \equiv {\widetilde a}_{n+1}(\mu^2) \right)
= \frac{(-1)^n}{\beta_0^n n!}
\frac{ \partial^n \A_1(\mu^2)}{\partial (\ln \mu^2)^n} \ ,
\qquad (n=1,2,\ldots) \ .
\label{tAn}
\ee
In order to understand the following construction of $\A_n$'s given below, 
it is convenient to consider first the corresponding
logarithmic derivatives in pQCD
\be
{\widetilde a}_{{\rm pt},n+1}(\mu^2)
\equiv \frac{(-1)^n}{\beta_0^n n!}
\frac{ \partial^n a_{\rm pt}(\mu^2)}{\partial (\ln \mu^2)^n} \ .
\qquad (n=1,2,\ldots) 
\label{tan}
\ee
These\footnote{
An expansion of the Adler function in terms of ${\widetilde a}_{{\rm pt},n+1}(\mu^2)$
is used in Ref.~\cite{CLMV} for an evaluation of $r_{\tau}$ in the context
of pQCD; this ``modified'' contour improved perturbation theory (mCIPT)
was shown there to have advantages over the standard (CIPT) approach,
most notably a lower RScl dependence of the result.}
are related to the powers $a_{\rm pt}^n$ via relations
involving the $c_j$ coefficients of the pQCD RGE Eq.~(\ref{btexp1})
\ba
{\widetilde a}_{{\rm pt},2} &=&
a_{\rm pt}^2 + c_1 a_{\rm pt}^3 + c_2 a_{\rm pt}^4 + \cdots \ ,
\label{ta2}
\\
{\widetilde a}_{{\rm pt},3} &=&
a_{\rm pt}^3 + \frac{5}{2} c_1 a_{\rm pt}^4 + \cdots \ ,
\label{ta3}
\\
{\widetilde a}_{{\rm pt},4} &=&
a_{\rm pt}^4 + \cdots \ , \qquad {\rm etc.}
\label{ta4}
\ea
The above relations are obtained by (repeatedly) applying the pQCD RGE.
The inverse relations are
\ba
a_{\rm pt}^2 & = & {\widetilde a}_{{\rm pt},2}
- c_1 {\widetilde a}_{{\rm pt},3} + \left( \frac{5}{2} c_1^2 - c_2 \right)
{\tilde a}_{{\rm pt},4} + \cdots \ ,
\label{a2}
\\
a_{\rm pt}^3 & = & {\widetilde a}_{{\rm pt},3}
- \frac{5}{2} c_1 {\widetilde a}_{{\rm pt},4} + \cdots \ ,
\label{a3}
\\
 a_{\rm pt}^4 & = & {\widetilde a}_{{\rm pt},4} + \cdots \ ,
\qquad {\rm etc.}
\label{a4}
\ea
Now we adopt the following replacement on the rhs of Eqs.~(\ref{a2})-(\ref{a4}):
\be
a_{\rm pt} \mapsto \A_1 \ , \quad
{\widetilde a}_{{\rm pt},n+1} \mapsto \tA_{n+1}
\quad (n=1,2,\ldots) \ ,
\label{anrule}
\ee
and use the generated expressions as definitions of $\A_n$, the
higher power analogs of pQCD powers $a_{\rm pt}^n$
\ba
\A_2 & = & \tA_2
- c_1 \tA_{3} + \left( \frac{5}{2} c_1^2 - c_2 \right) \tA_4 
+ \cdots \ ,
\label{A2}
\\
\A_3 & = & \tA_{3}
- \frac{5}{2} c_1 \tA_{4} + \cdots \ ,
\label{A3}
\\
\A_4 & = & \tA_{4} + \cdots \ , \qquad {\rm etc.}
\label{A4}
\ea
It is then straightforward to see that the 
analytic (``an'') series obtained from the perturbation series 
(\ref{PS})
via replacements $a_{\rm pt} \mapsto \A_1$, 
$a_{\rm pt}^n \mapsto \A_n$
\be
{\cal D}(Q^2)_{\rm an} = \A_1 +
d_1 \A_2 + d_2 \A_3 + \cdots \ ,
\label{an}
\ee
gives the corresponding truncated analytic series
\be
{\cal D}(Q^2)_{\rm an}^{[N]} = \A_1 +
d_1 \A_2 + \cdots d_{N-1} \A_N \ ,
\label{anN}
\ee
which really fulfills the condition (\ref{RSMATPSN}) of
increasingly good RS-independence, now in any anQCD model
\be
\frac{\partial {\cal D}(Q^2;{\rm RS})_{\rm an}^{[N]}}{\partial ({\rm RS})}
= k_N \A_{N+1} + {\cal O}(A_{N+2}) \ , 
\qquad ({\rm RS} = \ln \mu^2; c_2; c_3; \ldots) \ .
\label{RSanN}
\ee 
This relation continues to hold even if we truncate 
relations (\ref{A2})-(\ref{A4}) at the order $\sim \A_N$
(including the latter).

The above presentation suggests that, instead
of the perturbation series (\ref{PS}) in powers of $a_{\rm pt}$, 
a modified perturbation series in logarithmic derivatives 
${\widetilde a}_{\rm pt,n+1}$ (\ref{tan}) can be used
\be
{\cal D}(Q^2)_{\rm mpt} = a_{\rm pt} 
+ {\widetilde d}_1 {\widetilde a}_{{\rm pt},2}  
+ {\widetilde d}_2 {\widetilde a}_{{\rm pt},3} + 
\cdots \ ,
\label{mPS}
\ee
whose truncated form is
\be
{\cal D}(Q^2)_{\rm mpt}^{[N]} = a_{\rm pt} +
{\widetilde d}_1 {\widetilde a}_{{\rm pt},2} + \cdots 
{\widetilde d}_{N-1} {\widetilde a}_{{\rm pt},N} \ ,
\label{mTPSN}
\ee
where ``m'' in the subscript stands for ``modified,'' and the
modified coefficients ${\widetilde d}_j$ ($j=1, \ldots, N-1$)
are related to the original coefficients $d_j$
\ba
{\widetilde d}_1 & = & d_1 \ ,
\label{td1}
\\
{\widetilde d}_2 & = & d_2 - c_1 d_1 \ ,
\label{td2}
\\
{\widetilde d}_3 & = & d_3 - \frac{5}{2} c_1 d_2 + 
\left( \frac{5}{2} c_1^2 - c_2 \right) d_1 \ ,
\qquad {\rm etc.}
\label{td3}
\ea
When applying analytization to the modified perturbation series (\ref{mPS}),
via replacements (\ref{anrule}), we obtain modified analytic series
(``man'')
\be
{\cal D}(Q^2)_{\rm man} = \A_1 +
{\widetilde d}_1 \tA_2 + {\widetilde d}_2 \tA_3 + \cdots \ ,
\label{man}
\ee
whose truncated version is
\be
{\cal D}(Q^2)_{\rm man}^{[N]} = \A_1 +
{\widetilde d}_1 \tA_2 + \cdots 
{\widetilde d}_{N-1} \tA_N \ .
\label{manN}
\ee
Its RS dependence is
\be
\frac{\partial {\cal D}(Q^2;{\rm RS})_{\rm man}^{[N]}}{\partial ({\rm RS})}
= {\widetilde k}_N \tA_{N+1}  + {\cal O}(\tA_{N+2}) \quad (\sim \A_{N+1}) \ , 
\qquad ({\rm RS} = \ln \mu^2; c_2; c_3; \ldots) \ .
\label{RSmanN}
\ee 
It is interesting that in virtually all anQCD models [i.e., models
that define $\A_1(Q^2)$] holds the hierarchy $\A_1 > |\tA_2| > |\tA_3| > 
\cdots$ at (almost) all complex $Q^2$. Therefore, Eq.~(\ref{RSmanN})
signals an increasingly weak RS dependence of ${\cal D}(Q^2)_{\rm man}^{[N]}$
when $N$ increases, at any value of $Q^2$ and RScl $\mu^2$.

We stress that the analytic (``an'') and modified analytic (``man'')
series [Eqs.~(\ref{an}) and (\ref{man}), respectively],
if they converge, are identical to each other due to relations
(\ref{td1})-(\ref{td3}) and (\ref{A2})-(\ref{A4}).

In the specific case of MA, i.e., when $\A_1 = \A_1^{\rm (MA)}$
of Ref.~\cite{ShS}, it can be shown 
(using the results of Ref.~\cite{Magradze}) 
that the above procedure, Eqs.~(\ref{A2})-(\ref{A4}),
gives the same higher power analogs $\A_n^{\rm (MA)}$ as the analytization 
procedure of Ref.~\cite{MSS} (APT) that uses the MA-type dispersion 
relation involving ${\rm Im} a_{\rm pt}^n(Q^2=-\sigma - i \epsilon)$
\be
\A_n^{\rm (MA)}(Q^2) = \frac{1}{\pi} \int_0^{\infty} \; d \sigma
\frac{\rho_n^{\rm (pt)}(\sigma)}{\sigma + Q^2} \ ,
\label{dispnMA}
\ee
where $\rho_n^{\rm (pt)}(\sigma) = {\rm Im} a_{\rm pt}^n(-\sigma - i\epsilon)$
($n=2,\ldots$). We note that $\A_n^{\rm (MA)} \not= (\A_1^{\rm (MA)})^n$.
Furthermore, construction of $\A_n$ according to relations
(\ref{A2})-(\ref{A4}) in other models of anQCD (e.g., where $\A_1$ is 
constructed from a modified $\rho_1 \not= \rho_1^{\rm (pt)}$, 
e.g. Refs.~\cite{Nesterenko2,CV1,CV2}) also in general leads to
$\A_n \not= \A_1^n$. However, if analytic $A_1(Q^2) \equiv a(Q^2)$ 
is constructed from RGE with beta function $\beta(a)$ analytic at
$a=0$, as is the case in the present work and Ref.~\cite{CKV1},
it is straightforward to see that construction
(\ref{A2})-(\ref{A4}) gives 
\be
\A_n = a^n \quad (n=1,2,\ldots) \ .
\label{ourAn}
\ee 

In those anQCD models of analytic $\A_1(Q^2)$ where the aforedescribed
construction gives $\A_n \not= \A_1^n$ for $n \geq 2$ (such models do not 
appear in the present work), using $\A_1^n$ instead of $\A_n$ is not a
good idea for at least two reasons: 
(1) such a construction is formally not linear in $\A_1$ 
[see the discussion before Eq.~(\ref{tAn})]; 
(2) the RS dependence of the resulting truncated ``power'' analytic
series 
\be
{\cal D}_{\rm pan}(Q^2)^{[N]} = \A_1 + d_1 \A_1^2 + \cdots d_{N-1} \A_1^N
\label{panN}
\ee
is not entirely analogous to Eq.~(\ref{RSanN}) or Eq.~(\ref{RSmanN}),
but is rather
\be
\frac{\partial {\cal D}(Q^2;{\rm RS})_{\rm pan}^{[N]}}{\partial ({\rm RS})}
= k_N \A_1^{N+1}  + {\cal O}(\A_1^{N+2}) + {\rm NP}_{(N)} \ ,
\label{RSpanN}
\ee
where ${\rm NP}_{(N)}$ is an increasingly complicated expression of 
nonperturbative terms (such as $1/Q^{2n}$) when $N$ increases, and
$|{\rm NP}_{(N)}|$ in general does not decrease when $N$ increases.

\section{Leading-$\beta_0$ (skeleton-motivated) resummation in anQCD}
 \label{app:LB}

First we summarize here the resummation formalism for the leading-$\beta_0$
(LB) part of inclusive spacelike QCD observables in anQCD models, as presented
in \cite{CV1,CV2}. Subsequently, we present application of this
formalism to LB resummation for the Bjorken polarized sum rule (BjPSR)
$d_{\rm Bj}(Q^2)$ and, in a newly modified form, to the $\tau$ decay ratio
$r_{\tau}$.

Massless spacelike QCD observables ${\cal D}(Q^2)$, in canonical form, have
the pQCD (``pt'') expansion (\ref{PS}) in powers of $a_{\rm pt}$,
where $a_{\rm pt} = a_{\rm pt}(\mu^2; c_2,\ldots)$ is
defined at a given RScl $\mu$ and in a given RSch ($c_2, c_3,\dots$).
In the scaling definition of $\mu$ we use the convention
$\Lambda = {\overline \Lambda}$, which is the ${\overline {\rm MS}}$
reference scale for RScl's $\mu$ [the so-called $V$ scheme
$\Lambda_V$ is related to ${\overline \Lambda}$ via
${\overline \Lambda}^2 = \Lambda_V^2 \exp({\overline {\cal C}})$,
where ${\overline {\cal C}} = -5/3$].
The considered RSch classes will be such that the RSch 
coefficients $\beta_k \equiv \beta_0 c_k$ ($k \geq 2$) 
are polynomials in $n_f$, and consequently in $\beta_0 = (11 - 2 n_f/3)/4$
\be
\beta_k \equiv \beta_0 c_k = \sum_{j=0}^k b_{kj} \beta_0^j,
\qquad (k=2,3,\ldots)
\label{betak}
\ee
We recall that $\beta_0=(11 - 2 n_f/3)/4$ and $\beta_1=(102-38 n_f/3)/16$
are both universal (RSch-independent) parameters.
RSch's ${\overline {\rm MS}}$ and 't Hooft are clearly special
cases of such RSch's. The RSch independence of ${\cal D}(Q^2)$
implies a specific dependence of coefficients $d_n$ on the
RSch parameters \cite{Stevenson}; this and relations (\ref{betak})
imply that the coefficients $d_n$ have specific expansions in
powers of $\beta_0$
\be
d_1 = c^{(1)}_{11} \beta_0 + c^{(1)}_{10},
\qquad d_n = \sum_{k=-1}^{n} c^{(1)}_{nk} \beta_0^k \ .
\label{dns}
\ee
We note that $c^{(1)}_{1,-1}=0$. In ${\overline {\rm MS}}$ RSch, 
the negative power term $\propto 1/\beta_0$ does not appear.
Relations (\ref{dns}) and (\ref{td1})-(\ref{td3}) imply that
the modified perturbation (``mpt'') expansion (\ref{mPS}) of ${\cal D}(Q^2)$ 
in logarithmic
derivatives ${\widetilde a}_{\rm pt,n+1}$ of Eq.~(\ref{tan})
have coefficients ${\widetilde d}_n$ of a form similar to (\ref{dns})
\be
{\widetilde d}_n = \sum_{k=-1}^{n} {\widetilde c}^{(1)}_{nk} \beta_0^k \ .
\label{tdns}
\ee
Specifically, the leading-$\beta_0$ terms in Eqs.~(\ref{dns}) and
(\ref{tdns}) coincide\footnote{
Note that $\beta_1=b_{10} + b_{11} \beta_0$ (with: $b_{10}=-107/16$
and $b_{11}=19/4$); therefore, $c_1 \equiv \beta_1/\beta_0$ is
$\sim \beta_0^0$ in the leading-$\beta_0$ (LB) limit.}
\be
{\widetilde c}^{(1)}_{nn}=c^{(1)}_{nn}
\label{cnn}
\ee
The LB resummation of the inclusive spacelike ${\cal D}(Q^2)$ is obtained
in pQCD via integration of $a_{\rm pt}(\mu^2)$ over various
scales $\mu^2 = t Q^2 \exp({\overline {\cal C}})$
and weighted with a characteristic function  
$F_{\cal D}^{\cal {E}}(t)$ according\footnote{The superscript
${\cal {E}}$ indicates here that the observable is Euclidean,
i.e., spacelike.}
to formalism of Ref.~\cite{Neubert}
\be
{\cal D}^{\rm (LB)}_{\rm pt}(Q^2) \equiv 
\int_0^\infty \frac{dt}{t}\: F_{\cal D}^{\cal {E}}(t) \: 
a_{\rm pt}(t Q^2 e^{\overline {\cal C}}) \ .
\label{LBpt}
\ee
The integration cannot be performed unambiguously, 
due to the Landau poles of $a_{\rm pt}$ at low values of $t$.
In anQCD $a_{\rm pt}$ here is simply replaced by
analytic $\A_1$ ($\equiv a$)
\be
{\cal D}^{\rm (LB)}_{\rm an}(Q^2) \equiv 
\int_0^\infty \frac{dt}{t}\: F_{\cal D}^{\cal {E}}(t) \: 
\A_1(t Q^2 e^{\overline {\cal C}}) \ ,
\label{LB}
\ee
where now the integration is unambiguous since there
are no Landau poles. Expansion of the analytic
coupling $\A_1(t Q^2 e^{\overline {\cal C}})$ around
the RScl scale $\mu^2$, i.e., Taylor expansion
in powers of ${\cal L}=\ln[\mu^2/(t Q^2 e^{\overline {\cal C}})]$,
gives
\be
{\cal D}^{\rm (LB)}_{\rm an}(Q^2) =
\A_1 + \sum_{n=1}^{\infty} c^{(1)}_{nn} \beta_0^n
\tA_{n+1} \ .
\label{LBexp}
\ee
We thus see that integral (\ref{LB}), in anQCD, represents exactly
the leading-$\beta_0$ (LB) part of the modified analytic
(``man'') expansion (\ref{man}) in Appendix \ref{app:hoanQCD}.
The truncated series of the latter is given in Eq.~(\ref{manN}). 
We stress that the above expansion
is performed at a given RScl $\mu$ and in a given RSch
[$c_2, c_3, \ldots$ -- cf.~Eq.~(\ref{betak})]. In anQCD it
is convenient to perform explicitly the LB resummation
(\ref{LB}) since the integral there is finite, unambiguous,
and RScl independent.

The characteristic function $F_{\cal D}^{\cal {E}}(t)$
for the BjPSR ${\cal D}(Q^2) = d_{\rm Bj}(Q^2)$ was calculated
and used in Ref.~\cite{CV1} (on the basis of the known \cite{Broadhurst:1993ru}
coefficients $c^{(1)}_{nn}$ for it), and was presented in Ref.~\cite{CV2}
\ba
F_{\rm Bj}(t) = 
\left\{
\begin{array}{ll}
(8/9) t \left[ 1 - (5/8) t \right] 
& t \leq 1 \\ 
(4/(9 t)) \left[ 1 - 1/(4t) \right]
& t \geq 1
\end{array}
\right\} \ .
\label{FBj}
\ea

The (nonstrange massless) canonical\footnote{
Canonical form, in the sense that its pQCD expansion is
$r_{\tau}=a_{\rm pt}+{\cal O}(a_{\rm pt}^2)$.}
semihadronic $\tau$ decay ratio 
$r_{\tau} \equiv r_{\tau}(\Delta S=0,m_q=0)$ is
a timelike quantity, and can be expressed in terms of
the massless current-current correlation function
(V-V or A-A, both equal since massless) \cite{Tsai:1971vv}
\be
r_{\tau} = \frac{2}{\pi} \int_0^{m^2_{\tau}} \ \frac{d s}{m^2_{\tau}}
\left( 1 - \frac{s}{m^2_{\tau}} \right)^2 
\left(1 + 2 \frac{s}{m^2_{\tau}} \right) {\rm Im} \Pi(Q^2=-s) \ .
\label{rtauPi}
\ee
Use of the Cauchy theorem in the $Q^2$ plane and then integration by parts 
leads to the following contour integral form
\cite{Braaten:1988hc,Beneke:2008ad}:
\be
r_{\tau} = \frac{1}{2 \pi} \int_{-\pi}^{+ \pi}
d \phi \ (1 + e^{i \phi})^3 (1 - e^{i \phi}) \
d_{\rm Adl} (Q^2=m_{\tau}^2 e^{i \phi}) \ ,
\label{rtaucont}
\ee
with $d_{\rm Adl} (Q^2) = - d \Pi(Q^2)/d \ln Q^2$ being the
massless Adler function. In pQCD, use of the Cauchy theorem
to the expression (\ref{rtauPi}) is formally not allowed. This is so 
because $\Pi_{\rm pt}(Q^2)$, being a power series in
$a_{\rm pt}(Q^2)$ [or: $a_{\rm pt}(\kappa Q^2)$], has Landau
singularities along the positive axis $0 < Q^2 \leq \Lambda^2$.
In pQCD, expressions (\ref{rtauPi}) and (\ref{rtaucont}) are
two different quantities; in anQCD models they are always
the same. 

The massless Adler function $d_{\rm Adl} (Q^2)$ is a spacelike
(quasi)observable. On the basis of the known coefficients
$c^{(1)}_{nn}$ for it \cite{Broadhurst,Beneke}, its characteristic function
$F_{\rm Adl}(\tau)$ was obtained in Ref.~\cite{Neubert},
and from it and using relation (\ref{rtaucont}) the
characteristic function for $r_{\tau}$ 
was obtained in Ref.~\cite{Neubert2}, in the timelike LB form
\be
r_{\tau}(\Delta S=0,m_q=0)^{\rm (LB)} = 
\int_0^\infty \frac{dt}{t}\: F_{r}^{\cal {M}}(t) \: 
\tlA_1 (t e^{\cal C} m_{\tau}^2) \ .
\label{LBrt}
\ee
Here, the superscript ${\cal {M}}$ indicates that these
are Minkowskian (timelike) quantities;
$\tlA_1$ is the timelike coupling
\ba
\tlA_1(s) &=& \frac{1}{\pi} 
\int_s^{\infty} \frac{d \sigma}{\sigma}
{\rho}_1(\sigma) \ ;
\label{tlA1}
\ea
and the characteristic function $F_{r}^{\cal {M}}(t)$ was obtained
in \cite{Neubert2}.\footnote{
In fact, the quantity $W_{\tau}$ of Ref.~\cite{Neubert2}
is related to $F_{r}^{\cal {M}}$ here via: $F_{r}^{\cal {M}}(t) = (t/4)W_{\tau}(t)$.
Full expression for $F_{r}^{\cal {M}}(t)$ is given in Eqs.~(C10)-(C11) of
Ref.~\cite{CV2}; however, a typo appears in the last line of Eq.~(C11)
there: in a parenthesis there, the term $+3$ should be written as $3 t^2$;
the correct expression was used in calculations in Refs.~\cite{CV1,CV2}.}

It turns out that, in the calculations in the present work, it
is inconvenient to calculate the LB-contribution to $r_{\tau}$
by using formula (\ref{LBrt}) which involves function $\tlA_1(s)$. 
This inconvenience consists in the following: in this work, RGE
(\ref{RGEz}) [$\Leftrightarrow$ Eqs.~(\ref{RGEx})-(\ref{RGEy})]
is integrated in the entire physical stripe in the complex
$z$ plane, and as a result of this we numerically obtain,
among other things, the quantity 
$\rho_1(\sigma) = {\rm Im} a(Q^2=-\sigma - i \epsilon) = 
{\rm Im} F(z=|z| - i \pi)$
[with: $|z|=\ln(\sigma/\mu_{\rm in}^2)$]; to obtain the quantity
$\tlA_1(s)$, yet another numerical integration (\ref{tlA1}) is needed,
and then we go with this $\tlA_1(s)$ into the integration (\ref{LBrt}).
There are too many successive numerical integrations involved,
and the precision of calculation is expected to be low.

\begin{figure}[htb] %\unitlength=1mm
\begin{minipage}[b]{.49\linewidth}
% \centering\includegraphics[width=85mm]{tFtauplt.jpg}
%\centering\includegraphics[width=85mm]{figxl13a.eps}
\centering\includegraphics[width=85mm]{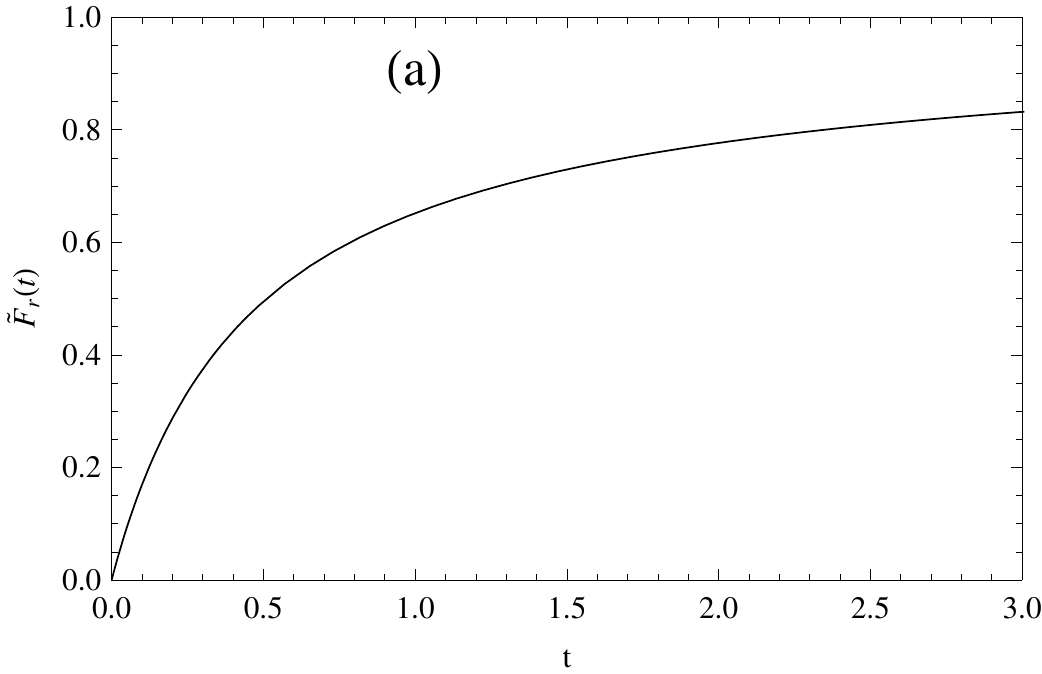}
%\centering\includegraphics[width=85mm]{figxl13a.jpg}
\end{minipage}
\begin{minipage}[b]{.49\linewidth}
% \centering\includegraphics[width=85mm]{tFtauplz.jpg}
%\centering\includegraphics[width=85mm]{figxl13b.eps}
\centering\includegraphics[width=85mm]{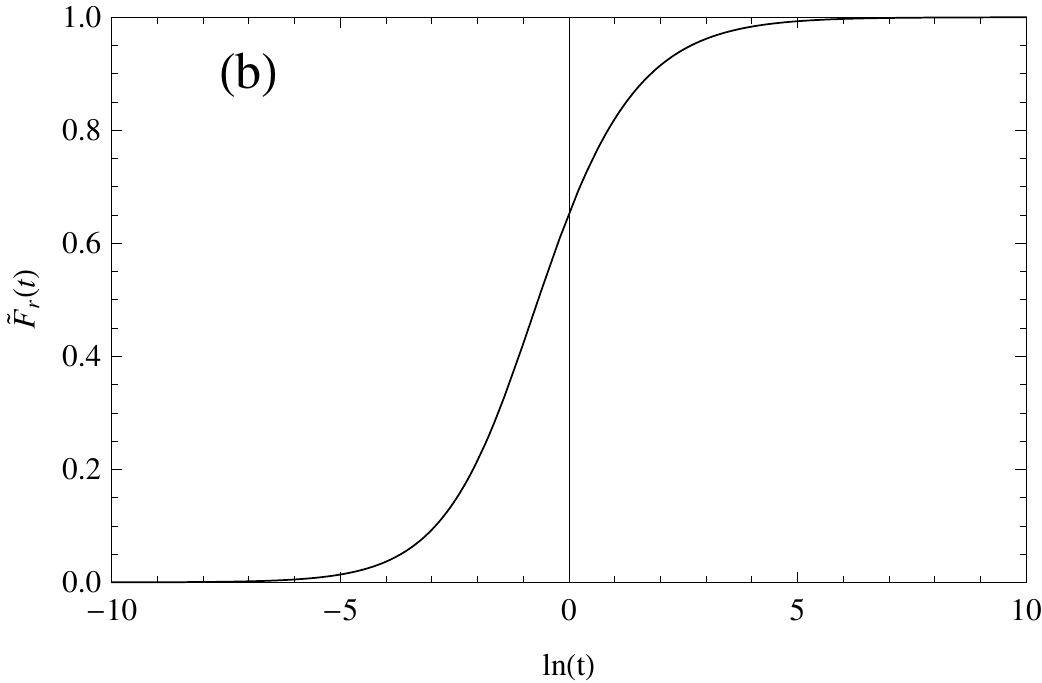}
%\centering\includegraphics[width=85mm]{figxl13b.jpg}
\end{minipage}
\vspace{-0.4cm}
 \caption{\footnotesize Characteristic function ${\widetilde F}_{r}(t)$
which appears in the LB integral (\ref{LBrt2}) of $r_{\tau}$:
(a) as a function of $t$; (b) as a function of $\ln t$.}
\label{tFtauplz}
 \end{figure}

Therefore, we perform in integral (\ref{LBrt}) integration by parts,
using relation $d \tlA_1(s)/d \ln s = -\rho_1(s)/\pi$ [cf.~Eq.~(\ref{tlA1})],
and we obtain the expression of $r_{\tau}^{\rm (LB)}$ in terms of
the discontinuity function $\rho_1(s)$:
\be
r_{\tau}^{\rm (LB)} = 
\frac{1}{\pi} \int_0^\infty \frac{dt}{t}\: {\widetilde F}_{r}(t) \: 
\rho_1(t e^{\cal {\overline C}} m_{\tau}^2) \ ,
\label{LBrt2}
\ee
where 
\be
{\widetilde F}_{r}(t) = \int_0^t \frac{dt'}{t'}\: F_{r}^{\cal {M}}(t') \ .
\label{tFtau}
\ee 
Integration in (\ref{tFtau}) can be performed analytically,
and the result for ${\widetilde F}_{r}(t)$ is ($C_F=4/3$):
\ba
{\widetilde F}_{r}(t)/(4 C_F) &=&
-\frac{1}{12} \text{Li}_2(-t) \left(t^4+6 t^3+18 t^2+10 t-12 t
   \ln (t)-3\right)-2 t \text{Li}_3(-t)
\nonumber\\
&&
+\frac{1}{1728} {\Big \{} 
-72 \ln (t) \left[t
   \left(-2 t^2-47 t+6\right)+2 \left(t^4+6 t^3+18 t^2+10
   t-3\right) \ln (t+1)\right]
\nonumber\\
&&
-259 t^4-600 t^3 -6948 t^2-5184 t \zeta (3)+7344 t
+72 (t+6) t^3 \ln ^2(t) {\Big \}}  \quad (t \leq 1), 
\label{tFtlt1}
\\
{\widetilde F}_{r}(t)/(4 C_F) &=&
\frac{1}{432}  {\bigg \{}
-36 t ( t^3 +6 t^2 +18 t-2 )
   \text{Li}_2\left(-\frac{1}{t}\right)
   -108 \text{Li}_2(-t)+864 t
   \text{Li}_3\left(-\frac{1}{t}\right)
\nonumber\\
&&
   +432 t
   \text{Li}_2\left(-\frac{1}{t}\right)
   (\ln (t)-1)-9 \left[ (t^2+8 t +36)
   t^2+96\right] \ln (t+1)-t \left[ 9 t (4 t+23)+598 \right]
\nonumber\\
&&
-18 \left[ 2 t (t^3 + 6 t^2 + 18 t+22)-3 \right] \ln ^2(t)
\nonumber\\
&&
+3 \left[ (3 t^4 + 12 t^3 + 42 t^2-184 t + 111)+12 (t^2 + 4 t+9) (t+1)^2 \ln
   (t+1)\right] \ln (t)
\nonumber\\
&&
+9 t (t^3 + 8 t^2 +36 t-96) \ln
   \left(\frac{1}{t}+1\right)
\nonumber\\
&&
+432 (\ln(t)-2) \left[ t \ln (t)-(t+1) \ln (t+1) \right]
+648 \zeta (3)-114 \pi^2+841 {\bigg \}}
\nonumber\\
&&
-\frac{3 \zeta(3)}{2}+\frac{2 \pi^2}{9}-\frac{463}{1728}  
\quad (t \geq 1).
\label{tFtgt1}
\ea
The function ${\widetilde F}_{r}(t)$ is continuous and 
monotonously increases when $t$ increases.
Its value is zero at $t=0$, and one at $t=+\infty$.
It is depicted in Figs.~\ref{tFtauplz} as a function of $t$ and $\ln t$.

\section{Inclusion of beyond-the-leading-$\beta_0$ (bLB) terms in anQCD}
 \label{app:bLB}

In pQCD, perturbation expansion of any massless
spacelike observable ${\cal D}(Q^2)$
can be written in the form (\ref{PS}) or (\ref{mPS}). 
In the considered (large) RSch classes (\ref{betak}), the coefficients
$d_n$ and ${\widetilde d}_n$ can be written in the form (\ref{dns}) and
(\ref{tdns}), respectively. 
Leading-$\beta_0$ (LB) resummation (\ref{LB}) reproduces
one part of these terms, Eq.~(\ref{LBexp}). In practice,
for inclusive spacelike observables only the leading-$\beta_0$
parts $c^{(1)}_{nn} \beta_0^n$ of coefficients $d_n$ and ${\widetilde d}_n$
are known for all $n$ [cf.~also Eq.~(\ref{cnn})], while the coefficients
known in their entirety are only the first two or three: $d_1, d_2, d_3$
[$\Leftrightarrow \ {\widetilde d}_1, {\widetilde d}_2, {\widetilde d}_3$,
cf.~Eqs.~(\ref{td1})-(\ref{td3})]. For this reason, the most that
one can include in the evaluation of any such observable in
anQCD are all the LB contributions, Eq.~(\ref{LBexp}), and
the beyond-the-leading-$\beta_0$ (bLB) terms of order $a^2, a^3,$ and$a^4$
($\Leftrightarrow$ of order ${\widetilde a}_2, {\widetilde a}_3, {\widetilde a}_4$].

In practice, the coefficients $d_1, d_2, d_3$ and $c^{(1)}_{nn} \beta_0^n$
are calculated and given in the literature in the ${\overline {\rm MS}}$
RSch [$c_2({\overline {\rm MS}}), c_3({\overline {\rm MS}}),\ldots$]
and with\footnote{
Sometimes, $c^{(1)}_{nn}$'s are calculated and given in the
literature at RScl $\mu^2 = Q^2 \exp( {\overline {\cal C}} ) =
Q^2 \exp(-5/3)$.} 
RScl $\mu^2=Q^2$; we will denote such quantities with
the bar over them.  In general, the evaluations are performed in 
another RSch ($c_2, c_3, \ldots$) (e.g., in the present work the 
RSch as dictated by the chosen $\beta$ function used), and
another RScl
\be
\mu^2 = Q^2 \exp({\cal C}) \qquad ({\cal C} \sim 1) \ .
\label{RSclC}
\ee
The LB contribution (\ref{LB}) is RScl independent; 
however, it depends on the RSch. The truncated bLB contribution still has
some remnant RScl dependence due to truncation, and is
RSch dependent.

The dependence of the coefficients ${\widetilde d}_j$ on RScl and RSch
can be deduced systematically, by the requirement of RScl and
RSch independence of the observable ${\cal D}$ and using the 
known RScl and RSch dependence of the pQCD coupling 
$a_{\rm pt}({\cal C}; c_2, c_3,\ldots)$ \cite{Stevenson}. The resulting
dependence of ${\widetilde d}_j$ is
\ba
{\widetilde d}_1 &=& {\overline {\widetilde d}}_1 + \beta_0 {\cal C} 
\quad (= d_1) \ ,
\label{td1gen}
\\
{\widetilde d}_2 &=& {\overline {\widetilde d}}_2 + 
\left[ 2 \beta_0 {\cal C} {\overline {\widetilde d}}_1 +
\beta_0^2 {\cal C}^2 \right] - (c_2 - {\overline c}_2) \ , 
\label{td2gen}
\\
{\widetilde d}_3 &=&  {\overline {\widetilde d}}_3 + 
\left[ 3 \beta_0 {\cal C} {\overline {\widetilde d}}_2 +
3 \beta_0^2 {\cal C}^2 {\overline {\widetilde d}}_1 + \beta_0^3 {\cal C}^3
\right] + \left[ - 3 ({\overline {\widetilde d}}_1 + \beta_0 {\cal C}) +
\frac{5}{2} c_1 \right] (c_2 - {\overline c}_2) 
- \frac{1}{2} (c_3 - {\overline c}_3)
\ , 
\label{td3gen}
\ea
etc. On the other hand, the RScl independence of LB contribution
(\ref{LB})-(\ref{LBexp}) implies for the LB coefficients
(\ref{cnn}) the following RScl dependence (they are RSch independent)
\be
c^{(1)}_{nn} = \sum_{k=0}^n \left(
\begin{array}{c}
n \\
k
\end{array}
\right)
{\overline c}^{(1)}_{kk} {\cal C}^{n-k} \ ,
\label{cnngen}
\ee
where ${\overline c}^{(1)}_{00}=1$ by definition.
When we subtract from the modified analytic (``man'') series 
(\ref{man}) the LB contribution
(\ref{LBexp}), we obtain the bLB contribution separately
\ba
{\cal D}_{\rm man}^{\rm (LB+bLB)}(Q^2) & = &
{\cal D}^{\rm (LB)}_{\rm an}(Q^2) + {\cal D}^{\rm (bLB)}_{\rm man}(Q^2)
\nonumber\\
&= &
\int_0^\infty \frac{dt}{t}\: F_{\cal D}^{\cal {E}}(t) \: 
\A_1(t Q^2 e^{\overline {\cal C}}) +
\sum_{n=1}^{\infty} (T_{\cal D})_n \tA_{n+1} \ ,
\label{man2}
\ea
where ${\overline {\cal C}}=-5/3$ as mentioned earlier in 
Appendix \ref{app:LB},
$\tA_{n+1}$ are in RSch ($c_2, c_3, \ldots$) and 
at RScl $\mu^2 = Q^2 \exp({\cal C})$, and the coefficients
$(T_{\cal D})_n$ are
\be
(T_{\cal D})_n = {\widetilde d}_n - c^{(1)}_{nn} \beta_0^n \ ,
\label{TDn}
\ee
where ${\widetilde d}_n$ and $c^{(1)}_{nn}$ are related with
the corresponding (bar) quantities in ${\overline {\rm MS}}$ RSch
and RScl $\mu^2=Q^2$ via relations (\ref{td1gen})-(\ref{td3gen})
and (\ref{cnngen}). This, and application of relations 
(\ref{td1})-(\ref{td3}) in ${\overline {\rm MS}}$ RSch
and RScl $\mu^2=Q^2$, allows us to obtain the first
three coefficients $(T_{\cal D})_n$ by knowing
the first three coefficients ${\overline d}_n$ ($n=1,2,3$)
(all ${\overline c}^{(1)}_{kk}$ are known).

Another variant of evaluation of ${\cal D}$ in anQCD
is not to perform the LB resummation (\ref{LB}) in (\ref{man2}), 
but rather use its expanded form (\ref{LBexp}). This leads to
\ba
{\cal D}_{\rm man}(Q^2) & = &
\A_1 + \sum_{n=1}^{\infty} {\widetilde d}_n \tA_{n+1} \ ,
\label{man3}
\ea  
where $a \equiv \A_1 \equiv \A_1(Q^2 \exp({\cal C}); c_2, \ldots)$. 
Series (\ref{man3}) was obtained in Appendix \ref{app:hoanQCD}
in Eq.~(\ref{man}).

In principle, both series (\ref{man2}) and (\ref{man3})
must lead to the same result if the series are convergent.
However, in practice, only the first three terms in the
sums there ($n=1,2,3$) are known. Hence the series
(\ref{man2}) and (\ref{man3}) truncated at $n=3$ 
\ba
{\cal D}_{\rm man}^{\rm (LB+bLB)}(Q^2)^{[4]} & = &
\int_0^\infty \frac{dt}{t}\: F_{\cal D}^{\cal {E}}(t) \: 
\A_1(t Q^2 e^{\overline {\cal C}}) +
\sum_{n=1}^{3} (T_{\cal D})_n \tA_{n+1} \ ,
\label{man2tr}
\\
{\cal D}_{\rm man}(Q^2)^{[4]} & = &
\A_1 + \sum_{n=1}^{3} {\widetilde d}_n \tA_{n+1} \ ,
\label{man3tr}
\ea
will give in general somewhat different results, the
difference being $\sim \tA_5 (\sim \A_5)$. In theory, the
LB-resummed truncated version (\ref{man2tr}) is better since it
includes more contributions than the simple truncated version
(\ref{man3tr}). Which of the two is better in practice,
in the case of a specific considered inclusive observable ${\cal D}(Q^2)$,
can be decided numerically, e.g., by establishing which
of the two truncated series has weaker variation under
the variation of the RScl 
($\Leftrightarrow$ under the variation of ${\cal C}$).   
If ${\cal D}(Q^2)$ is not an inclusive observable (e.g.,
jet observables, etc.), LB resummation cannot be performed
since $F_{\cal D}^{\cal {E}}(t)$ does not exist, and only the
expression (\ref{man3tr}) is applicable in such a case.

The bLB part of expression (\ref{man2}), and the sum over $\tA_{n+1}$
in Eq.~(\ref{man3}), can be reorganized
into sums over $\A_{n+1}$'s as defined in Eqs.~(\ref{A2})-(\ref{A4})
[$\A_{n+1} = a^{n+1}$ in our paper since $\beta(a)$ is analytic in $a=0$,
Eq.~(\ref{ourAn})]. In such a case, the truncated analytic expressions
analogous to (\ref{man2tr})-(\ref{man3tr}) are
\ba
{\cal D}_{\rm an}^{\rm (LB+bLB)}(Q^2)^{[4]} & = &
\int_0^\infty \frac{dt}{t}\: F_{\cal D}^{\cal {E}}(t) \: 
\A_1(t Q^2 e^{\overline {\cal C}}) +
\sum_{n=1}^{3} (t_{\cal D})_n \A_{n+1} \ ,
\label{an2tr}
\\
{\cal D}_{\rm an}(Q^2)^{[4]} & = &
\A_1 + \sum_{n=1}^{3} d_n \A_{n+1} \ .
\label{an3tr}
\ea
The truncated series (\ref{an3tr}) was obtained in Appendix \ref{app:hoanQCD}
in Eq.~(\ref{anN}). Again, theoretically, the truncated expansion
(\ref{an2tr}) is better than (\ref{an3tr}).
All the truncated expansions (\ref{man2tr}), (\ref{man3tr}), (\ref{an2tr}),
(\ref{an3tr}) differ from each other by $\sim \tA_5 \sim \A_5$.
Our numerically preferred version of evaluation will be 
the truncated expansion (\ref{man2tr}).

Expressions for bLB coefficients $(T_{\cal D})_n$ 
($n=1,2,3$), appearing in Eqs.~(\ref{TDn}) and (\ref{man2tr}),
are obtained from the (usually known) coefficients
${\overline d}_j$ ($j=1,2,3$) via successive use of
Eqs.~(\ref{td1})-(\ref{td3}) 
[${\overline d}_j \mapsto {\overline {\widetilde d}}_j$]; 
Eqs.~(\ref{td1gen})-(\ref{td3gen})
[${\overline {\widetilde d}}_j \mapsto {\widetilde d}_j$];
Eq.~(\ref{cnngen}) [${\overline c}^{(1)}_{jj} \mapsto c^{(1)}_{jj}$];
and Eq.~(\ref{TDn}).

It turns out that these coefficients are equal to
the coefficients ${\widetilde t}_{n+1}$ as derived in Appendix A
of Ref.~\cite{CV2}, ${\widetilde t}_{n+1} = (T_{\cal D})_n$,
as it should be.\footnote{
In Eq.~(A18) for ${\widetilde t}_{4} =(T_{\cal D})_3$ 
of Ref.~\cite{CV2} there is a typo: in the first line
the last term should be 
$- \delta b_{21} 3 ({\overline c}_{11}^{(1)} + {\cal C})$
instead of $- \delta b_{21} 3 {\overline c}_{11}^{(1)}$.
The correct formula was used in the calculations there;
Eqs.~(89)-(92) in Ref.~\cite{CV2}, 
which follow from Eq.~(A18) there, are correct. In terms of the quantities
of Ref.~\cite{CV2}, Eq.~(A18) there (without the typo) can be rewritten in the form:
\ba
{\widetilde t}_4 &=& (T_{\cal D})_3 = {\overline {\widetilde t}}_4
- (1/2) (c_3 - {\overline c}_3) - (c_2 - {\overline c}_2)
\left[ 3 {\overline c}^{(1)}_{10} 
+ 3 ({\overline c}_{11}^{(1)} + {\cal C}) \beta_0 
- (5/2) c_1 \right] + 3 {\cal C} \beta_0 {\overline {\widetilde t}}_3 +
3 {\cal C}^2 \beta_0^2 {\overline c}^{(1)}_{10} \ .
\label{tt4}
\ea} 
The bLB coefficients $(t_{\cal D})_n$ ($n=1,2,3$) appearing in
Eq.~(\ref{an2tr}), on the other hand,
turn out to be equal to expressions $t_{n+1} = t_{n+1}^{(2)} +
\cdots t_{n+1}^{(n+1)}$ of Appendix A of Ref.~\cite{CV2} when
the RScl parameters ${\cal C}_k$ there are all set equal to ${\cal C}$.

In our evaluations of BjPSR and $r_{\tau}$, we will use
${\overline d}_n$ ($n=1,2,3$) coefficients (in ${\overline {\rm MS}}$
RSch with RScl $\mu^2=Q^2$)
for massless BjPSR
${\cal D}(Q^2) = d_{\rm Bj}(Q^2)$ and massless Adler function 
${\cal D}(Q^2) = d_{\rm Adl}(Q^2)$.

Coefficients ${\overline d}_1$ and ${\overline d}_2$ for 
massless BjPSR were obtained in Ref.~\cite{LV}, 
\ba
({\overline d}_{\rm Bj})_1 & = & - \frac{11}{12} + 2 \beta_0 \ , 
\label{d1Bj}
\\
({\overline d}_{\rm Bj})_2  & = &
-35.7644 + 10.5048 \beta_0 + 6.38889 \beta_0^2 \ ,
\label{d2Bj} 
\ea
and ${\overline d}_3$ was estimated in Ref.~\cite{KS}
\be
({\overline d}_{\rm Bj})_3 \approx 130 \qquad (n_f=3) \ .
\label{d3Bj}
\ee
The leading-$\beta_0$ coefficients $c^{(1)}_{nn}$ for BjPSR were
calculated in Ref.~\cite{Broadhurst:1993ru} in the
${\overline {\rm MS}}$ RSch and at RScl $\mu^2=Q^2 \exp({\overline {\cal C}})$
(where: ${\overline {\cal C}} = -5/3$). When changing RScl to
$\mu^2=Q^2$ using an ``inverted'' version of relations (\ref{cnngen})
(with $c_{nn}^{(1)} \mapsto {\overline c}^{(1)}_{nn}$,
${\overline c}^{(1)}_{kk} \mapsto c_{kk}^{(1)}$, and
${\cal C} \mapsto - {\overline {\cal C}} = +5/3$),
we obtain ${\overline c}^{(1)}_{11}=2$ [cf.~Eq.~(\ref{d1Bj})];
${\overline c}^{(1)}_{22}=115/18 (\approx 6.38889)$ [cf.~Eq.~(\ref{d2Bj})];
and ${\overline c}^{(1)}_{33}=605/27 (\approx 22.4074)$.

Coefficients ${\overline d}_n$ ($n=1,2,3$) for the massless Adler function
were obtained in Refs.~\cite{d1,d2,d3}, respectively
\ba
({\overline d}_{\rm Adl})_1 & = & \frac{1}{12} + 0.691772 \beta_0 \ ,
\label{d1Adl}
\\
({\overline d}_{\rm Adl})_2 &=& -27.849 + 8.22612 \beta_0 + 3.10345 \beta_0^2
\ ,
\label{d2Adl}
\\
({\overline d}_{\rm Adl})_3 &=& 
32.727 - 115.199 \beta_0 + 49.5237 \beta_0^2 + 2.18004 \beta_0^3 \ .
\label{d3Adl}
\ea
The light-by-light contributions are not included in these
coefficients; however, they are zero when $n_f=3$, and the value
$n_f=3$ is used in the evaluation of $d_{\rm Adl}(Q^2)$ and
subsequently in the evaluation of $r_{\tau}$. The latter
observable (with $\Delta S=0$ and the mass effects subtracted)
is calculated by using the massless Adler function 
$d_{\rm Adl}(Q^2 = m_{\tau}^2 \exp(i \phi))$ in the
contour integration (\ref{rtaucont}).
Specifically, applying this contour integration to 
the analytic expansion (\ref{man2}) of the Adler function, we obtain
\be
(r_{\tau})_{\rm man}^{\rm (LB+nLB)} = r_{\tau}^{\rm (LB)} + 
\sum_{n=1}^{\infty} \ (T_{\rm Adl})_n I(\tA_{n+1},{\cal C}) \ ,
\label{rtman2}
\ee
where
\be
I(\tA_{n+1},{\cal C}) =
\frac{1}{2 \pi} \int_{-\pi}^{+ \pi}
d \phi \ (1 + e^{i \phi})^3 (1 - e^{i \phi}) \
\tA_{n+1}(e^{\cal C} m_{\tau}^2 e^{i \phi}) \ ,
\label{IanC}
\ee
and $r_{\tau}^{\rm (LB)}$ is given in Eq.~(\ref{LBrt2}).
In practical evaluation, the sum in (\ref{rtman2}) is truncated
at $n=3$
\be
(r_{\tau})_{\rm man}^{\rm (LB+nLB),[4]} = 
\frac{1}{\pi} \int_0^\infty \frac{dt}{t}\: {\widetilde F}_{r}(t) \: 
\rho_1(t e^{\cal {\overline C}} m_{\tau}^2) +
\sum_{n=1}^{3} \ (T_{\rm Adl})_n I(\tA_{n+1},{\cal C}) \ .
\label{rtman2tr}
\ee
The other three analytic versions of evaluation are obtained by 
contour-integrating, via (\ref{rtaucont}), the analytic
truncated series (\ref{man3tr}), (\ref{an2tr}) and (\ref{an3tr})
of massless Adler function ${\cal D}(Q^2) = d_{\rm Adl}(Q^2)$:
\ba
(r_{\tau})_{\rm man}^{[4]} &=& I(\A_1,{\cal C}) +
\sum_{n=1}^{3} \ ({\widetilde d}_{\rm Adl})_n I(\tA_{n+1},{\cal C}) \ ,
\label{rtman3tr}
\\
(r_{\tau})_{\rm an}^{\rm (LB+nLB),[4]} &=& 
\frac{1}{\pi} \int_0^\infty \frac{dt}{t}\: {\widetilde F}_{r}(t) \: 
\rho_1(t e^{\cal {\overline C}} m_{\tau}^2) +
\sum_{n=1}^{3} \ (t_{\rm Adl})_n I(\A_{n+1},{\cal C}) \ ,
\label{rtan2tr}
\\
(r_{\tau})_{\rm an}^{[4]} &=& I(\A_1,{\cal C}) +
\sum_{n=1}^{3} \ (d_{\rm Adl})_n I(\A_{n+1},{\cal C}) \ .
\label{rtan3tr}
\ea
Again, all four versions of the anQCD evaluation of
$r_{\tau}$ differ from each other by $\sim \tA_5 \sim \A_5$.
The truncated expansion (\ref{rtman2tr}) is our numerically
preferred version.


\begin{thebibliography}{99}

\bibitem{BS}
N.N.~Bogoliubov and D.V.~Shirkov, 
{\it Introduction to the Theory of Quantum Fields\/}, New York, Wiley, 1959); (1980), 3rd ed.
  %%CITATION = NONE;%%

\bibitem{SDEs}
%\cite{Alkofer:2004it}
  R.~Alkofer, C.~S.~Fischer and F.~J.~Llanes-Estrada,
  %``Vertex functions and infrared fixed point in Landau gauge SU(N)  Yang-Mills
  %theory,''
  Phys.\ Lett.\  B {\bf 611}, 279 (2005)
  [Erratum-ibid.\  {\bf 670}, 460 (2009)]
  [arXiv:hep-th/0412330];
  %%CITATION = PHLTA,B611,279;%%
%\cite{Aguilar:2008xm}
  A.~C.~Aguilar, D.~Binosi and J.~Papavassiliou,
  %``Gluon and ghost propagators in the Landau gauge: Deriving lattice results
  %from Schwinger-Dyson equations,''
  Phys.\ Rev.\  D {\bf 78}, 025010 (2008)
  [arXiv:0802.1870 [hep-ph]];
  %%CITATION = PHRVA,D78,025010;%%
%\cite{Aguilar:2009nf}
  A.~C.~Aguilar, D.~Binosi, J.~Papavassiliou and J.~Rodriguez-Quintero,
  %``Non-perturbative comparison of QCD effective charges,''
  Phys.\ Rev.\  D {\bf 80}, 085018 (2009)
  [arXiv:0906.2633 [hep-ph]].
  %%CITATION = PHRVA,D80,085018;%%


\bibitem{latt}
%\cite{Cucchieri:2007rg}
  A.~Cucchieri and T.~Mendes,
  %``Constraints on the IR behavior of the gluon propagator in Yang-Mills
  %theories,''
  Phys.\ Rev.\ Lett.\  {\bf 100}, 241601 (2008)
  [arXiv:0712.3517 [hep-lat]];
  %%CITATION = PRLTA,100,241601;%%
%\cite{Bogolubsky:2009dc}
  I.~L.~Bogolubsky, E.~M.~Ilgenfritz, M.~Muller-Preussker and A.~Sternbeck,
  %``Lattice gluodynamics computation of Landau gauge Green's functions in the
  %deep infrared,''
  Phys.\ Lett.\  B {\bf 676}, 69 (2009)
  [arXiv:0901.0736 [hep-lat]].
  %%CITATION = PHLTA,B676,69;%%


\bibitem{ShS} 
%\cite{Shirkov:1996cd}
  D.~V.~Shirkov and I.~L.~Solovtsov,
  %``Analytic QCD running coupling with finite IR behaviour 
%and universal ${\bar \alpha}_s(0)$ value,''
  hep-ph/9604363;
  %%CITATION = HEP-PH 9604363;%%
%\cite{Shirkov:1997wi}
%  D.~V.~Shirkov and I.~L.~Solovtsov,
  %``Analytic model for the QCD running coupling with universal  
%alpha(s)-bar(0) value,''
  Phys.~Rev.~Lett.~{\bf 79}, 1209 (1997)
[arXiv:hep-ph/9704333].
  %%CITATION = HEP-PH 9704333;%%

\bibitem{MSS}
  K.~A.~Milton, I.~L.~Solovtsov and O.~P.~Solovtsova,
  %``Analytic perturbation theory and inclusive tau decay,''
  Phys.\ Lett.\ B {\bf 415}, 104 (1997)
[arXiv:hep-ph/9706409].
  %%CITATION = HEP-PH 9706409;%%

\bibitem{Sh}
%\cite{Shirkov:2000qv}
  D.~V.~Shirkov,
  %``Analytic perturbation theory for QCD observables,''
  Theor.\ Math.\ Phys.\  {\bf 127}, 409 (2001)
[hep-ph/0012283];
  %%CITATION = HEP-PH 0012283;%%
%\cite{Shirkov:2001sm}
%  D.~V.~Shirkov,
%``Analytic perturbation theory in analyzing some QCD observables,''
  Eur.\ Phys.\ J.\ C {\bf 22}, 331 (2001)
  [hep-ph/0107282].
  %%CITATION = HEP-PH 0107282;%%

%\cite{Webber:1998um}
\bibitem{Webber:1998um}
  B.~R.~Webber,
  %``{QCD} power corrections from a simple model for the running coupling,''
  JHEP {\bf 9810}, 012 (1998)
  [arXiv:hep-ph/9805484].
  %%CITATION = JHEPA,9810,012;%%

%\cite{Srivastava:2001ts}
\bibitem{Srivastava:2001ts}
  Y.~Srivastava, S.~Pacetti, G.~Pancheri and A.~Widom,
  %``Dispersive techniques for alpha(s), R(had) and instability of the
  %perturbative vacuum,''
%in {\it Proc. of the $e^+ e^-$ Physics at Intermediate Energies Conference } ed. Diego Bettoni,
{\it In the Proceedings of $e^+ e^-$ Physics at Intermediate Energies, SLAC, Stanford, CA, USA, 30 April - 2 May 2001, pp T19}
  [arXiv:hep-ph/0106005].
  %%CITATION = ECONF,C010430,T19;%%


\bibitem{Simonov}
Yu.~A.~Simonov,
  %``Perturbative theory in the nonperturbative QCD vacuum,''
  Phys.\ Atom.\ Nucl.\  {\bf 58}, 107 (1995)
  [Yad.\ Fiz.\  {\bf 58}, 113 (1995)]
  [arXiv:hep-ph/9311247];
  %%CITATION = YAFIA,58,113;%%
A.~M.~Badalian and Yu.~A.~Simonov,
  %``Freezing of alpha(s)(Q**2) in e+ e- annihilation,''
  Phys.\ Atom.\ Nucl.\  {\bf 60}, 630 (1997)
  [Yad.\ Fiz.\  {\bf 60}, 714 (1997)];
  %%CITATION = YAFIA,60,714;%%
  Yu.~A.~Simonov,
  %``Perturbative expansions in QCD and analytic properties of alpha(s),''
  Phys.\ Atom.\ Nucl.\  {\bf 65}, 135 (2002)
  [Yad.\ Fiz.\  {\bf 65}, 140 (2002)]
  [arXiv:hep-ph/0109081];
  %%CITATION = YAFIA,65,140;%%
{\it ibid\/},
  %``Novel solutions of RG equations for alpha(s) and beta(alpha) in the  large
  %N(c) limit,''
  Phys.\ Atom.\ Nucl.\ {\bf 66}, 764 (2003)
  [Yad.\ Fiz.\  {\bf 66}, 796 (2003)]
  [arXiv:hep-ph/0109159];
  %%CITATION = YAFIA,66,796;%%
  %``RG solutions for alpha(s) at large N(c) in d = 3+1 QCD,''
  J.\ Nonlin.\ Math.\ Phys.\  {\bf 12}, S625 (2005)
  [arXiv:hep-ph/0409265].
  %%CITATION = 00154,12,S625;%%


\bibitem{Nesterenko}
%\cite{Nesterenko:1999dx}
  A.~V.~Nesterenko,
  %``Quark antiquark potential in the analytic approach to QCD,''
  Phys.\ Rev.\ D {\bf 62}, 094028 (2000);
  %%CITATION = HEP-PH 9912351;%%
%\cite{Nesterenko:2001st}
%  A.~V.~Nesterenko,
  %``New analytic running coupling in spacelike and timelike regions,''
  Phys.\ Rev.\ D {\bf 64}, 116009 (2001);
  %%CITATION = HEP-PH 0102124;%%
%\cite{Nesterenko:2003xb}
%  A.~V.~Nesterenko,
  %``Analytic invariant charge in QCD,''
  Int.\ J.\ Mod.\ Phys.\ A {\bf 18}, 5475 (2003);
  %%CITATION = HEP-PH 0308288;%%
%\cite{Nesterenko:2004tg}

\bibitem{Nesterenko2}
  A.~V.~Nesterenko and J.~Papavassiliou,
  %``The massive analytic invariant charge in QCD,''
  Phys.\ Rev.\ D {\bf 71}, 016009 (2005);
  %%CITATION = HEP-PH 0410406;%%
%\cite{Aguilar:2005sb}
  A.~C.~Aguilar, A.~V.~Nesterenko and J.~Papavassiliou,
  %``Infrared enhanced analytic coupling and chiral symmetry breaking in  QCD,''
  J.\ Phys.\ G {\bf 31}, 997 (2005).
  %%CITATION = HEP-PH 0504195;%%
%\bibitem{Nesterenko:2005wh}
%  A.~V.~Nesterenko and J.~Papavassiliou,
  %``Infrared behavior of the Adler function from a novel dispersion
  %relation,''
  J.\ Phys.\ G {\bf 32}, 1025 (2006)
  [arXiv:hep-ph/0511215];
  %%CITATION = JPHGB,G32,1025;%%
%\cite{Nesterenko:2007fm}
%\bibitem{Nesterenko:2007fm}
  A.~V.~Nesterenko,
  %``Adler function in the analytic approach to QCD,''
  arXiv:0710.5878 [hep-ph].
  %%CITATION = ARXIV:0710.5878;%%

%\cite{Alekseev:2005he}
\bibitem{Alekseev:2005he}
  A.~I.~Alekseev,
  %``Synthetic running coupling of QCD,''
  Few Body Syst.\  {\bf 40}, 57 (2006)
  [arXiv:hep-ph/0503242].
  %%CITATION = FBSYE,40,57;%%

\bibitem{CV1}
%\bibitem{Cvetic:2006mk}
  G.~Cveti\v c and C.~Valenzuela,
  %``An approach for evaluation of observables in analytic versions of QCD,''
  J.\ Phys.\ G {\bf 32}, L27 (2006)
  [arXiv:hep-ph/0601050].
  %%CITATION = JPHGB,G32,L27;%%

\bibitem{CV2}
%\bibitem{Cvetic:2006gc}
  G.~Cveti\v c and C.~Valenzuela,
  %``Various versions of analytic QCD and skeleton-motivated evaluation of
  %observables,''
  Phys.\ Rev.\  D {\bf 74}, 114030 (2006)
  [arXiv:hep-ph/0608256].
  %%CITATION = PHRVA,D74,114030;%%

\bibitem{CCEM}
  C.~Contreras, G.~Cveti\v{c}, O.~Espinosa and H.~Mart\'{\i}nez,
  %``Simple analytic QCD model with perturbative QCD behavior at high momenta,''
  Phys.\ Rev.\  D {\bf 82}, 074005 (2010)
  [arXiv:1006.5050 [hep-ph]].
  %%CITATION = PHRVA,D82,074005;%%

\bibitem{mes1}
%\cite{Baldicchi:2002qm}
%\bibitem{Baldicchi:2002qm}
  M.~Baldicchi and G.~M.~Prosperi,
  %``Infrared behavior of the running coupling constant and bound states in
  %QCD,''
  Phys.\ Rev.\  D {\bf 66}, 074008 (2002);
%  [arXiv:hep-ph/0202172];
  %%CITATION = PHRVA,D66,074008;%%
%\cite{Baldicchi:2004wj}
%\bibitem{Baldicchi:2004wj}
%  M.~Baldicchi and G.~M.~Prosperi,
  %``Running coupling constant and masses in QCD, the meson spectrum,''
  AIP Conf.\ Proc.\  {\bf 756}, 152 (2005)
  [arXiv:hep-ph/0412359].
  %%CITATION = APCPC,756,152;%%

\bibitem{mes2}
%\cite{Baldicchi:2007ic}
%\bibitem{Baldicchi:2007ic}
  M.~Baldicchi, A.~V.~Nesterenko, G.~M.~Prosperi, D.~V.~Shirkov and C.~Simolo,
  %``Bound state approach to the QCD coupling at low energy scales,''
  Phys.\ Rev.\ Lett.\  {\bf 99}, 242001 (2007);
%  [arXiv:0705.0329 [hep-ph]];
  %%CITATION = PRLTA,99,242001;%%
%\cite{Baldicchi:2007zn}
%\bibitem{Baldicchi:2007zn}
%  M.~Baldicchi, A.~V.~Nesterenko, G.~M.~Prosperi, D.~V.~Shirkov and C.~Simolo,
  %``QCD coupling below 1-GeV from quarkonium spectrum,''
  Phys.\ Rev.\  D {\bf 77}, 034013 (2008).
%  [arXiv:0705.1695 [hep-ph]].
  %%CITATION = PHRVA,D77,034013;%%

\bibitem{Bakulev}
  A.~P.~Bakulev, S.~V.~Mikhailov and N.~G.~Stefanis,
  %``QCD analytic perturbation theory: From integer powers 
% to any power of the running coupling,''
  Phys.\ Rev.\ D {\bf 72}, 074014 (2005)
  [Erratum-ibid.\ D {\bf 72}, 119908 (2005)];
%[hep-ph/0506311];
  %%CITATION = HEP-PH 0506311;%%
%\bibitem{Bakulev:2006ex}
%  A.~P.~Bakulev, S.~V.~Mikhailov and N.~G.~Stefanis,
  %``Fractional Analytic Perturbation Theory in Minkowski space and
  %application to Higgs boson decay into a b\bar{b} pair,''
  Phys.\ Rev.\  D {\bf 75}, 056005 (2007);
%  [arXiv:hep-ph/0607040];
  %%CITATION = PHRVA,D75,056005;%%
%\bibitem{Bakulev:2005fp}
  A.~P.~Bakulev, A.~I.~Karanikas and N.~G.~Stefanis,
  %``Analyticity properties of three-point functions 
%in QCD beyond leading order,''
  Phys.\ Rev.\ D {\bf 72}, 074015 (2005).
%[hep-ph/0504275].
  %%CITATION = HEP-PH 0504275;%%

%\cite{Prosperi:2006hx}
\bibitem{Prosperi:2006hx}
  G.~M.~Prosperi, M.~Raciti and C.~Simolo,
  %``On the running coupling constant in QCD,''
  Prog.\ Part.\ Nucl.\ Phys.\  {\bf 58}, 387 (2007)
  [arXiv:hep-ph/0607209].
  %%CITATION = PPNPD,58,387;%%

%\cite{Shirkov:2006gv}
\bibitem{Shirkov:2006gv}
  D.~V.~Shirkov and I.~L.~Solovtsov,
  %``Ten years of the analytic perturbation theory in QCD,''
  Theor.\ Math.\ Phys.\  {\bf 150}, 132 (2007)
  [arXiv:hep-ph/0611229].
  %%CITATION = TMPHA,150,132;%%

%\cite{Cvetic:2008bn}
\bibitem{Cvetic:2008bn}
  G.~Cveti\v c and C.~Valenzuela,
  %``Analytic QCD - a short review,''
  Braz.\ J.\ Phys.\  {\bf 38}, 371 (2008)
  [arXiv:0804.0872 [hep-ph]].
  %%CITATION = BJPHE,38,371;%%

\bibitem{Shirkov:1999hm}
  D.~V.~Shirkov,
  %``Unitary mechanism of infrared freezing in QCD with massive gluons,''
  Phys.\ Atom.\ Nucl.\  {\bf 62}, 1928 (1999)
  [Yad.\ Fiz.\  {\bf 62}, 2082 (1999)]
  [arXiv:hep-ph/9903431].
  %%CITATION = YAFIA,62,2082;%%


\bibitem{DMW}
  Y.~L.~Dokshitzer, G.~Marchesini and B.~R.~Webber,
  %``Dispersive Approach to Power-Behaved Contributions 
%in QCD Hard Processes,''
  Nucl.\ Phys.\ B {\bf 469}, 93 (1996)
  [arXiv:hep-ph/9512336].
  %%CITATION = HEP-PH 9512336;%%

%\cite{Cvetic:2007ad}
\bibitem{Cvetic:2007ad}
  G.~Cveti\v c and C.~Valenzuela,
  %``Exponentially Modified QCD Coupling,''
  Phys.\ Rev.\  D {\bf 77}, 074021 (2008)
  [arXiv:0710.4530 [hep-ph]].
  %%CITATION = PHRVA,D77,074021;%%


%\cite{Shifman:1978bx}
\bibitem{Shifman:1978bx}
  M.~A.~Shifman, A.~I.~Vainshtein and V.~I.~Zakharov,
  %``QCD And Resonance Physics. Sum Rules,''
  Nucl.\ Phys.\  B {\bf 147}, 385 (1979);
  %%CITATION = NUPHA,B147,385;%%
%\cite{Shifman:1978by}
%\bibitem{Shifman:1978by}
%  M.~A.~Shifman, A.~I.~Vainshtein and V.~I.~Zakharov,
  %``QCD And Resonance Physics: Applications,''
  Nucl.\ Phys.\  B {\bf 147}, 448 (1979).
  %%CITATION = NUPHA,B147,448;%%

\bibitem{Dyson1952}
F.~J.~Dyson,
  %``Divergence Of Perturbation Theory In Quantum Electrodynamics,''
  Phys.\ Rev.\  {\bf 85}, 631 (1952).
  %%CITATION = PHRVA,85,631;%%


\bibitem{KSh1980}
D.~I.~Kazakov and D.~V.~Shirkov,
  %``Asymptotic Series Of Quantum Field Theory And Their Summation,''
  Fortsch.\ Phys.\  {\bf 28}, 465 (1980).
  %%CITATION = FPYKA,28,465;%%

\bibitem{MSSY}
  K.~A.~Milton, I.~L.~Solovtsov, O.~P.~Solovtsova and V.~I.~Yasnov,
  %``Renormalization scheme and higher loop stability in hadronic tau decay
  %within analytic perturbation theory,''
  Eur.\ Phys.\ J.\ C {\bf 14}, 495 (2000)
  [arXiv:hep-ph/0003030].
  %%CITATION = HEP-PH 0003030;%%

%\cite{Milton:2001mq}
\bibitem{Milton:2001mq}
  K.~A.~Milton, I.~L.~Solovtsov and O.~P.~Solovtsova,
  %``The Adler function for light quarks in analytic perturbation theory,''
  Phys.\ Rev.\  D {\bf 64}, 016005 (2001)
  [arXiv:hep-ph/0102254].
  %%CITATION = PHRVA,D64,016005;%%

%\cite{Geshkenbein:2001mn}
\bibitem{Geshkenbein:2001mn}
  B.~V.~Geshkenbein, B.~L.~Ioffe and K.~N.~Zyablyuk,
  %``The check of QCD based on the tau-decay data analysis in the complex
  %q^2-plane,''
  Phys.\ Rev.\  D {\bf 64}, 093009 (2001)
  [arXiv:hep-ph/0104048].
  %%CITATION = PHRVA,D64,093009;%%

\bibitem{CKV1}
  G.~Cveti\v{c}, R.~K\"ogerler and C.~Valenzuela,
  %``Analytic QCD coupling with no power terms in UV regime,''
  J.\ Phys.\ G {\bf 37}, 075001 (2010)
  [arXiv:0912.2466 [hep-ph]].
  %%CITATION = JPHGB,G37,075001;%%

\bibitem{Raczka}
P.~A.~R\c{a}czka, 
  %``Stable perturbative QCD predictions at moderate energies with a modified
  %couplant,''
  Nucl.\ Phys.\ Proc.\ Suppl.\  {\bf 164}, 211 (2007)
  [arXiv:hep-ph/0512339];
  %%CITATION = NUPHZ,164,211;%%
% ``Improving reliability of perturbative QCD predictions at moderate
 %energies,''
hep-ph/0602085;
%%CITATION = HEP-PH 0602085;%%
  %``Towards more reliable perturbative QCD predictions at moderate energies,''
hep-ph/0608196.
  %%CITATION = HEP-PH 0608196;%%

%\cite{Buras:1977qg}
\bibitem{Buras:1977qg}
  A.~J.~Buras, E.~G.~Floratos, D.~A.~Ross and C.~T.~Sachrajda,
  %``Asymptotic Freedom Beyond The Leading Order,''
  Nucl.\ Phys.\ B {\bf 131}, 308 (1977).
  %%CITATION = NUPHA,B131,308;%%

%\cite{Bardeen:1978yd}
\bibitem{Bardeen:1978yd}
  W.~A.~Bardeen, A.~J.~Buras, D.~W.~Duke and T.~Muta,
%   ``Deep Inelastic Scattering Beyond The Leading Order 
%In Asymptotically Free
  %Gauge Theories,''
  Phys.\ Rev.\ D {\bf 18}, 3998 (1978).
  %%CITATION = PHRVA,D18,3998;%%

\bibitem{BW}
  W.~Wetzel,
  %``Minimal Subtraction And The Decoupling Of Heavy Quarks For Arbitrary Values
  %Of The Gauge Parameter,''
  Nucl.\ Phys.\  B {\bf 196}, 259 (1982);
  %%CITATION = NUPHA,B196,259;%%
  W.~Bernreuther and W.~Wetzel,
  %``Decoupling of heavy quarks in the minimal subtraction scheme,''
  Nucl.\ Phys.\  B {\bf 197}, 228 (1982)
  [Erratum-ibid.\  B {\bf 513}, 758 (1998)];
  %%CITATION = NUPHA,B197,228;%%
  W.~Bernreuther,
  %``Decoupling Of Heavy Quarks In Quantum Chromodynamics,''
  Annals Phys.\  {\bf 151}, 127 (1983);
  %%CITATION = APNYA,151,127;%%
%``Heavy Quark Effects On The Parameters Of Quantum Chromodynamics Defined By
  %Minimal Subtraction,''
  Z.\ Phys.\  C {\bf 20}, 331 (1983).
  %%CITATION = ZEPYA,C20,331;%%

\bibitem{RS}
  G.~Rodrigo and A.~Santamaria,
  %``QCD matching conditions at thresholds,''
  Phys.\ Lett.\  B {\bf 313}, 441 (1993)
  [arXiv:hep-ph/9305305].
  %%CITATION = PHLTA,B313,441;%%

\bibitem{LRV}
  S.~A.~Larin, T.~van Ritbergen and J.~A.~M.~Vermaseren,
  %``The Large quark mass expansion of Gamma (Z0 $\to$ hadrons) and Gamma (tau-
  %$\to$ tau-neutrino + hadrons) in the order alpha-s**3,''
  Nucl.\ Phys.\  B {\bf 438}, 278 (1995)
  [arXiv:hep-ph/9411260].
  %%CITATION = NUPHA,B438,278;%%

\bibitem{CKS}
  K.~G.~Chetyrkin, B.~A.~Kniehl and M.~Steinhauser,
  %``Strong coupling constant with flavour thresholds at four loops in the
  %MS-bar scheme,''
  Phys.\ Rev.\ Lett.\  {\bf 79}, 2184 (1997)
  [arXiv:hep-ph/9706430].
  %%CITATION = PRLTA,79,2184;%%

\bibitem{PDG2008}
%\bibitem{Amsler:2008zzb}
  C.~Amsler {\it et al.}  [Particle Data Group],
  %``Review of particle physics,''
  Phys.\ Lett.\  B {\bf 667}, 1 (2008).
  %%CITATION = PHLTA,B667,1;%%

\bibitem{CK1}
 G.~Cveti\v c and R.~K\"ogerler,
  %``Scale- and scheme-independent extension of Pade approximants: Bjorken
  %polarized sum rule as an example,''
  Phys.\ Rev.\  D {\bf 63}, 056013 (2001)
  [arXiv:hep-ph/0006098].
  %%CITATION = PHRVA,D63,056013;%%

\bibitem{math}
MATHEMATICA 7.0.1, Wolfram Co.
  %%CITATION = NONE;%%

%\cite{Cvetic:2009mq}
\bibitem{Cvetic:2009mq}
  G.~Cveti\v{c} and H.~E.~Mart\'{\i}nez,
  %``Rational approximations in Analytic QCD,''
  J.\ Phys.\ G {\bf 36}, 125006 (2009)
  [arXiv:0907.0033 [hep-ph]].
  %%CITATION = JPHGB,G36,125006;%%

\bibitem{MS}
  A.~C.~Mattingly and P.~M.~Stevenson,
  %``Optimization Of R(E+ E-) And 'Freezing' Of The QCD Couplant At
  %Low-Energies,''
  Phys.\ Rev.\  D {\bf 49}, 437 (1994)
  [arXiv:hep-ph/9307266].
  %%CITATION = PHRVA,D49,437;%%

\bibitem{Stevenson}
P.~M.~Stevenson, Phys. Rev. D {\bf 23}, 2916 (1981).
%%CITATION = PHRVA,D23,2916;%%


\bibitem{Deur2008}
A.~Deur {\it et al.},
  %``Experimental study of isovector spin sum rules,''
  Phys.\ Rev.\  D {\bf 78}, 032001 (2008)
  [arXiv:0802.3198 [nucl-ex]].
  %%CITATION = PHRVA,D78,032001;%%

%\cite{Kataev:1994gd}
\bibitem{Kataev:1994gd}
  A.~L.~Kataev,
  %``The Ellis-Jaffe sum rule: The Estimates of the next to next-to-leading
  %order QCD corrections,''
  Phys.\ Rev.\  D {\bf 50}, 5469 (1994)
  [arXiv:hep-ph/9408248];
  %%CITATION = PHRVA,D50,5469;%%
  %``Infrared renormalons and the relations between the Gross-Llewellyn  Smith
  %and the Bjorken polarized and unpolarized sum rules,''
  JETP Lett.\  {\bf 81}, 608 (2005)
  [Pisma Zh.\ Eksp.\ Teor.\ Fiz.\  {\bf 81}, 744 (2005)]
  [arXiv:hep-ph/0505108];
  %%CITATION = ZFPRA,81,744;%%
%``Renormalons at the boundaries between perturbative and non-perturbative
  %QCD,''
  Mod.\ Phys.\ Lett.\  A {\bf 20}, 2007 (2005)
  [arXiv:hep-ph/0505230].
  %%CITATION = MPLAE,A20,2007;%%



\bibitem{Pase1}
R.~S.~Pasechnik, D.~V.~Shirkov and O.~V.~Teryaev,
  %``Bjorken Sum Rule and pQCD frontier on the move,''
  Phys.\ Rev.\  D {\bf 78}, 071902 (2008)
  [arXiv:0808.0066 [hep-ph]].
  %%CITATION = PHRVA,D78,071902;%%

\bibitem{Pase2}
R.~S.~Pasechnik, D.~V.~Shirkov, O.~V.~Teryaev, O.~P.~Solovtsova and V.~L.~Khandramai,
  %``Nucleon spin structure and pQCD frontier on the move,''
  Phys.\ Rev.\  D {\bf 81}, 016010 (2010)
  [arXiv:0911.3297].
  %%CITATION = PHRVA,D81,016010;%%


%\cite{Gardi:1998qr}
\bibitem{Gardi:1998qr}
  E.~Gardi, G.~Grunberg and M.~Karliner,
  %``Can the {QCD} running coupling have a 
%causal analyticity structure?,''
  JHEP {\bf 9807}, 007 (1998)
  [hep-ph/9806462];
  %%CITATION = HEP-PH 9806462;%%
B.~A.~Magradze,
  %``The gluon propagator in analytic perturbation theory,''
  arXiv:hep-ph/9808247.
  %%CITATION = HEP-PH/9808247;%%

%\cite{Braaten:1988hc}
\bibitem{Braaten:1988hc}
  E.~Braaten,
  %``QCD Predictions for the Decay of the tau Lepton,''
  Phys.\ Rev.\ Lett.\  {\bf 60}, 1606 (1988);
  %%CITATION = PRLTA,60,1606;%%
%\bibitem{Narison:1988ni}
  S.~Narison and A.~Pich,
  %``QCD Formulation of the tau Decay and Determination of Lambda (MS),''
  Phys.\ Lett.\  B {\bf 211}, 183 (1988);
  %%CITATION = PHLTA,B211,183;%%
%\bibitem{Braaten:1992qm}
E.~Braaten, S.~Narison, and A.~Pich,
%``QCD analysis of the tau hadronic width,''
Nucl.\ Phys.\ B {\bf 373}, 581 (1992);
%%CITATION = NUPHA,B373,581;%%
%\bibitem{Pich:1998yn}
A.~Pich and J.~Prades,
%``Perturbative quark mass corrections to the tau hadronic width,''
JHEP{\bf 9806}, 013 (1998)
[hep-ph/9804462].
  %%CITATION = JHEPA,9806,013;%%

%\cite{Pivovarov:1991rh}
\bibitem{Pivovarov:1991rh}
  A.~A.~Pivovarov,
  %``Renormalization group analysis of the tau-lepton decay within QCD,''
  Z.\ Phys.\  C {\bf 53}, 461 (1992)
  [Sov.\ J.\ Nucl.\ Phys.\  {\bf 54}, 676 (1991\ YAFIA,54,1114.1991)]
  [arXiv:hep-ph/0302003].
  %%CITATION = YAFIA,54,1114;%%

%\cite{Le Diberder:1992te}
\bibitem{Le Diberder:1992te}
  F.~Le Diberder and A.~Pich,
  %``The Perturbative QCD Prediction To R(Tau) Revisited,''
  Phys.\ Lett.\  B {\bf 286}, 147 (1992).
  %%CITATION = PHLTA,B286,147;%%

%\cite{Ball:1995ni}
\bibitem{Ball:1995ni}
  P.~Ball, M.~Beneke and V.~M.~Braun,
  %``Resummation of (beta0 alpha-s)**n corrections in QCD: Techniques and
  %applications to the tau hadronic width and the heavy quark pole mass,''
  Nucl.\ Phys.\  B {\bf 452}, 563 (1995)
  [arXiv:hep-ph/9502300].
  %%CITATION = NUPHA,B452,563;%%

\bibitem{ALEPH2}
%\bibitem{Schael:2005am}
  S.~Schael {\it et al.}  [ALEPH Collaboration],
%   ``Branching ratios and spectral functions of tau decays: Final ALEPH
  %measurements and physics implications,''
  Phys.\ Rept.\  {\bf 421}, 191 (2005)
  [hep-ex/0506072];
  %%CITATION = HEP-EX 0506072;%%

\bibitem{ALEPH3}
%\bibitem{Davier:2005xq}
  M.~Davier, A.~H\"ocker and Z.~Zhang,
  %``The physics of hadronic tau decays,''
 Rev.\ Mod.\ Phys.\  {\bf 78}, 1043 (2006)
  [arXiv:hep-ph/0507078].
  %%CITATION = RMPHA,78,1043;%%


\bibitem{DDHMZ}
  M.~Davier, S.~Descotes-Genon, A.~H\"ocker, B.~Malaescu and Z.~Zhang,
  %``The Determination of $\alpha_s$ from $\tau$ Decays Revisited,''
  Eur.\ Phys.\ J.\  C {\bf 56}, 305 (2008)
  [arXiv:0803.0979 [hep-ph]].
  %%CITATION = EPHJA,C56,305;%%

\bibitem{Ioffe}
  B.~L.~Ioffe,
  %``QCD at low energies,''
  Prog.\ Part.\ Nucl.\ Phys.\  {\bf 56}, 232 (2006)
  [arXiv:hep-ph/0502148].
  %%CITATION = PPNPD,56,232;%%

\bibitem{MY}
K.~Maltman and T.~Yavin,
  %``Alpha_s(M_Z) from hadronic tau decays,''
  Phys.\ Rev.\  D {\bf 78}, 094020 (2008)
  [arXiv:0807.0650 [hep-ph]].
  %%CITATION = PHRVA,D78,094020;%%

%\cite{Beneke:2008ad}
\bibitem{Beneke:2008ad}
  M.~Beneke and M.~Jamin,
  %``$\alpha_s$ and the $\tau$ hadronic width: fixed-order, contour-improved and
  %higher-order perturbation theory,''
  JHEP {\bf 0809}, 044 (2008)
  [arXiv:0806.3156 [hep-ph]].
  %%CITATION = JHEPA,0809,044;%%


%\cite{Caprini:2009vf}
\bibitem{Caprini:2009vf}
  I.~Caprini and J.~Fischer,
  %``$\alpha_s$ from $\tau$ decays: contour-improved versus fixed-order
  %summation in a new QCD perturbation expansion,''
  Eur.\ Phys.\ J.\  C {\bf 64}, 35 (2009)
  [arXiv:0906.5211 [hep-ph]].
  %%CITATION = EPHJA,C64,35;%%

%\cite{DescotesGenon:2010cr}
\bibitem{DescotesGenon:2010cr}
  S.~Descotes-Genon and B.~Malaescu,
  %``A note on renormalon models for the determination of alpha_s(M_tau),''
  arXiv:1002.2968.
  %%CITATION = ARXIV:1002.2968;%%


\bibitem{wp}
G.~Cveti\v{c}, R.~K\"ogerler and C.~Valenzuela,
work in progress.
  %%CITATION = NONE;%%

%\cite{Braaten:1990ef}
\bibitem{Braaten:1990ef}
E.~Braaten and C.~Li,
%``Electroweak Radiative Corrections To The Semihadronic Decay Rate
% Of The Tau Lepton,''
Phys.\ Rev.\ D {\bf 42}, 3888 (1990).
%%CITATION = PHRVA,D42,3888;%%


\bibitem{Magradze}
%\cite{Kurashev:2003pt}
  D.~S.~Kurashev and B.~A.~Magradze,
  %``Explicit expressions for timelike and spacelike observables of quantum
  %chromodynamics in analytic perturbation theory,''
  Theor.\ Math.\ Phys.\  {\bf 135}, 531 (2003);
  %%CITATION = TMPHA,135,531;%%
%\cite{Kourashev:2001kd}
%  D.S.~Kourashev and B.A.~Magradze,
  %``Explicit expressions for Euclidean and Minkowskian QCD observables in
  %analytic perturbation theory,''
  hep-ph/0104142.


\bibitem{Shirkov:1999np}
D.~V.~Shirkov,
  %``Renormalization group, causality, and nonpower perturbation expansion  in
  %QFT,''
  Theor.\ Math.\ Phys.\  {\bf 119}, 438 (1999)
  [Teor.\ Mat.\ Fiz.\  {\bf 119}, 55 (1999)]
  [arXiv:hep-th/9810246];
  %%CITATION = TMFZA,119,55;%%
%``Nonpower 'spectral' perturbation expansion in QFT,''
  Lett.\ Math.\ Phys.\  {\bf 48}, 135 (1999).
  %%CITATION = LMPHD,48,135;%%

\bibitem{CLMV}
  G.~Cveti\v{c}, M.~Loewe, C.~Mart\'{\i}nez and C.~Valenzuela,
  %``Modified Contour-Improved Perturbation Theory,''
  Phys.\ Rev.\  D {\bf 82}, 093007 (2010)
  [arXiv:1005.4444 [hep-ph]].
  %%CITATION = PHRVA,D82,093007;%%


\bibitem{Neubert}
  M.~Neubert,
  %``Scale setting in QCD and the momentum flow in Feynman diagrams,''
  Phys.\ Rev.\ D {\bf 51}, 5924 (1995)
  [hep-ph/9412265].
  %%CITATION = HEP-PH 9412265;%%


%\cite{Broadhurst:1993ru}
\bibitem{Broadhurst:1993ru}
  D.~J.~Broadhurst and A.~L.~Kataev,
  %``Connections between deep inelastic and annihilation 
%processes at next to next-to-leading order and beyond,''
  Phys.\ Lett.\ B {\bf 315}, 179 (1993)
[hep-ph/9308274].
  %%CITATION = HEP-PH 9308274;%%

%\cite{Tsai:1971vv}
\bibitem{Tsai:1971vv}
  Y.~S.~Tsai,
  %``Decay Correlations Of Heavy Leptons In E+ E- $\to$ Lepton+ Lepton-,''
  Phys.\ Rev.\  D {\bf 4} (1971) 2821
  [Erratum-ibid.\  D {\bf 13} (1976) 771].
  %%CITATION = PHRVA,D4,2821;%%


\bibitem{Broadhurst}
  D.~J.~Broadhurst,
  %``Large N expansion of QED: Asymptotic photon propagator and contributions to
  %the muon anomaly, for any number of loops,''
  Z.\ Phys.\  C {\bf 58}, 339 (1993).
  %%CITATION = ZEPYA,C58,339;%%

\bibitem{Beneke}
  M.~Beneke,
  %``Renormalization scheme invariant large order perturbation theory and
  %infrared renormalons in QCD,''
  Phys.\ Lett.\  B {\bf 307}, 154 (1993);
  %%CITATION = PHLTA,B307,154;%%
  %``Large order perturbation theory for a physical quantity,''
  Nucl.\ Phys.\  B {\bf 405}, 424 (1993).
  %%CITATION = NUPHA,B405,424;%%

\bibitem{Neubert2}
M.~Neubert, hep-ph/9502264.
  %%CITATION = HEP-PH 9502264;%%

\bibitem{LV}
%\bibitem{GLZN}
S.~G.~Gorishny and S.~A.~Larin, Phys. Lett. B {\bf 172}, 109 (1986);
  %%CITATION = PHLTA,B172,109;%%
E.~B.~Zijlstra and W.~Van Neerven, Phys. Lett. B {\bf 297}, 377 (1992);
  %%CITATION = PHLTA,B297,377;%%
%\bibitem{LV}
S.~A.~Larin and J.~A.~M. Vermaseren, Phys. Lett. B {\bf 259}, 
345 (1991).
  %%CITATION = PHLTA,B259,345;%%

\bibitem{KS}
%\bibitem{Kataev:1995vh}
  A.~L.~Kataev and V.~V.~Starshenko,
%   ``Estimates Of The Higher Order QCD Corrections To R(S), R(Tau) And Deep
  %Inelastic Scattering Sum Rules,''
  Mod.\ Phys.\ Lett.\ A {\bf 10}, 235 (1995)
  [hep-ph/9502348].
  %%CITATION = HEP-PH 9502348;%%

\bibitem{d1}
%\cite{Chetyrkin:1979bj}
%\bibitem{Chetyrkin:1979bj}
  K.~G.~Chetyrkin, A.~L.~Kataev and F.~V.~Tkachov,
%``Higher Order Corrections To Sigma-T (E+ E- $\to$ Hadrons) 
% In Quantum Chromodynamics,''
  Phys.\ Lett.\ B {\bf 85}, 277 (1979);
  %%CITATION = PHLTA,B85,277;%%
%\cite{Dine:1979qh}
%\bibitem{Dine:1979qh}
  M.~Dine and J.~R.~Sapirstein,
  %``Higher Order QCD Corrections In E+ E- Annihilation,''
  Phys.\ Rev.\ Lett.\  {\bf 43}, 668 (1979);
  %%CITATION = PRLTA,43,668;%%
%\cite{Celmaster:1979xr}
%\bibitem{Celmaster:1979xr}
  W.~Celmaster and R.~J.~Gonsalves,
%``An Analytic Calculation Of Higher Order Quantum Chromodynamic 
%Corrections In E+ E- Annihilation,''
  Phys.\ Rev.\ Lett.\  {\bf 44}, 560 (1980).
  %%CITATION = PRLTA,44,560;%%

\bibitem{d2}
%\cite{Gorishnii:1990vf}
%\bibitem{Gorishnii:1990vf}
  S.~G.~Gorishnii, A.~L.~Kataev and S.~A.~Larin,
%   ``The O (alpha-s**3) corrections to 
%sigma-tot (e+ e- $\to$ hadrons) and Gamma
  %(tau- $\to$ tau-neutrino + hadrons) in QCD,''
  Phys.\ Lett.\ B {\bf 259}, 144 (1991);
  %%CITATION = PHLTA,B259,144;%%
%\cite{Surguladze:1990tg}
%\bibitem{Surguladze:1990tg}
  L.~R.~Surguladze and M.~A.~Samuel,
%   ``Total Hadronic Cross-Section In E+ E- Annihilation 
%At The Four Loop Level Of Perturbative QCD,''
  Phys.\ Rev.\ Lett.\  {\bf 66}, 560 (1991)
  [Erratum-ibid.\  {\bf 66}, 2416 (1991)].
  %%CITATION = PRLTA,66,560;%%

\bibitem{d3}
  P.~A.~Baikov, K.~G.~Chetyrkin and J.~H.~K\"uhn,
  %``Order $\alpha^4_s$ QCD Corrections to $Z$ and $\tau$ Decays,''
  Phys.\ Rev.\ Lett.\  {\bf 101}, 012002 (2008)
  [arXiv:0801.1821 [hep-ph]].
  %%CITATION = PRLTA,101,012002;%%

\end{thebibliography}
\end{document}